\documentclass[12pt]{article}

\usepackage{amsfonts}
\usepackage{amsmath}
\usepackage{amssymb}
\usepackage{epsfig}
\usepackage{latexsym}
\usepackage{graphicx}
\usepackage{epstopdf}
\usepackage{color}
\usepackage{amssymb}
\usepackage{framed}
\usepackage{braketmod}

\numberwithin{equation}{section}
\allowdisplaybreaks

\def\spa#1{\phantom{\fbox{\rule[-#1cm]{0cm}{0cm}}}}

\def\be{\begin{equation}}
\def\ee{\end{equation}}
\def\bea{\begin{eqnarray}}
\def\eea{\end{eqnarray}}
\def\bequ{\begin{equation}}
\def\eequ{\end{equation}}

\renewcommand{\thefootnote}{\fnsymbol{footnote}}

\newcommand{\eq} {equation}
\newcommand{\eqa} {eqnarray}
\newcommand{\NN} {\nonumber}

\global\long\def\bra#1{\Bra{#1}}%
\global\long\def\bbra#1{\Bbra{#1}}%
\global\long\def\ket#1{\Ket{#1}}%
\global\long\def\kket#1{\Kket{#1}}%
\global\long\def\braket#1{\Braket{#1}}%
\global\long\def\bbraket#1{\Bbraket{#1}}%
\global\long\def\brakket#1{\Brakket{#1}}%
\global\long\def\bbrakket#1{\Bbrakket{#1}}%
\global\long\def\Left{\mathrm{L}}%
\global\long\def\Right{\mathrm{R}}%
\global\long\def\HW{\mathrm{HW}}%
\global\long\def\FIL{\mathrm{FIL}}%

\begin{document}
\hfuzz=100pt
\title{\LARGE
Non-perturbative Tests of Duality Cascades\\ 
in Three Dimensional Supersymmetric Gauge Theories }
\author{
 Masazumi Honda\footnote{masazumi.honda(at)yukawa.kyoto-u.ac.jp}, \quad
 Naotaka Kubo\footnote{naotaka.kubo(at)yukawa.kyoto-u.ac.jp} 
  \spa{0.5} \\
\\
{\small{\it Center for Gravitational Physics, Yukawa Institute for Theoretical Physics,}}\\ 
{\small{\it Kyoto University, Sakyo-ku, Kyoto 606-8502, Japan}} \\
}
\date{\small{October 2020}}
\maketitle
\thispagestyle{empty}
\centerline{}

\begin{abstract}
It has been conjectured that
duality cascade occurs 
in the $\mathcal{N}=3$ supersymmetric Yang-Mills Chern-Simons theory 
with the gauge group $U(N )_k \times U(N+M )_{-k}$
coupled to two bi-fundamental hypermultiplets.
The brane picture suggests that
this duality cascade can be generalized to a class of 3d $\mathcal{N}=3$ supersymmetric quiver gauge theories coming from so-called Hanany-Witten type brane configurations.
In this paper
we perform non-perturbative tests of the duality cascades using supersymmetry localization.
We focus on $S^3$ partition functions and 
prove predictions from the duality cascades.
We also discuss that
our result can be applied to generate new dualities for more general theories which include less supersymmetric theories and theories without brane constructions.
\end{abstract}
\vfill
\noindent
YITP-20-128

\renewcommand{\thefootnote}{\arabic{footnote}}
\setcounter{footnote}{0}

\newpage
\setcounter{page}{1}
\tableofcontents

\section{Introduction}
Duality plays a prominent role in theoretical physics.
It relates apparently different theories in nontrivial ways.
One of the most famous dualities for gauge theories is Seiberg duality \cite{Seiberg:1994pq},
which says that two 4d supersymmetric (SUSY) QCD's with different ranks flow to the same theory in IR.
There are also various other dualities in diverse dimensions.
While duality in a narrow sense is a relation between two theories, 
there is a more exotic phenomenon called {\it duality cascade} for a class of gauge theories.
It was first discovered by Klebanov and Strassler \cite{Klebanov:2000hb}
in the 4d $\mathcal{N}=1$ $SU(N)\times SU(N+M)$ SUSY gauge theory
with (anti-)bi-fundamental chiral multiplets.
It has been discussed in \cite{Klebanov:2000hb} that along the RG flow, 
the theory flows to another theory with $N\rightarrow N-M$, and
repeating this gives rise to the sequence of the Seiberg dualities 
(see \cite{Strassler:2005qs} for a review).
The gravity dual of the Klebanov-Strassler theory has been also useful
to construct models of string inflation \cite{Giddings:2001yu,Kachru:2003aw,Kachru:2003sx}.

In this paper we mainly study duality cascades 
in a class of 3d $\mathcal{N}= 3$ SUSY Yang-Mills Chern-Simons (YMCS) theories.
The simplest case is  the $\mathcal{N}=3$ SUSY YMCS theory 
with the gauge group $U(N )_k \times U(N+M )_{-k}$ coupled to two bi-fundamental hypermultiplets. 
This theory with $|M | \leq |k|$
is known to flow 
to the famous $\mathcal{N}=6$ superconformal Chern-Simons theory 
called the ABJ theory with the same gauge group\footnote{
It has $\mathcal{N}=8$ SUSY for some special cases: $(k,M)=(1,0), (2,0)$ and $(1,1)$.
} \cite{Aharony:2008ug,Aharony:2008gk}.
It is expected that
the ABJ theory enjoys the Seiberg-like duality \cite{Aharony:2008gk,Giveon:2008zn}.
For $M \geq 0$,
it is the duality 
between the theories with the gauge groups\footnote{
The opposite case $M \leq 0$ is simply obtained by $k\rightarrow -k$.
}
\begin{\eq}
U(N)_k \times U(N+M)_{-k}\quad {\rm and} \quad U(N+|k|-M)_{k} \times U(N)_{-k} .
\label{eq:Seiberg}
\end{\eq}
We can easily see that 
acting the duality transformation twice, we come back to the original theory.
This duality essentially comes from the Hanany-Witten effect \cite{Hanany:1996ie} of the brane configuration and 
has been tested in various ways \cite{Aharony:2008gk,Kapustin:2010mh,Awata:2012jb,Honda:2013pea,Matsumoto:2013nya,Honda:2014npa,Hirano:2014bia}.

It has been expected\footnote{
It was argued in \cite{Aharony:2008gk} that the ABJ theory, which is the superconformal theory, with $M>|k|$ does not exist as a unitary theory.
However, this does not necessarily mean that the YMCS theory with $M>|k|$ never has a supersymmetric vacua.
In \cite{Aharony:2009fc,Evslin:2009pk}, the authors predicted that the duality cascade occurs, and consequently, the SUSY breaking condition is different from $M>|k|$.
We will review this point below and in sec.~\ref{sec:review}.
} in \cite{Aharony:2009fc,Evslin:2009pk}
that the duality (\ref{eq:Seiberg}) is extended to the case 
with $M>|k|$.
In this case, 
the big difference from the previous case is that
the lower rank of the two unitary gauge groups changes:
it decreases from $N$ to $N+|k|-M$.
For this case, acting the duality transformation once more does not get back to the original theory and
we go to another dual theory.
This implies that 
the theory enjoys the following sequence of the dualities,
i.e.~the duality cascade as in the 4d case:
\begin{\eq}
 (N,M) 
\rightarrow (N^{(1)} ,M^{(1)} ) \rightarrow (N^{(2)} ,M^{(2)} )
\rightarrow (N^{(3)} ,M^{(3)} )  \rightarrow \cdots ,
\label{eq:CascEx1}
\end{\eq}
where $N^{(j)}$ and $M^{(j)}\geq 0$ denote the lower rank and the difference of ranks of the two unitary gauge groups in the theory obtained by applying the duality transformation $j$ times\footnote{
We have omitted the label of Chern-Simons level for simplicity of explanation.
In the main text, 
we will explicitly write it.
}.
The values of $(N^{(j)} ,M^{(j)})$ are explicitly given by
\begin{\eq}
N^{(j)} = N^{(j-1)}+|k|-M^{(j-1)},\quad 
M^{(j)} = M^{(j-1)}-|k| \quad
{\rm for }\ M^{(j-1)} > |k|, 
\end{\eq}
with the initial condition
\begin{\eq}
(N^{(0)} ,M^{(0)}) =(N,M).
\end{\eq}
This sequence of the dualities last until we encounter
\begin{\eq}
N^{(n)} <M^{(n)}-|k| ,\quad {\rm or}\quad M^{(n)} \leq |k| .
\end{\eq}
When $N^{(n)} <M^{(n)}-|k|$, the brane interpretation implies SUSY breaking.
This is modified version of s-rule 
as we will review.
When $M^{(n)} \leq |k|$, the $n$th theory corresponds to the previous case, thus the theory finally flows to the ABJ theory with the gauge group $U(N^{(n)})_{k} \times U(N^{(n)} +M^{(n)})_{-k}$.
Since it
is the IR duality, 
all of the theories (\ref{eq:CascEx1}) should finally flow to the same ABJ theory.

The above duality cascade can be generalized to  
more general $\mathcal{N}=3$ SUSY theories coming 
from the Hanany-Witten brane configurations on a circle \cite{Hanany:1996ie}
although it has not been discussed explicitly in literature.
This class includes $\mathcal{N}=3$ YMCS theories with circular quivers (see e.g.~\cite{Gaiotto:2008sd,Hosomichi:2008jd,Imamura:2008dt}).
We discuss this generalization from viewpoints of 
brane constructions
in sec.~\ref{sec:generalization}.

In this paper,
we 
perform a non-perturbative test of the duality cascades
(and SUSY breaking)
using SUSY localization \cite{Pestun:2007rz}.
We focus on $S^3$ partition functions and prove that 
the problem of showing the duality cascades 
is reduced to show
a duality relation coming 
from a worldvolume theory of a building block of Hanany-Witten type brane configurations which we refer to as the local theory.  
Then we prove the fundamental duality relation for the local theory and
this amounts to prove the predictions from the duality cascades for the $S^3$ partition functions.
This provides the strong evidence that 
the conjectured duality cascades are indeed true
(although it would be also important to perform further checks by other observables). 
We also discuss that
our result can be applied to generate new dualities for more general theories which include less SUSY theories and theories without brane constructions.

This paper is organized as follows.
In sec.~\ref{sec:review} we review the duality cascades
and SUSY breaking
from the viewpoint of the brane constructions.
We also discuss that the duality cascade can be easily extended to 
more general $\mathcal{N}=3$ SUSY theories coming from the Hanany-Witten brane configurations on a circle.
In sec.~\ref{sec:test} we show that
the problem of showing the duality cascades and SUSY breaking for the $S^3$ partition functions 
is reduced to show a duality relation for the local theory.  
We then prove the the duality relation.
In sec.~\ref{sec:gluing} we discuss that 
our result can be used to generate new dualities.
Sec.~\ref{sec:conclusion} is devoted to conclusion and discussions.

\section{Review of the duality cascades in three dimensions}
\label{sec:review}

After the discovery of the Seiberg duality, 
various dualities which have natural interpretations from string theory have been found
(see e.g.~\cite{Hanany:1996ie,Elitzur:1997hc,Giveon:2008zn}).
It was pointed out in \cite{Aharony:2009fc,Evslin:2009pk} that the duality cascade for the ABJ theory also can be captured from the viewpoint of the brane constructions, especially using the Hanany-Witten effect.
Therefore, we start in sec.~\ref{subsec:setup} with explaining the brane configurations and their worldvolume theories.
We then review the Hanany-Witten effect in our setup in sec.~\ref{subsec:HWeffect}.
We also discuss a condition of the SUSY breaking known as the (modified) s-rule by using the Hanany-Witten effect.
Finally, in sec.~\ref{sec:cascade} we first review
the duality cascade (and SUSY breaking) for the ABJ theory from the viewpoint of the brane constructions.
We see that this viewpoint naturally leads to the duality cascade for general 3d SUSY gauge theories.

\subsection{Brane configuration and gauge theory} 
\label{subsec:setup}
\begin{figure}[t]
\begin{center}
  \begin{tabular}{|c||c|c|c|c|c|c|c|c|c|c|}
\hline      & 0          & 1          & 2           & 3         & 4        & 5          & 6 ($S^1 $) & 7 &8 & 9\\ \hline\hline
 D3-branes  &  $\circ$ & $\circ$ & $\circ$  &             &          &            & $\circ$    &     &   &   \\ \hline
  5-branes& $\circ$ & $\circ$ & $\circ$  & $\slash_{37}$ & $\slash_{48}$ & $\slash_{59}$&              & $\slash_{37}$ & $\slash_{48}$  & $\slash_{59}$ \\ \hline\hline
  \end{tabular}
\caption{
The ingredients of the type IIB brane configuration studied in this paper.
The 6th direction is taken to be a small circle.
The symbol ``$\slash_{ij}$'' for 5-branes means that
the 5-branes are oblique in the $\left(x_i,x_j\right)$ plane with appropriate angles dependent on $(\ell ,k)$.
}
\label{fig:brane}
\end{center}
\end{figure}
\begin{figure}[t]
\begin{center}
\includegraphics[width=90mm]{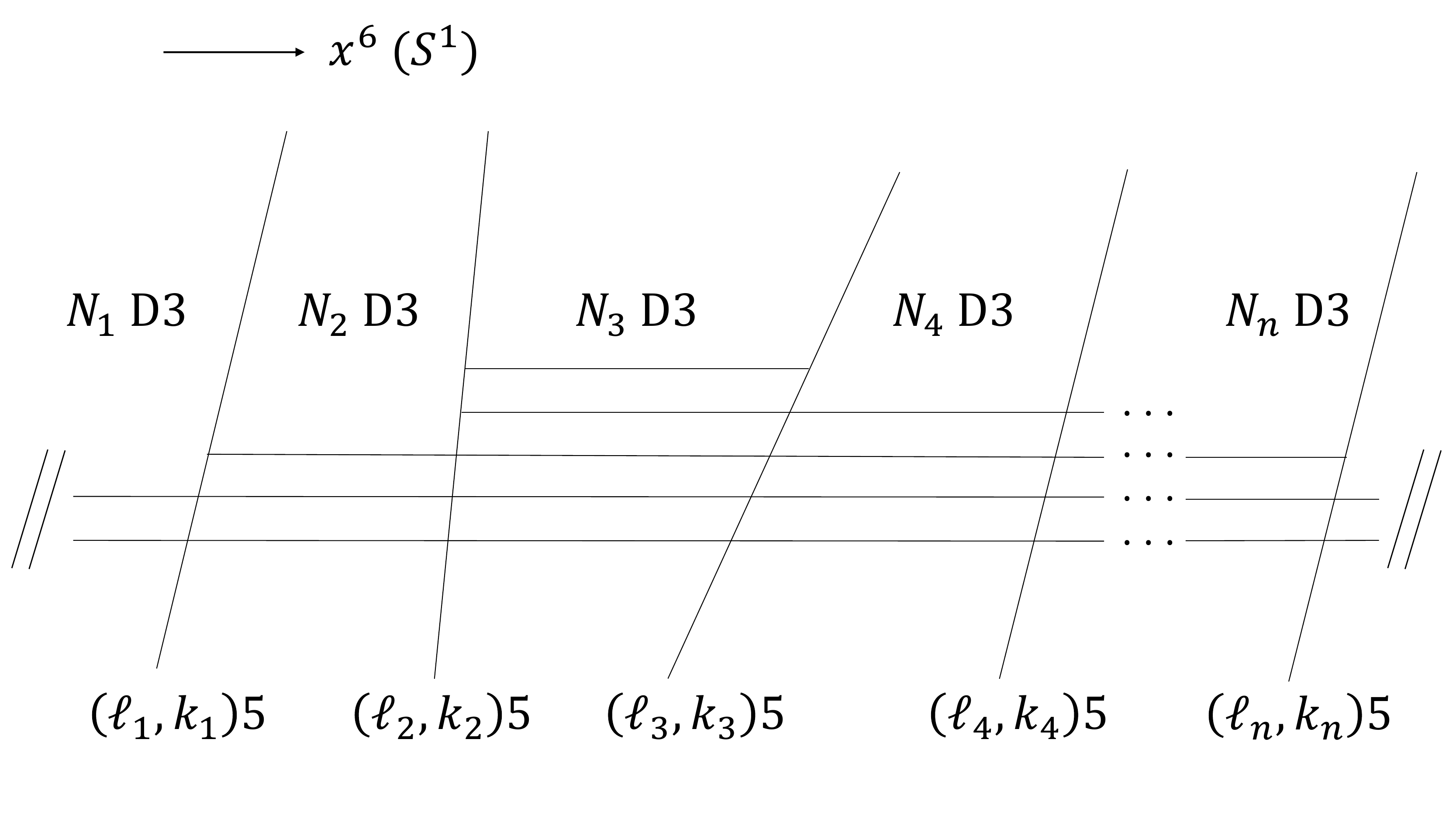}
\caption{
A generic Hanany-Witten type brane configuration.
The horizontal direction is the $S^1$ and therefore identified on the both ends.
When $\ell$'s are 1, 
the worldvolume theory is a circular quiver gauge theory
with the gauge group $U(N_1 )_{k_1 -k_n} \times U(N_2 )_{k_2 -k_1} \times \cdots \times U(N_n )_{k_n -k_{n-1}}$.
}
\label{fig:generic}
 \end{center}
\end{figure}

Let us consider a class of SUSY theories 
given by the Hanany-Witten type brane configurations \cite{Hanany:1996ie,Gaiotto:2008sd,Imamura:2008dt}
which are brane configurations in the type IIB superstring theory shown in fig.~\ref{fig:brane} and \ref{fig:generic}.
It is known that
this type of brane configurations preserves $\mathcal{N}=3$ SUSY at least.
Ingredients and their field theory interpretations are as follows:
\begin{itemize}
\item D3-brane \\
It gives vector multiplets with gauge group $U(N)$
when the number of D3-branes between adjacent two 5-branes is $N$.
Since one of the longitudinal direction is the small $S^1$,
the world volume theory is effectively three dimensional.

\item D5-brane\\
It gives $N_f$ fundamental hypermultiplets
when the number of D5-branes is $N_f$.

\item NS5-brane\\
It gives bi-fundamental hypermultiples
under the two gauge groups coming from neighbor D3-branes.

\item $(\ell ,k)$ 5-brane\\
It is a bound state of $\ell$ NS5-branes and $k$ D5-branes with  ${\rm gcd}(\ell ,k )=1$.
$(0,1)$ brane denotes single D5-brane while $(1,0)$ brane is single NS5-brane.
$(1,k)$ 5-brane gives bi-fundamental hypermultiples as NS 5-brane
and Chern-Simons term with Chern-Simons level $+k$ to left ($-k$ to right) gauge group.
When $(\ell ,k)$ is neither $(0,1)$, $(1,k)$ nor $(1 ,0)$,
it gives non-Lagrangian theory
meaning that we do not currently know its Lagrangian.

\end{itemize}

\begin{figure}[t]
\begin{center}
\includegraphics[width=75mm]{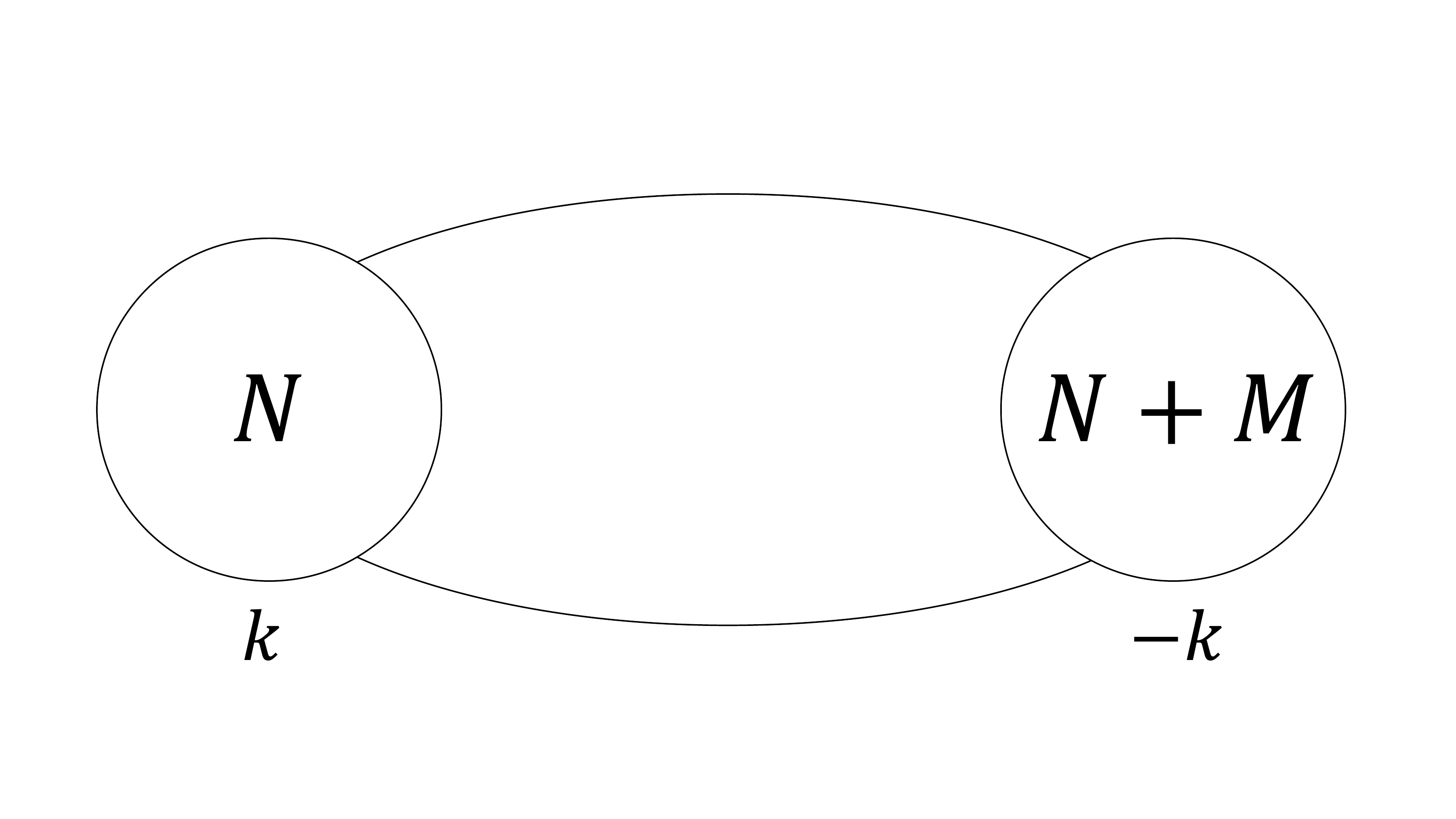}
\includegraphics[width=75mm]{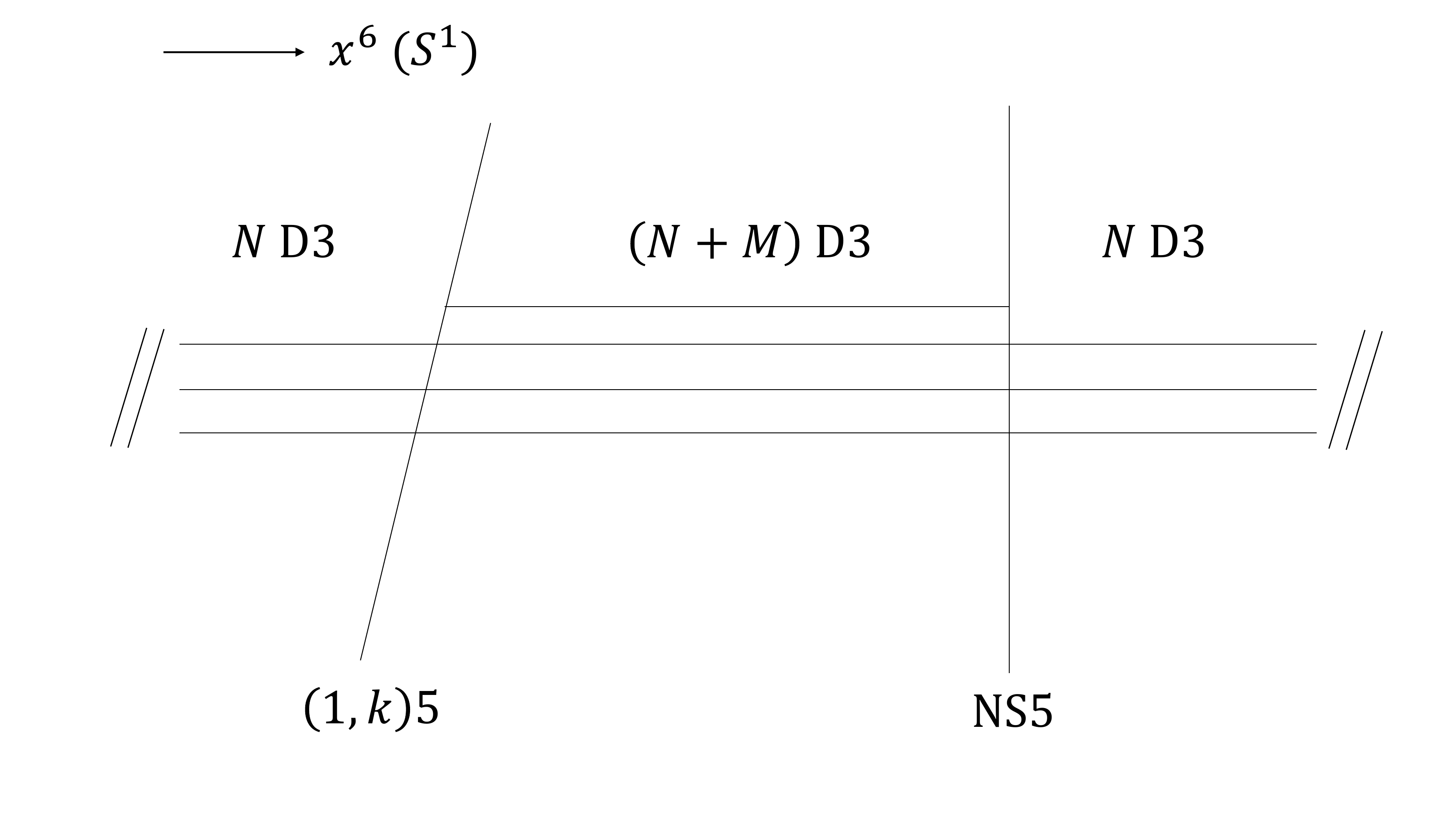}
\caption{
[Left] The quiver diagram of the $\mathcal{N}=3$ $U(N)_k \times U(N+M)_{-k}$ SUSY YMCS theory
coupled to two bi-fundamental hypermultiplets.
This theory flows to the ABJ theory for $|k| \geq M$.
[Right] The corresponding type IIB brane configuration 
($N=3$ and $M=1$ in the figure). 
The horizontal direction is the $S^1$ and therefore identified on the both ends.
}
\label{fig:ABJ}
 \end{center}
\end{figure}

As a simple example, fig.~\ref{fig:ABJ} shows the type IIB brane configuration
for the $\mathcal{N}=3$ $U(N)_k \times U(N+M)_{-k}$ SUSY YMCS theory with two bi-fundamental hypermultiplets,
which flows to the ABJ theory for $|k|\geq M$.
Fig.~\ref{fig:generic} shows a generic brane configuration in the class.
As we have mentioned,
the Chern-Simons levels in each gauge group are determined by differences 
between  ``$k$''s of the adjacent $(1 ,k)$-type 5-branes.

\subsection{Duality and Hanany-Witten effect}
\label{subsec:HWeffect}
It is known that
the Seiberg-like dualities 
in the worldvolume theories of the Hanany-Witten type brane configurations
can be understood from the so-called Hanany-Witten effect \cite{Hanany:1996ie,Kitao:1998mf}.
The Hanany-Witten effect tells us that
if two different types of 5-branes pass through each other, 
then it creates new D3-branes stretched between them. 
More precisely,
when 
$\left(\ell,k\right)$5 and $\left(\ell',k'\right)$5-branes 
with $\ell k'-\ell'k\neq0$ 
pass through each other,
\[
\left|\ell k'-\ell'k\right|\ {\rm D3-branes}
\]
are created. 
In addition, this move reverses the orientation of D3-branes that were initially stretched and thus change them to anti-D3-branes. 
If there are more D3-branes than the anti-D3-branes,
then the anti-D3-branes are annihilated with them and completely vanish. 
On the other hand, 
if the anti-D3-branes remain, then supersymmetry is broken. 
We will elaborate
this point shortly.

\begin{figure}[t]
\begin{center}
\includegraphics[width=75mm]{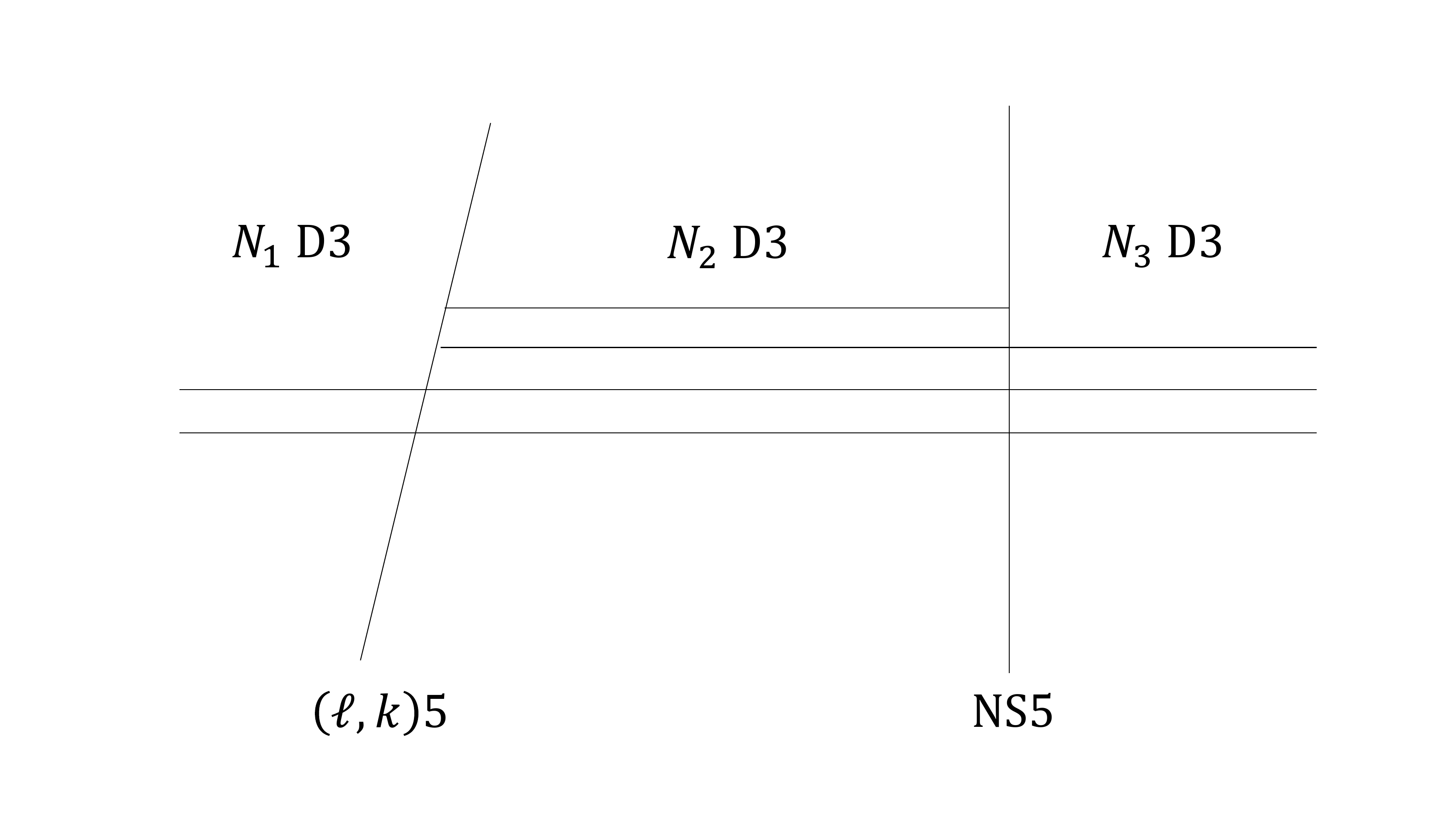}
\includegraphics[width=75mm]{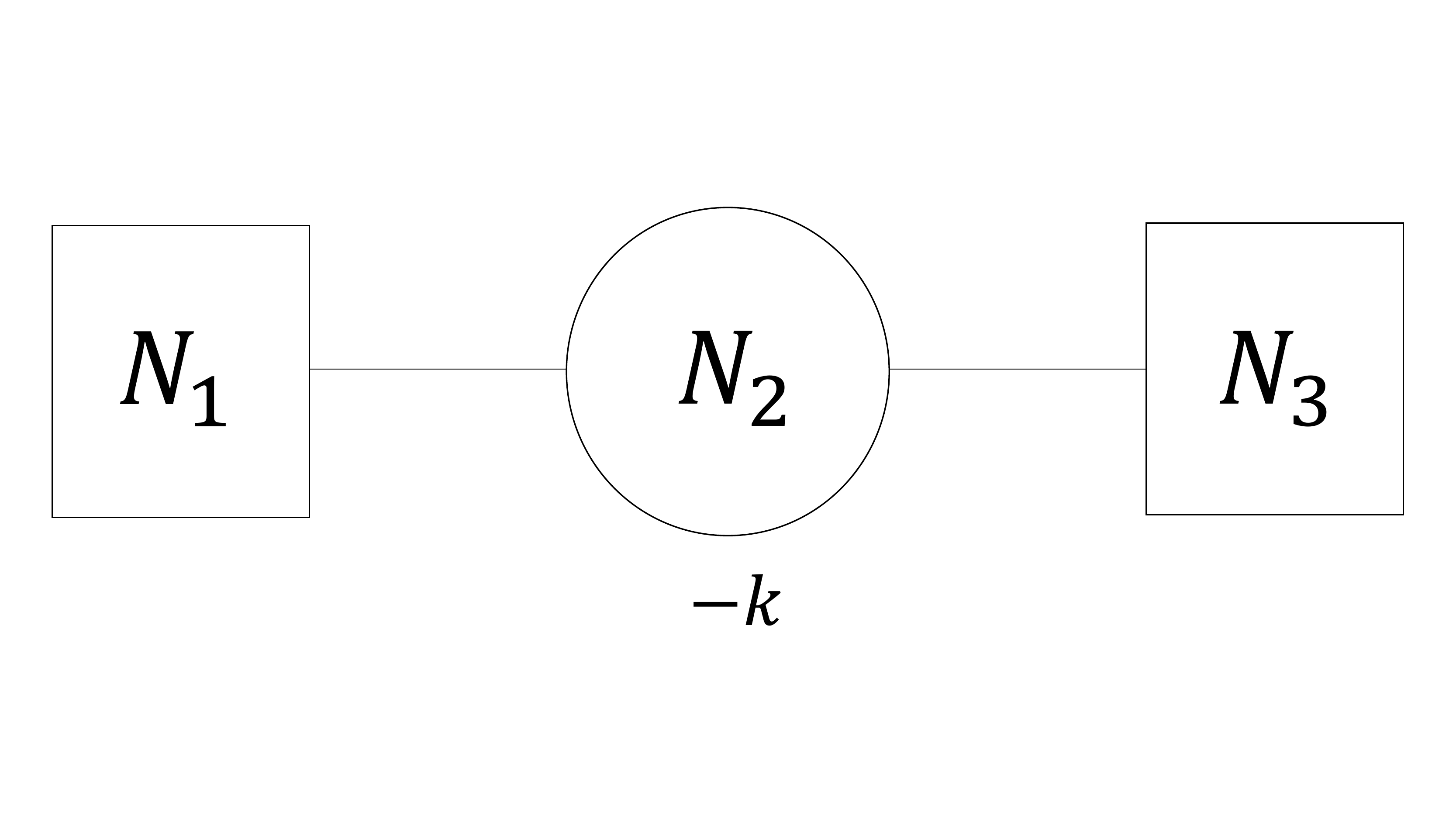}
\caption{
[Left] The brane configuration of the local theory.
Note that the horizontal direction is not identified on the both ends unlike fig.~\ref{fig:ABJ}.
[Right] The quiver diagram for $\ell =1$.
For $\ell >1$, it is non-Lagrangian theory.
}
\label{fig:local}
 \end{center}
\end{figure}
\begin{figure}[t]
\begin{center}
\includegraphics[width=75mm]{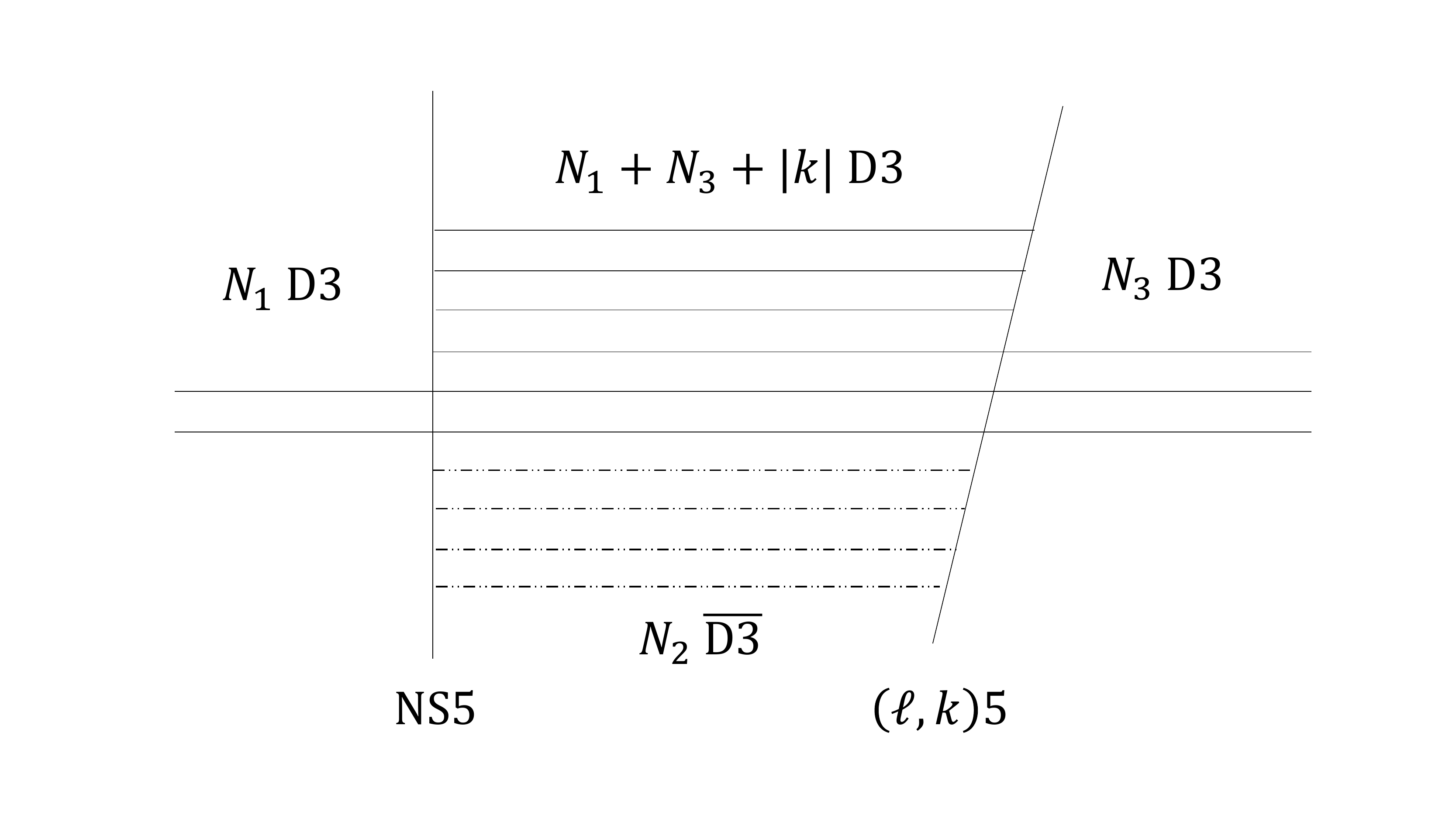}
\includegraphics[width=75mm]{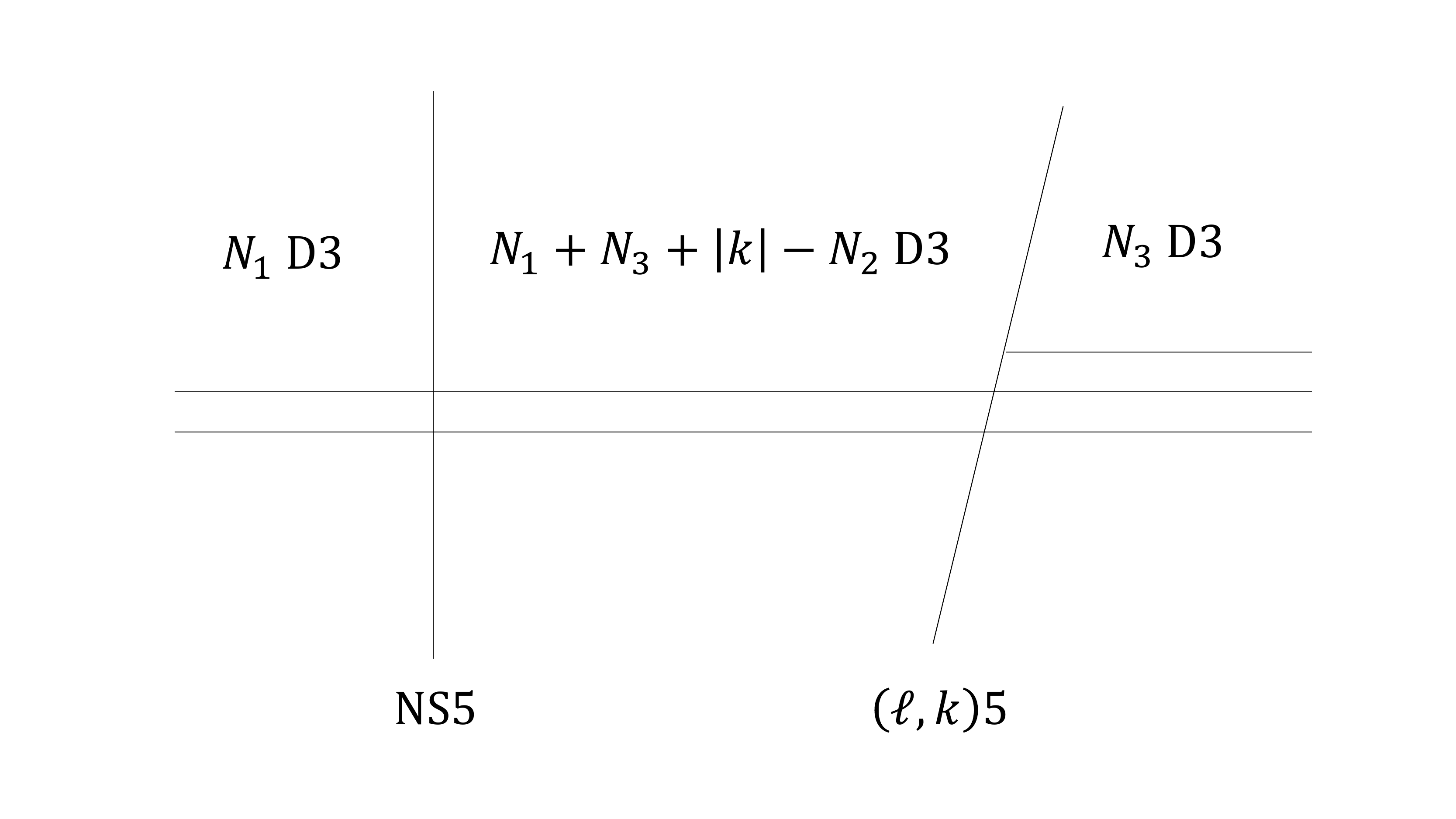}\\
\includegraphics[width=75mm]{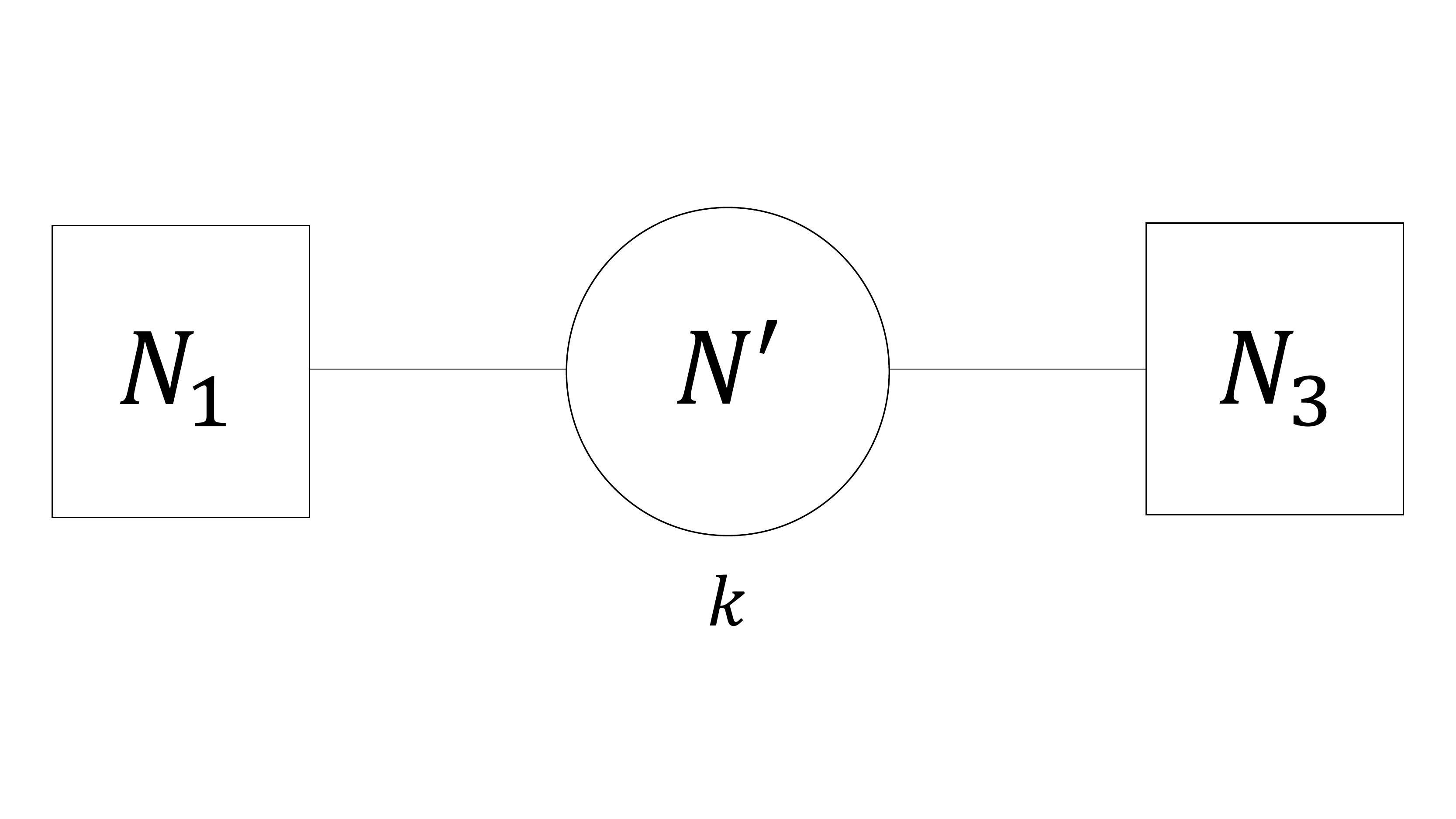}
\caption{
An illustration of The Hanany-Witten move for the local theory ($N_1 =2$, $N_2 =4$, $N_3 =3$, $k=1$ in this figure).
[Left] $N_2$ D3-branes become anti-D3-branes (dotted lines) and $|k|$ D3-branes are created.
[Right] The $N_2$ pairs of D3 and anti-D3-branes are annihilated.
[Center] The quiver diagram of the dual theory for $\ell =1$
with $N' =N_1 +N_3 +|k| -N_2$.
}
\label{fig:HW}
 \end{center}
\end{figure}

As an important example, let us consider the brane configuration
shown in fig.~\ref{fig:local},
which we denote as
\begin{equation}
\left\langle N_{1}\bullet N_{2}\circ N_{3}\right\rangle ,
\end{equation}
where $\bullet$ and  $\circ$ denote $\left(\ell ,k\right)$5-brane
and an NS5-brane respectively
while the numbers represent the ones of D3-branes.
In this setup, the Hanany-Witten effect tells us that
$\left|\ell \times 0-1\times k\right|$ $=$ $\left|k\right|$ D3-branes are created
after the move of the 5-branes
as illustrated in fig.~\ref{fig:HW}.
Therefore, 
taking annihilation of anti-D3-branes into account,
the brane configuration changes as 
\begin{align}
\left\langle N_{1}\bullet N_{2}\circ N_{3}\right\rangle  & \rightarrow\left\langle N_{1}\circ N_{1}+N_{3}-N_{2}+\left|k\right|\bullet N_{3}\right\rangle .
\label{eq:HW}
\end{align}
Similarly, 
if we start with the configuration $\left\langle N_{1}\circ N_{2}\bullet N_{3}\right\rangle$, then we have
\begin{align}
\left\langle N_{1}\circ N_{2}\bullet N_{3}\right\rangle  & \rightarrow\left\langle N_{1}\bullet N_{1}+N_{3}-N_{2}+\left|k\right|\circ N_{3}\right\rangle .
\label{eq:HW2}
\end{align}
Note that although we displayed the Hanany-Witten move using the arrow, 
the move is reversible. 
In other words, the combination of (\ref{eq:HW}) and (\ref{eq:HW2}) result 
in the original brane configuration.

\subsubsection{Supersymmetry breaking and s-rule}
\label{sRule}
So far we have focused on the brane configurations preserving supersymmetry.
However, as we have briefly mentioned,
there are sometimes cases where supersymmetry is broken
since some D3-branes have boundaries.
The condition whether or not the supersymmetry is broken
depends on the number of D3-branes 
stretched between 5-branes.
The Hanany-Witten transition can be used to determine whether the supersymmetry breaking occurs or not.
In particular,
$\mathcal{N}=3$ SUSY in our brane configurations is completely broken 
when there remain anti-D3-branes after the Hanany-Witten moves.
This leads to the so-called s-rule \cite{Hanany:1996ie,Bergman:1998ej,Bergman:1999na}:
if a brane configuration is related to another one including anti-D3-branes 
by 
the Hanany-Witten moves, 
neither configurations are supersymmetric.

Note that, when D3-branes do not have any compact directions, 
sometimes the s-rule becomes simpler.
For example, the Hanany-Witten effect relates $\left\langle 0\circ M\bullet0\right\rangle$ and $\left\langle 0\bullet |k|-M\circ0\right\rangle$. 
Therefore, the s-rule says that the number of D3-branes stretched between the NS5-brane and  $\left(\ell,k\right)$5-brane 
should be less than or equal to $k$ to preserve the supersymmetry.
On the other hand, when the D3-branes have compact directions, 
such a rephrasing is not easy.
For example, 
the case $\left\langle \circ~3|k|\bullet |k| \right\rangle$ with the periodic identification, namely the ABJ case, 
cannot be related to any configurations including anti-D3-branes though the number of D3-branes stretched between the NS5-brane and the $\left(\ell,k\right)$5-brane is $2|k|$.
In this paper, therefore, we do not paraphrase it further.
The s-rule in the case when compact directions exist is also called modified s-rule
and studied, for example, in \cite{Dasgupta:1999wx,Aharony:2009fc,Evslin:2009pk}.

\subsection{The duality cascades}
\label{sec:cascade}
\begin{figure}[t]
\begin{center}
\includegraphics[width=75mm]{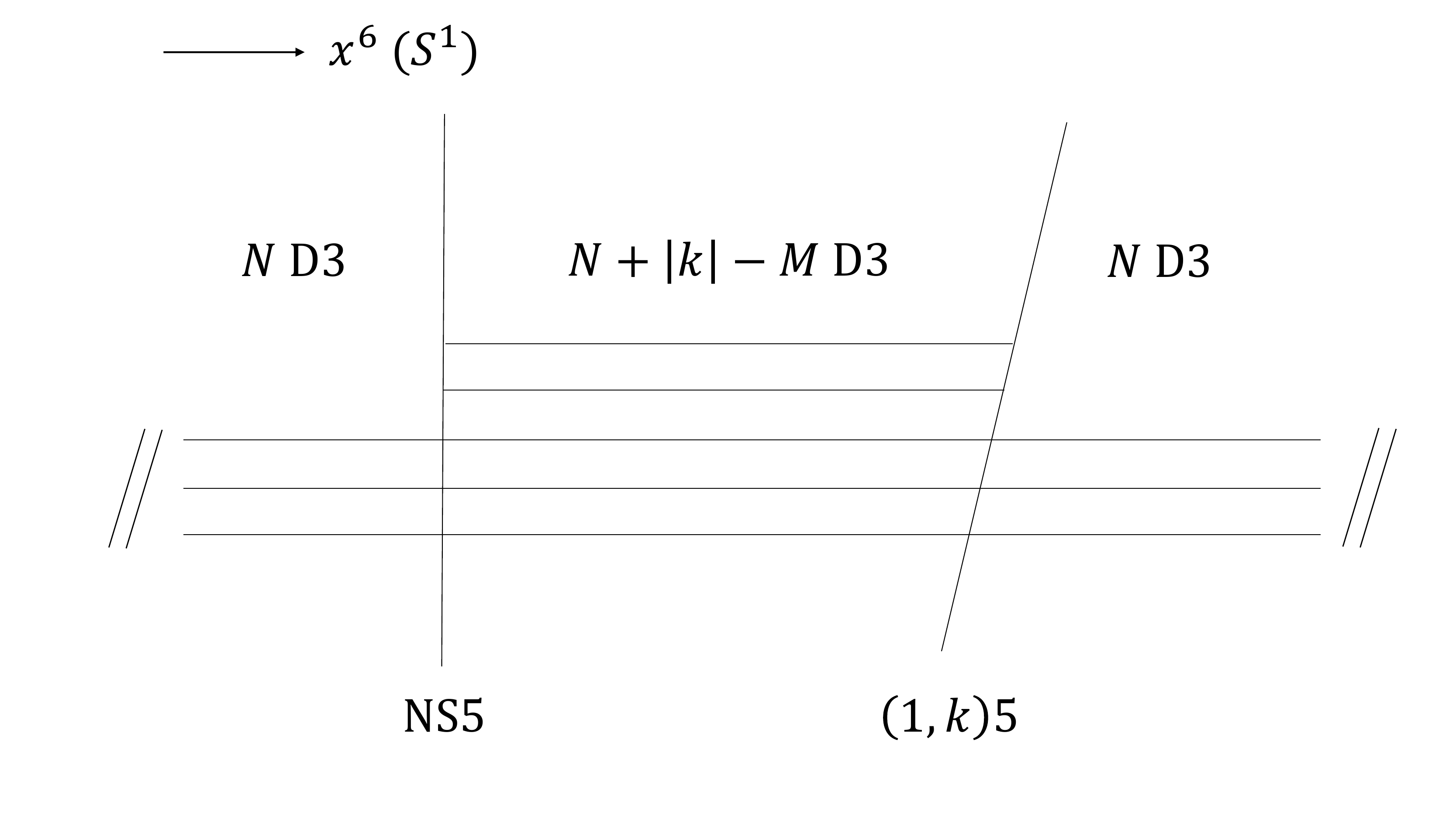}
\caption{
The brane configuration after applying the Hanany-Witten move 
to the brane configuration of fig.~\ref{fig:ABJ},
which corresponds to the $\mathcal{N}=3$ $U(N)_k \times U(N+M)_{-k}$ theory with $|k|\geq M$.
($N=3$, $M=1$, $k=3$ in this figure).
This implies the duality 
with the $\mathcal{N}=3$ $U(N+|k|-M)_{k} \times U(N)_{-k}$ theory.
}
\label{fig:ABJ2}
 \end{center}
\end{figure}

In this section we see how the duality cascade is motivated by the brane dynamics.
For simplicity of explanations,
we start with the $\mathcal{N}=3$ $U(N)_k \times U(N+M)_{-k}$ SUSY YMCS theory
shown in fig.~\ref{fig:ABJ}.
It is known that 
the theory for $|k| \geq M$ flows to the $\mathcal{N}=6$ superconformal theory called ABJ theory.
It is expected that
the ABJ theory enjoys the Seiberg-like duality \eqref{eq:Seiberg}
between the theories with the gauge groups
\[
U(N)_k \times U(N+M)_{-k}\quad {\rm and} \quad U(N+|k|-M)_{k} \times U(N)_{-k} ,
\]
The duality can be understood from the Hanany-Witten effect
as illustrated in fig.~\ref{fig:ABJ2}.
If we apply the Hanany-Witten move to the brane configuration in fig.~\ref{fig:ABJ},
then we obtain the brane configuration shown in fig.~\ref{fig:ABJ2},
which gives the $U(N+|k|-M)_{k} \times U(N)_{-k}$ theory.
If we parameterize the $U(N)_k \times U(N+M)_{-k}$ gauge theory or the corresponding brane configuration shown in fig.~\ref{fig:ABJ} as $(N,M,k)$ 
(more precisely, $N$ is the lowest rank, $M>0$ is the difference of ranks, and $k$ is the Chern-Simons level corresponding to the lowest rank),
the duality can be expressed as
\begin{\eq}
{\rm Duality}: (N,M,k) \rightarrow (N, |k|-M ,-k ) \quad {\rm for}\ M\leq |k| .
\label{eq:dual0}
\end{\eq}
We can easily see that 
acting the duality transformation (\ref{eq:dual0}) twice, we come back to the original theory.
This duality has been tested in various ways \cite{Aharony:2008gk,Kapustin:2010mh}.

What happens if we take $M>|k|$ \cite{Aharony:2009fc,Evslin:2009pk}?
Let us again consider the Hanany-Witten effect\footnote{
The Hanany-Witten move considered in (\ref{eq:dual0}) and (\ref{eq:dual1}) is tacitly for the sector where the number of D3-branes are largest, in other words, the D3-branes stretched between 5-branes exist.
}.
In this case, after the Hanany-Witten move, the lowest rank becomes not $N$ but $N+|k|-M$.
Therefore, the duality in the brane configuration is
\begin{\eq}
{\rm Duality}: (N,M,k) 
\rightarrow 
(N^{(1)} ,M^{(1)} ,k ) \quad {\rm for}\ M> |k|,
\label{eq:dual1}
\end{\eq}
where 
\begin{\eq}
N^{(1)} = N+|k|-M ,\quad M^{(1)} =M-|k|.
\end{\eq}
The duality (\ref{eq:dual1}) implies that, as the $M<|k|$ case, the corresponding SUSY YMCS theories with $(N,M,k)$ and $(N^{(1)},M^{(1)},k)$ is Seiberg-like (IR) dual each other.

Now we have two questions.
First, what happens if $N^{(1)}<0$?
In this case, there remain anti-D3-branes in the description of brane construction,
and this implies that supersymmetry is broken for the gauge theory with $(N,M,k)$.
This is nothing but the modified s-rule explained in sec.~\ref{sRule}.

\begin{figure}[t]
\begin{center}
\includegraphics[width=54mm]{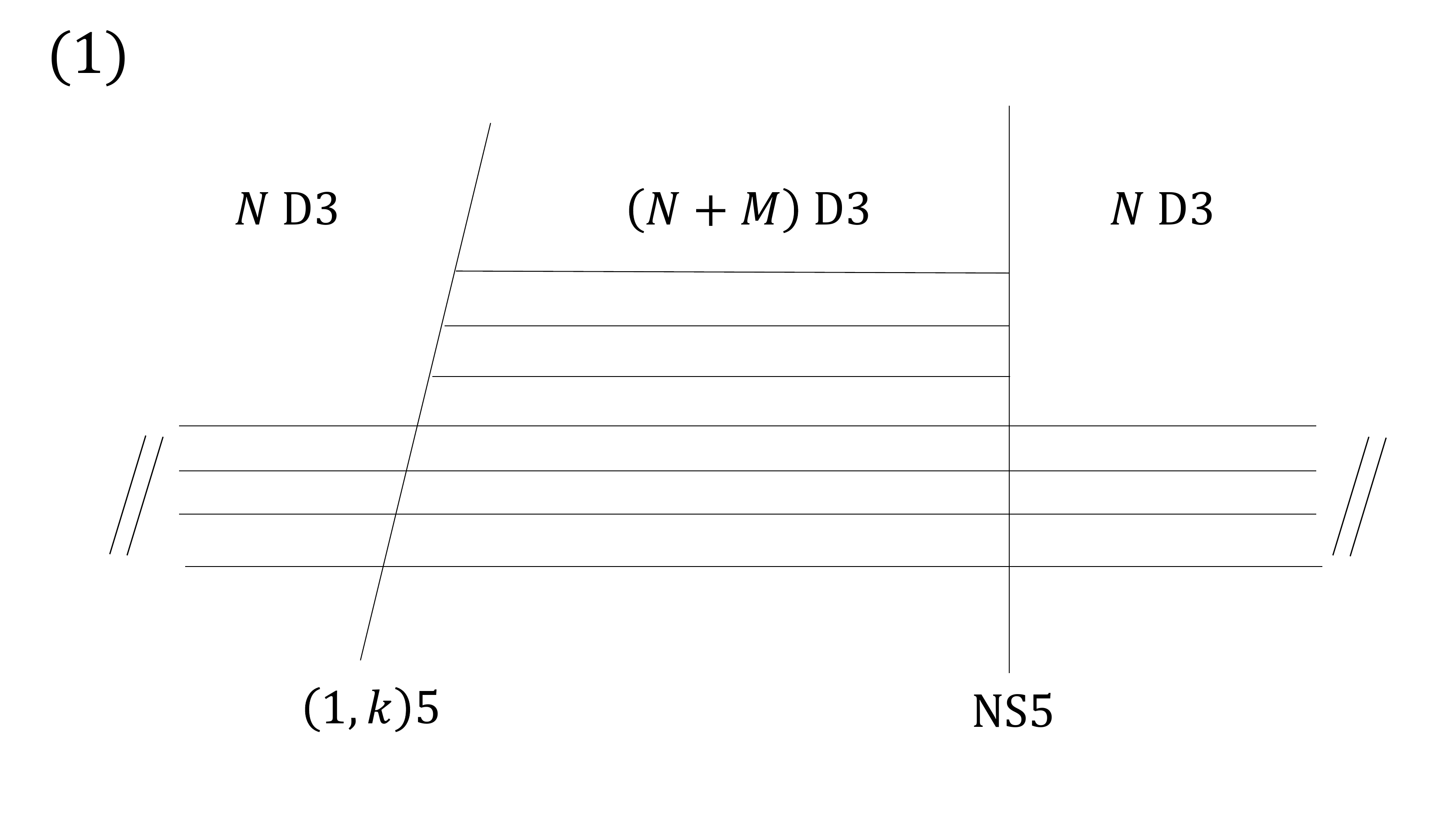}
\includegraphics[width=54mm]{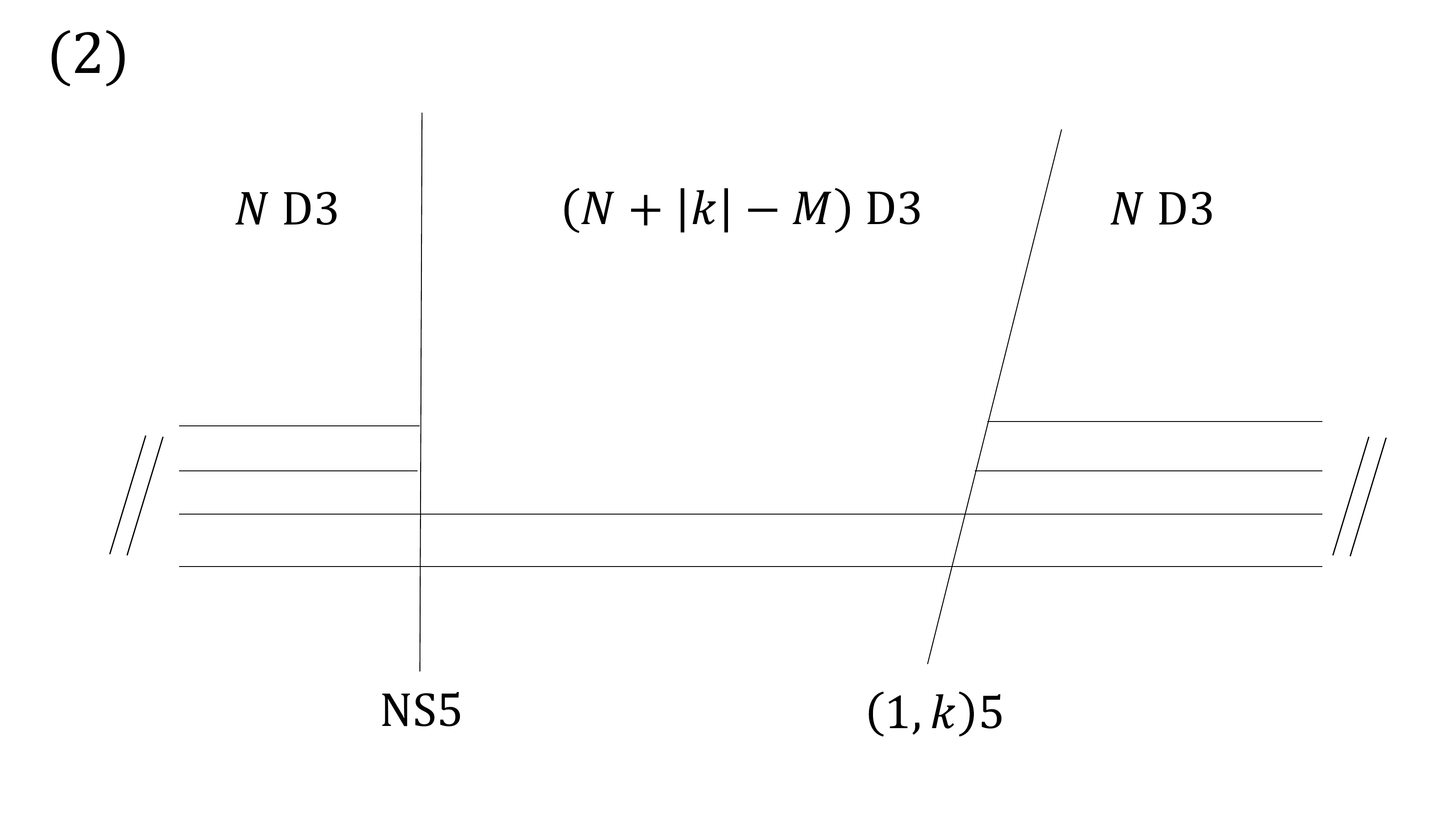}
\includegraphics[width=54mm]{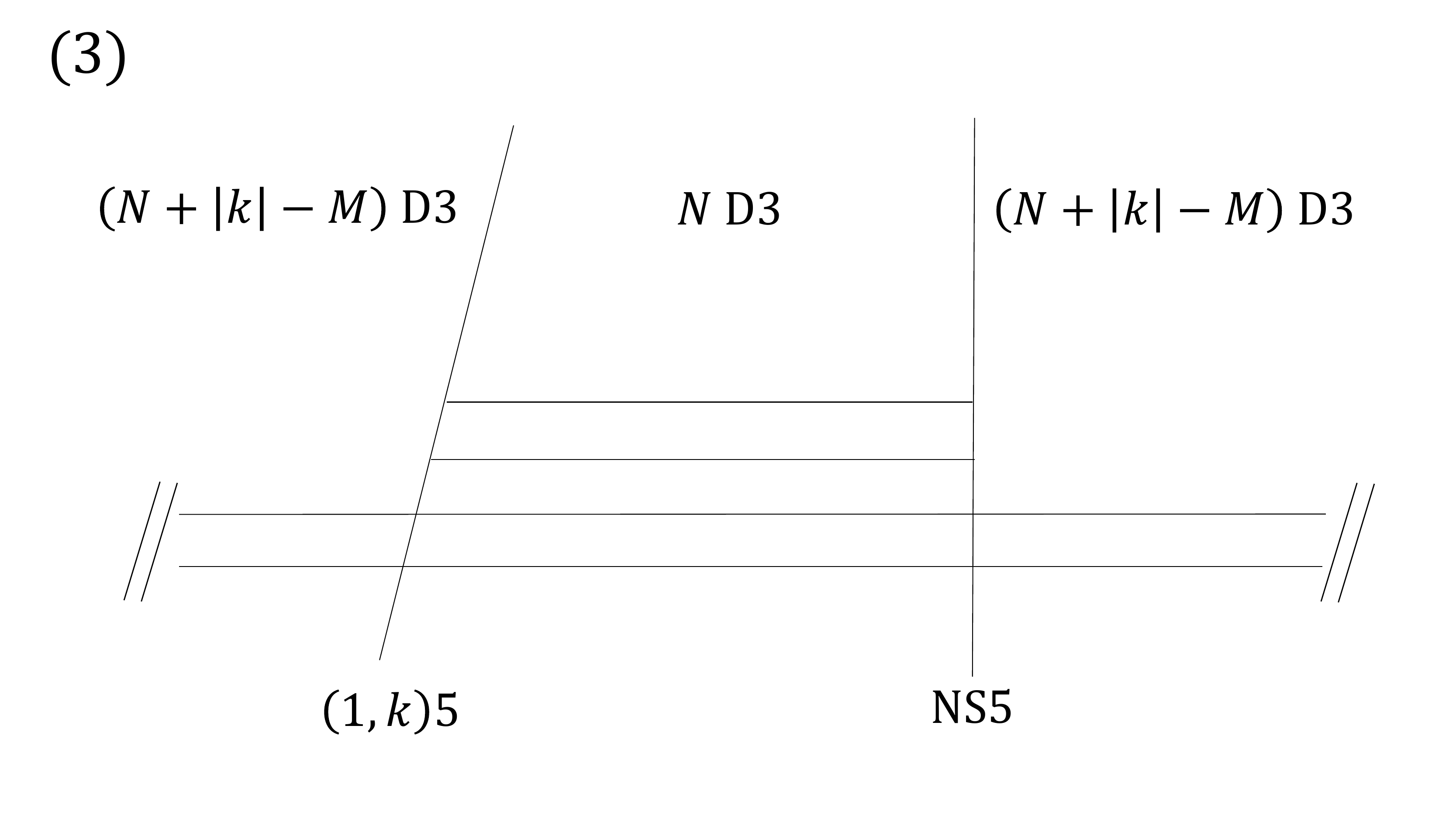}\\
\includegraphics[width=54mm]{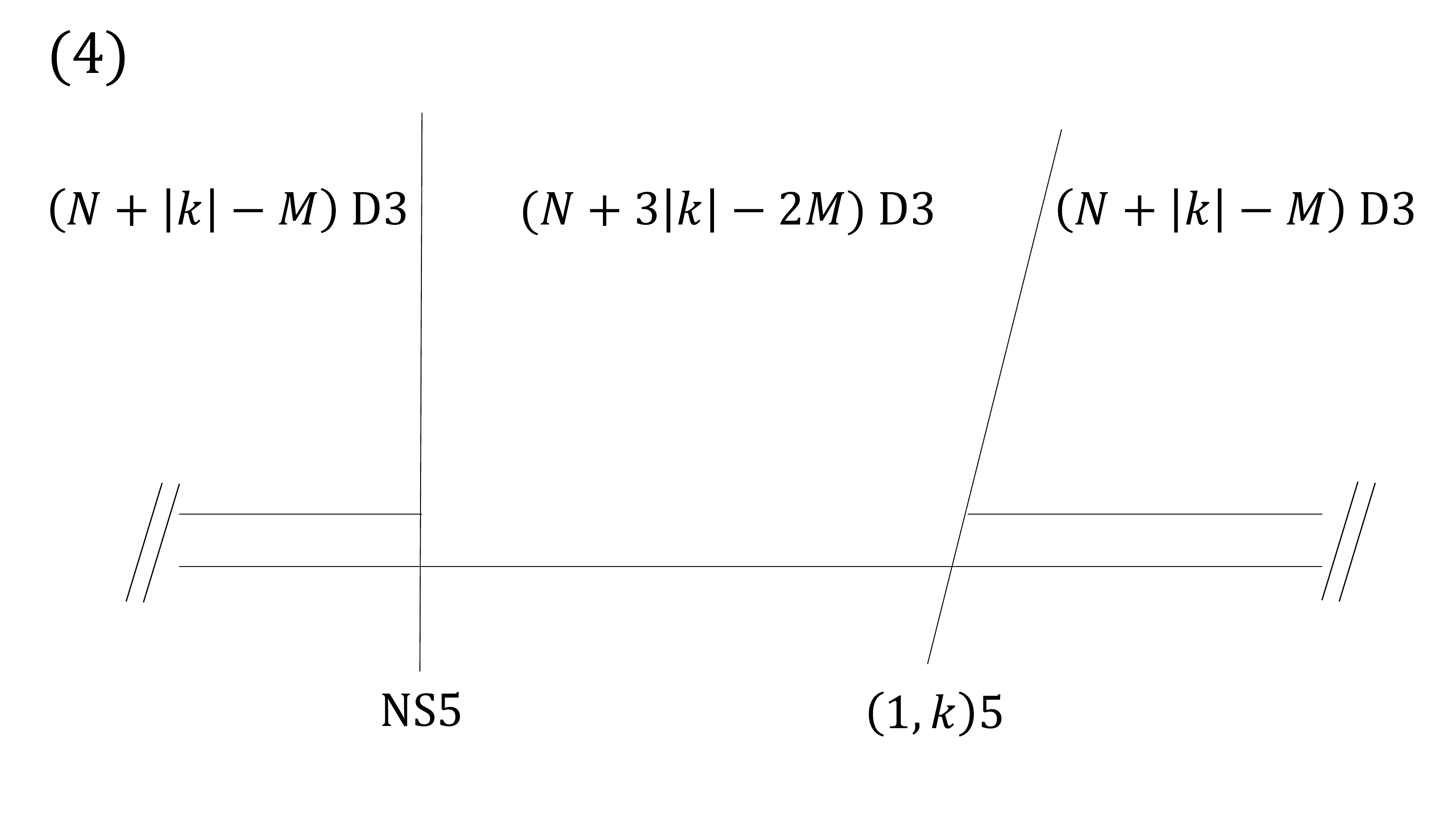}
\includegraphics[width=54mm]{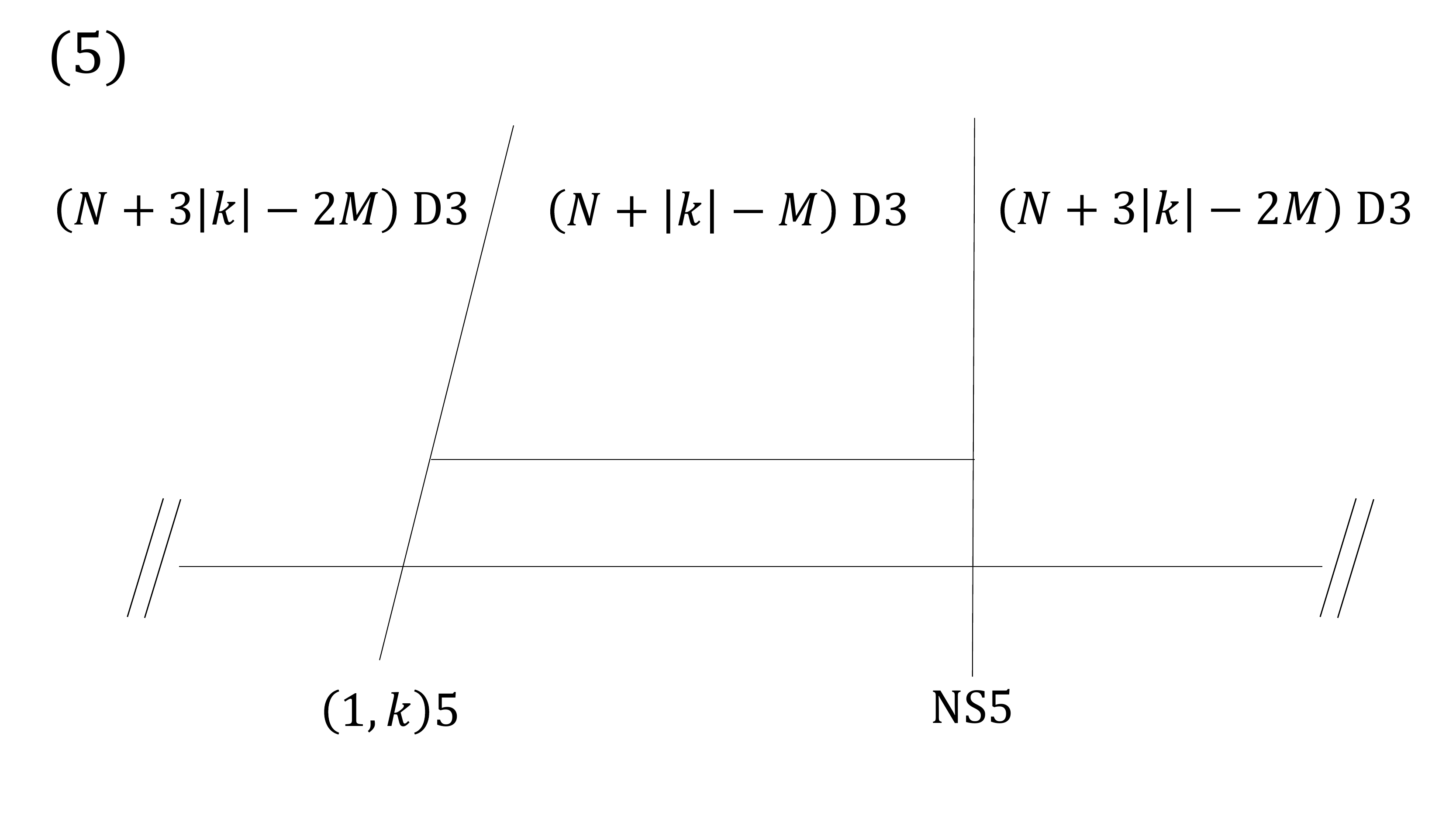}
\includegraphics[width=54mm]{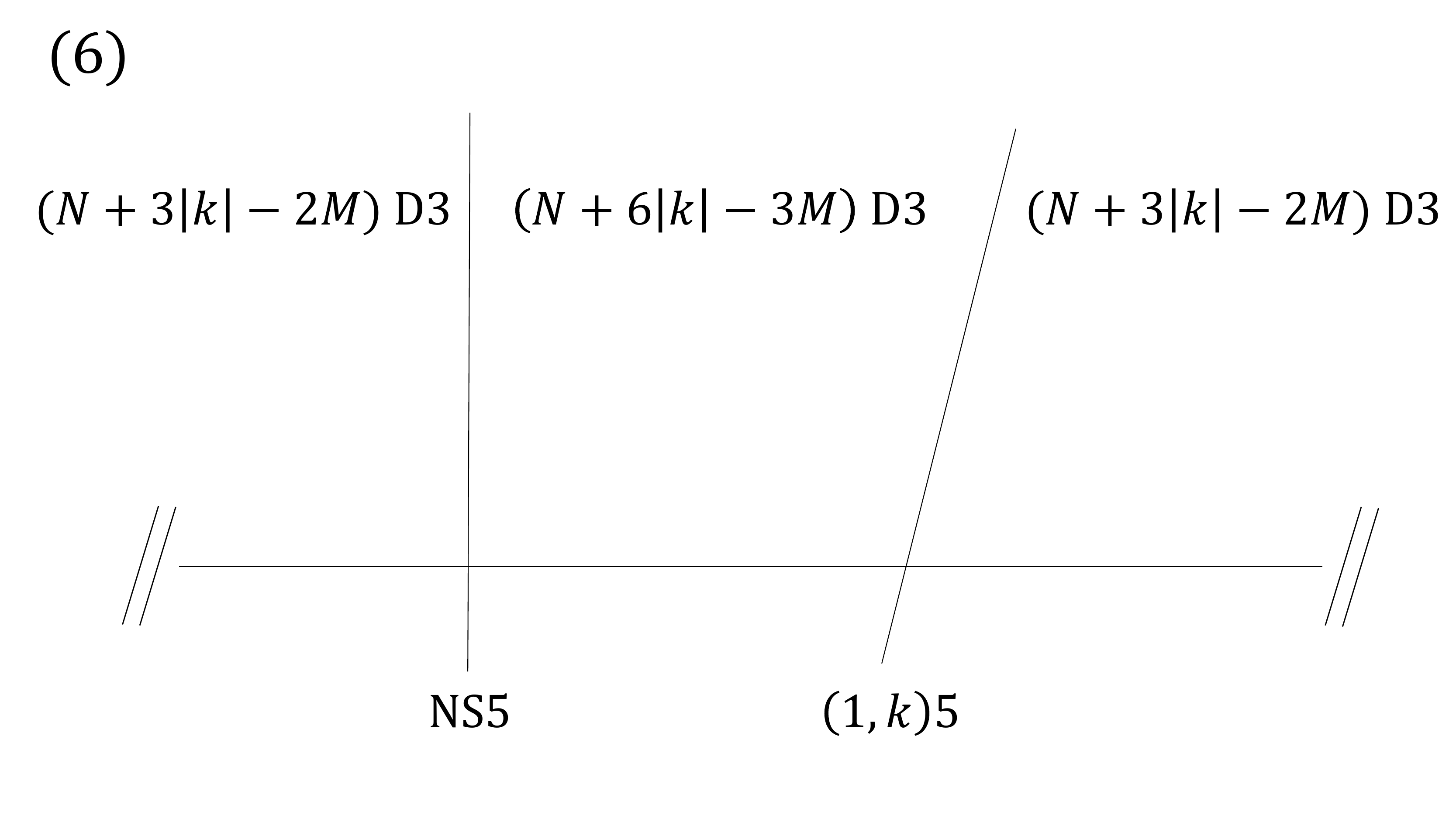}
\caption{
Illustration of the sequence of the dualities among apparently different four theories.
(1) The brane configuration corresponding 
to the $\mathcal{N}=3$ $U(N)_k \times U(N+M)_{-k}$ theory with $2|k|< M \leq 3|k|$
($N=4$, $M=3$, $k=1$ in this figure).
(2) After the 1st Hanany-Witten move.
(3) The same as the 2nd brane configuration since the both ends are identified.
(4) After the 2nd Hanany-Witten move.
(5) The same as the 4th brane configuration.
(6) After the 3rd Hanany-Witten move.
Applying the Hanany-Witten move once more gives the 4th brane configuration.
}
\label{fig:cascade}
 \end{center}
\end{figure}

Second, can we apply the duality transformation once more 
for $M^{(1)} > |k|$ and $N^{(1)} \geq 0$?
As in the first application of the duality transformation \eqref{eq:dual1},
there are differences on parameters $(N^{(1)} ,M^{(1)} ,k)$ and the duality acts as
\begin{\eq}
{\rm Duality}: (N^{(1)} ,M^{(1)} ,k) 
\rightarrow 
(N^{(2)} ,M^{(2)} ,k )  \quad {\rm for}\ M^{(1)}> |k|,
\end{\eq}
where 
\begin{\eq}
N^{(2)} = N^{(1)}+|k|-M^{(1)} ,\quad M^{(2)} = M^{(1)}-|k| .
\end{\eq}
When $N^{(2)}<0$, the modified s-rule again suggests SUSY breaking for the worldvolume theories with $(N ,M ,k)$ and $(N^{(1)} ,M^{(1)} ,k)$.
Otherwise, when $M^{(2)} > |k|$, we can apply the duality transformation further once more.
to relate it to another apparently different theory.
Repeating the above reasoning leads us to the conclusion that the theory has the duality cascade:
\begin{\eq}
 (N,M,k) 
\rightarrow (N^{(1)} ,M^{(1)} ,k) \rightarrow (N^{(2)} ,M^{(2)} ,k)
\rightarrow (N^{(3)} ,M^{(3)} ,k)  \rightarrow \cdots ,
\end{\eq}
where the parameters $(N^{(j)} ,M^{(j)} ,k^{(j)} )$ 
are given by
\begin{\eq}
N^{(j)} = N^{(j-1)}+|k|-M^{(j-1)},\quad 
M^{(j)} = M^{(j-1)}-|k| \quad
{\rm for }\ M^{(j-1)} > |k|, 
\end{\eq}
or equivalently,
\begin{\eq}
N^{(j)} = N+\frac{j(j+1)}{2}|k|-jM ,\quad 
M^{(j)} = M-j|k|.
\end{\eq}
This sequence of the dualities lasts until we encounter
\begin{\eq}
N^{(n)} <M^{(n)}-|k| ,\quad {\rm or}\quad M^{(n)} \leq |k| .
\end{\eq}
Note that when the ranks satisfy neither the first nor the second condition, the theory is still in the process of the duality cascade.
Furthermore, if the ranks satisfy the second condition, 
then the first condition is not satisfied.
Therefore, when the duality cascade lasts, the ranks always satisfy either of the two cases.

Let us again consider the worldvolume theories corresponding to the sequence.
For the former case $N^{(n)} <M^{(n)}-|k|$, the modified s-rule implies that SUSY breaking occurs for the worldvolume theories because of  $N^{(n+1)}=N^{(n)}+|k|-M^{(n)} <0$.
For the latter case $ M^{(n)} \leq |k|$, the theories finally flow to the superconformal theory,
i.e.~the ABJ theory with the gauge group $U(N^{(n)})_{k} \times U(N^{(n)} +M^{(n)})_{-k}$.
Note that in this case, the rule of the duality transformation we should apply finally changes to  (\ref{eq:dual0}).
This means that the sequence of the dualities actually ends since after acting (\ref{eq:dual0}) twice, the theory comes back to itself as explained.
The fig.~\ref{fig:cascade} illustrates the case for $2|k|<M\leq 3|k|$
where we apply the Hanany-Witten move twice
to go to the theory with $M\leq |k|$.
The third Hanany-Witten move in the figure shows the famous Seiberg-like duality of ABJ theories (\ref{eq:Seiberg}).

The above dualities are IR dualities
in the sense that apparently different theories flow to the same theory in IR.
Although in this paper we focus only on the IR limit,
there are also arguments on behaviors of RG flows of the above theories 
from the IIA string/M-theory viewpoint \cite{Aharony:2009fc,Evslin:2009pk}.
We have seen that
when the duality cascade seems to occur,
the Hanany-Witten effect decreases the lowest rank $N^{(j)}$ (and also the difference $M^{(j)}$) of the gauge group.
The string computation of effective Yang-Mills couplings 
implies that
the YMCS theory with $(N^{(j)},M^{(j)},k^{(j)} )$ flows to
another YMCS theory with $(N^{(j+1)},M^{(j+1)},k^{(j+1)} )$
before flowing to the superconformal CS theories.
Therefore, one RG flow step corresponds to one Hanany-Witten move.

\subsubsection{Generalization and local theories}
\label{sec:generalization}
\begin{figure}[t]
\begin{center}
\includegraphics[width=80mm]{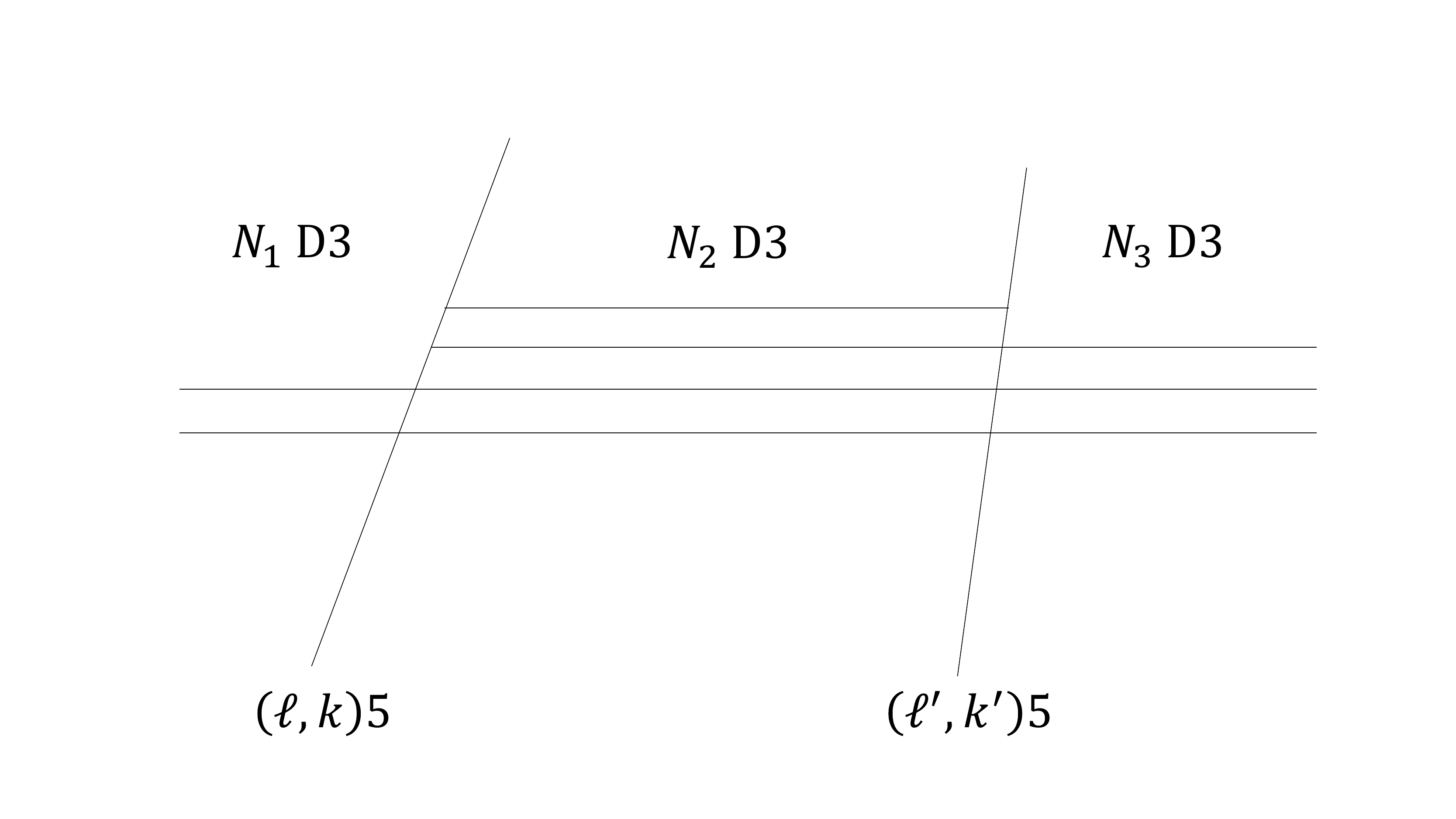}
\caption{
The brane configuration of the local theory for general 5-branes.
Note that the horizontal direction is not identified on the both ends unlike fig.~\ref{fig:ABJ}.
}
\label{fig:local2}
 \end{center}
\end{figure}

So far we focused on the $U(N)_k \times U(N+M)_{-k}$ gauge theory and the corresponding brane configuration in fig.~\ref{fig:ABJ}.
This brane configuration is merely a specific case of the generic Hanany-Witten type brane configurations discussed in sec.~\ref{subsec:setup}.
As explained above, the duality cascade with the modified s-rule can be clearly explained by using the brane construction.
Notice that we have essentially not used any specific property that the quiver has only two nodes.
Therefore, it is natural to propose that the same story should holds for the generic Hanany-Witten type brane configurations although it has not been explicitly discussed in literature\footnote{
For the 4d case,
the duality cascade was generalized to affine ADE quiver theories \cite{Cachazo:2001sg}.
}.
We write this proposition more explicitly.
The worldvolume theories of the generic Hanany-Witten type brane configurations
are $\mathcal{N}=3$ SUSY YMCS theories with corresponding 
quiver diagrams. 
If the theories are related by the Hanany-Witten moves, they are Seiberg-like dual, in other words, they flows to a same IR theory.

\begin{figure}[t]
\begin{center}
\includegraphics[width=54mm]{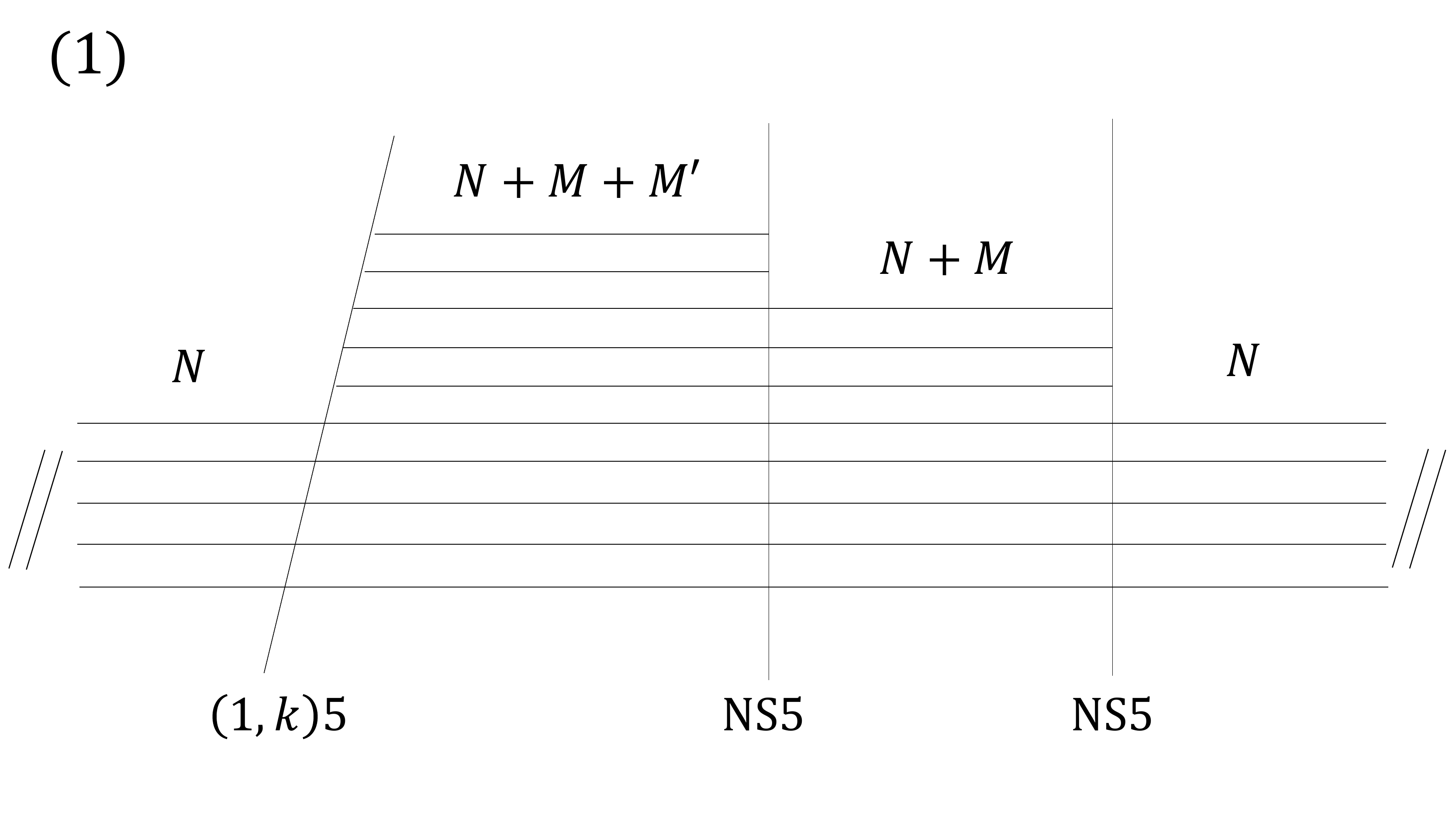}
\includegraphics[width=54mm]{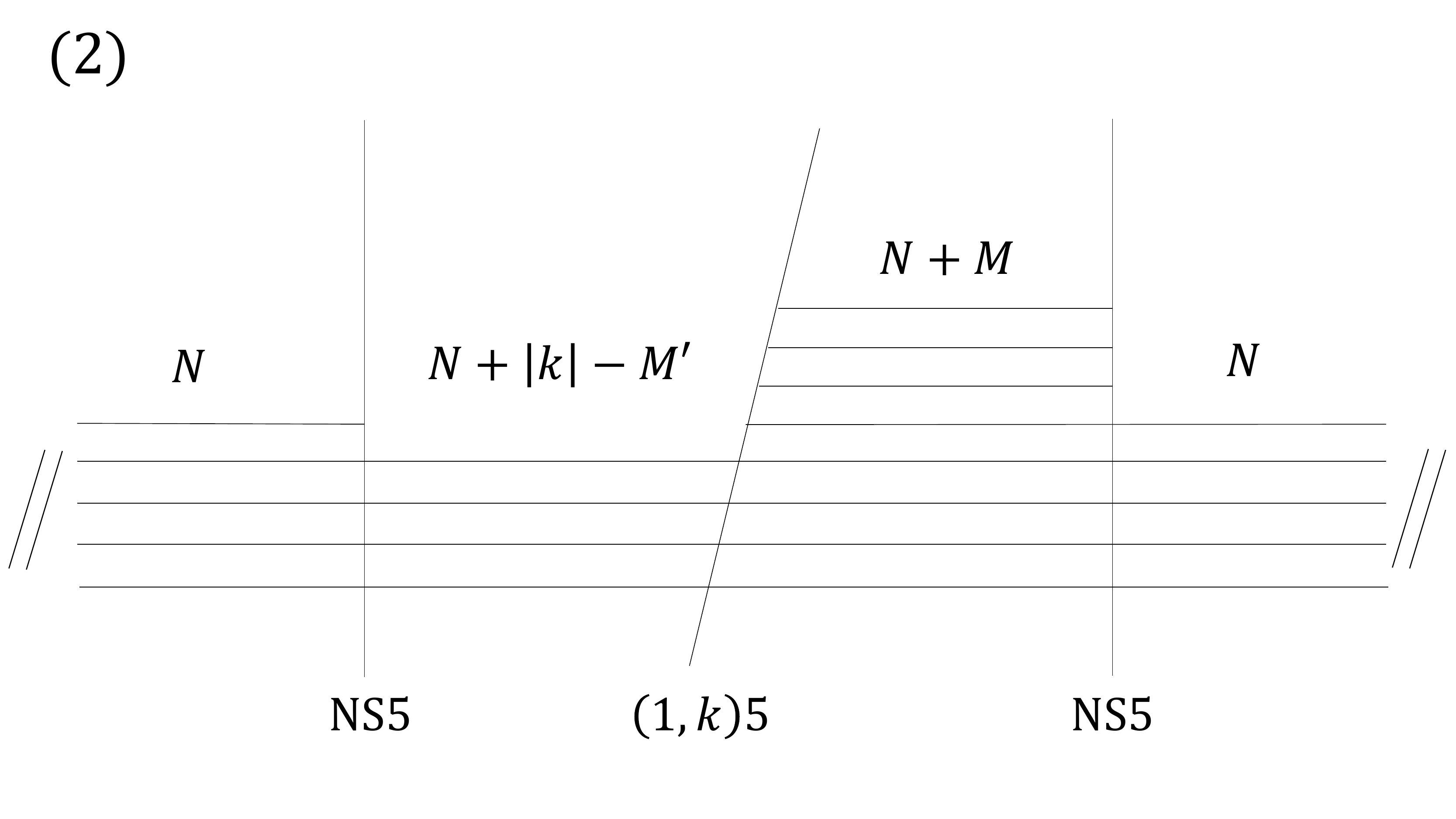}
\includegraphics[width=54mm]{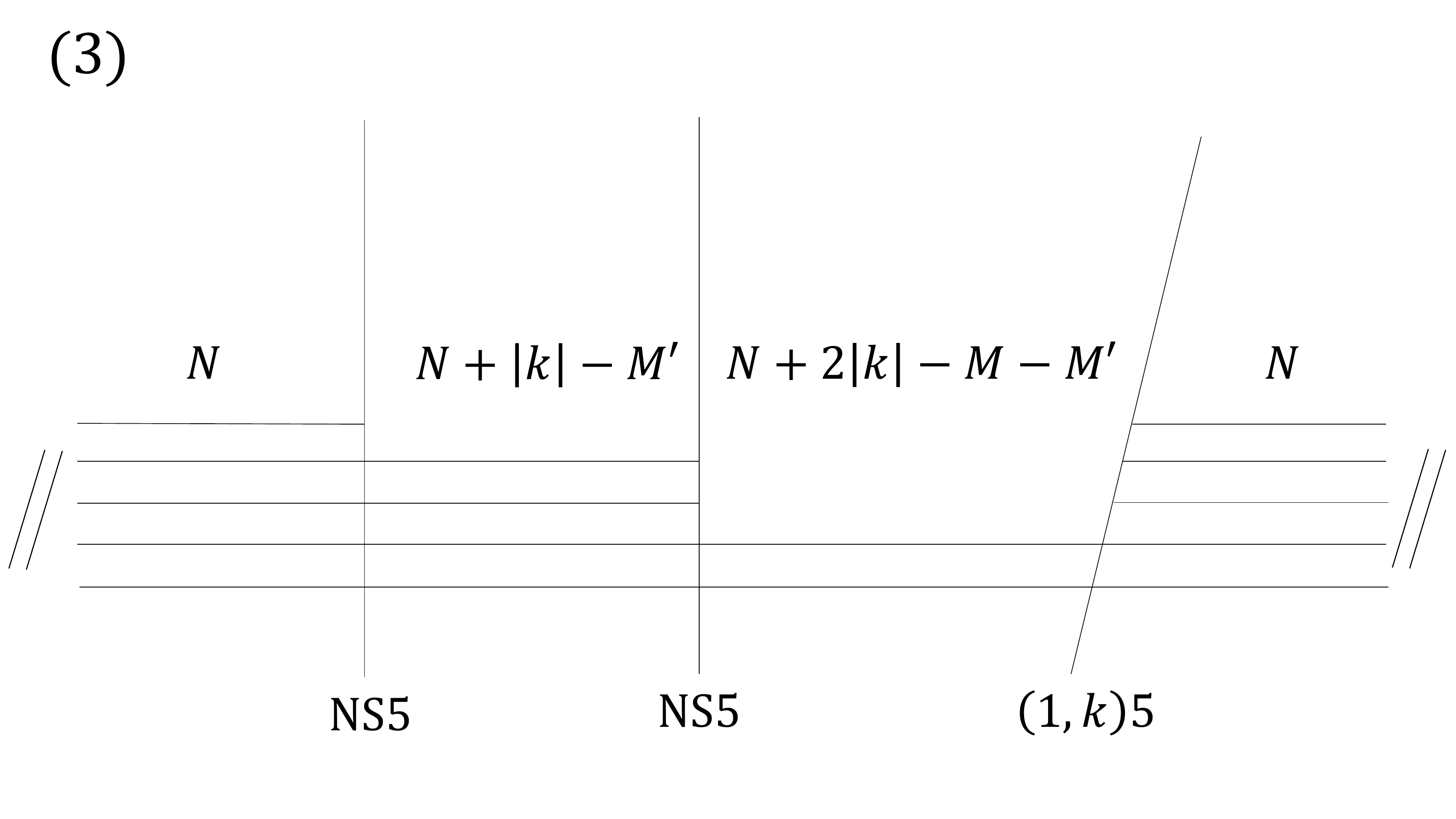}\\
\includegraphics[width=54mm]{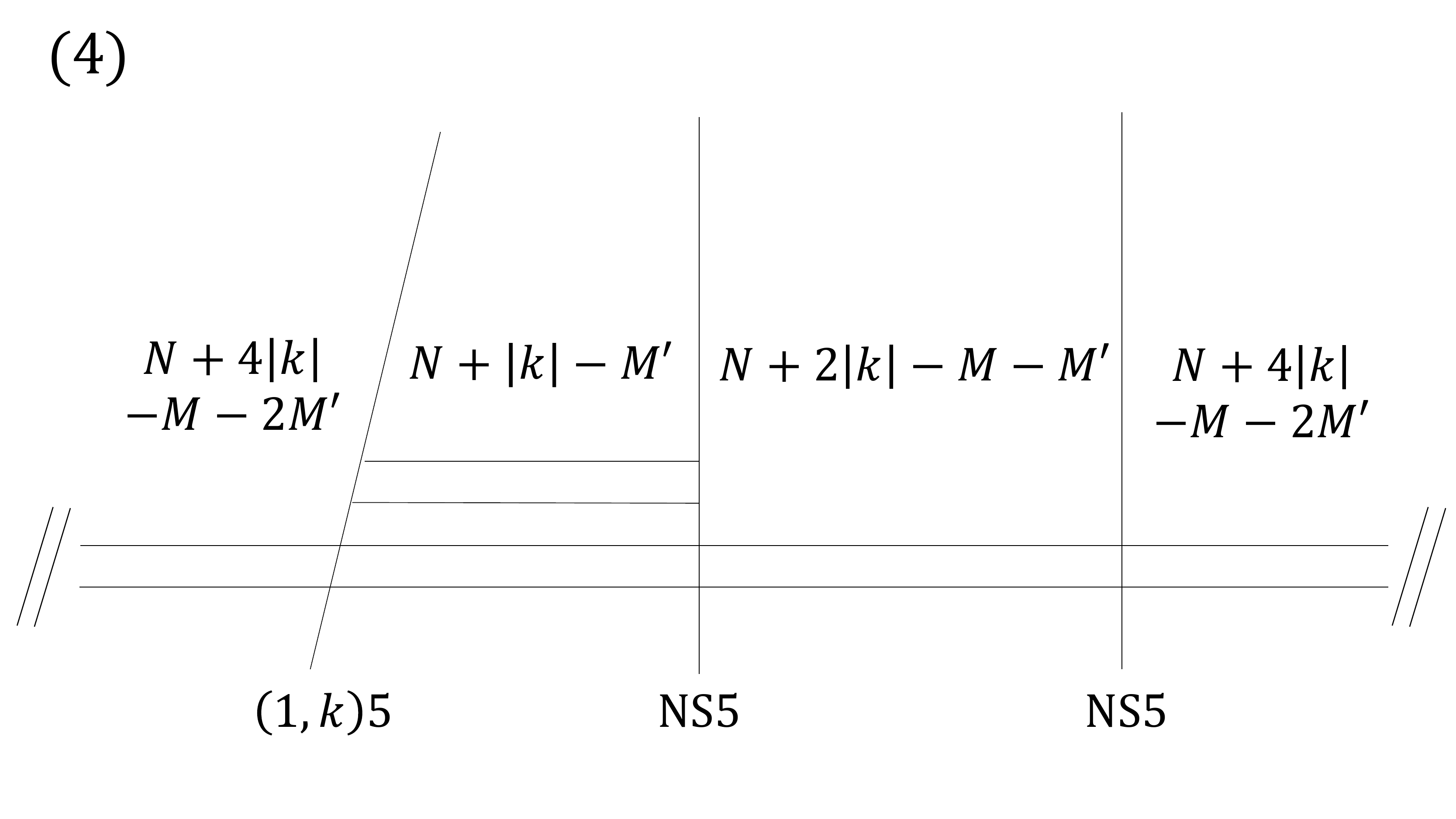}
\includegraphics[width=54mm]{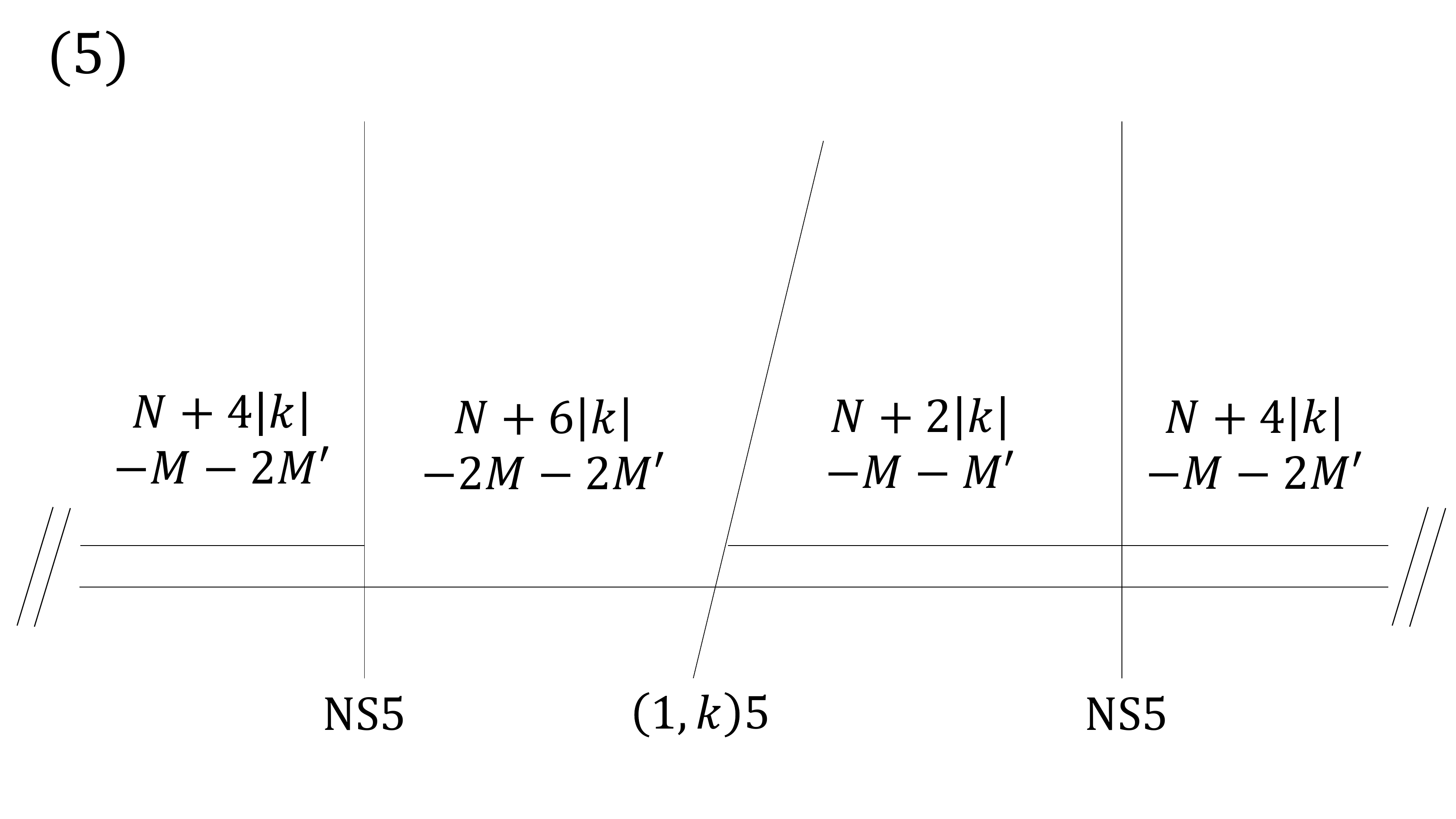}
\includegraphics[width=54mm]{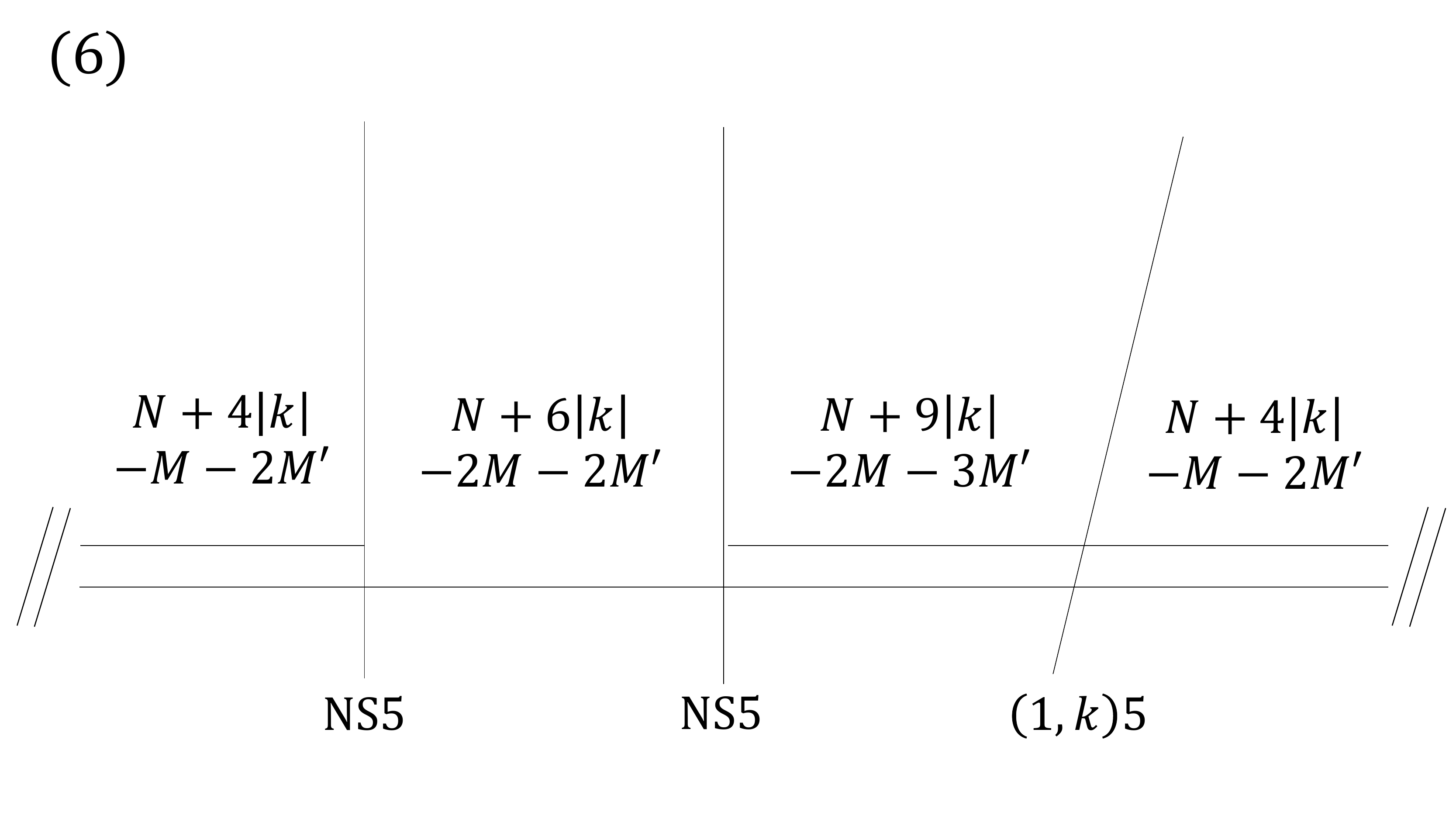}\\
\includegraphics[width=54mm]{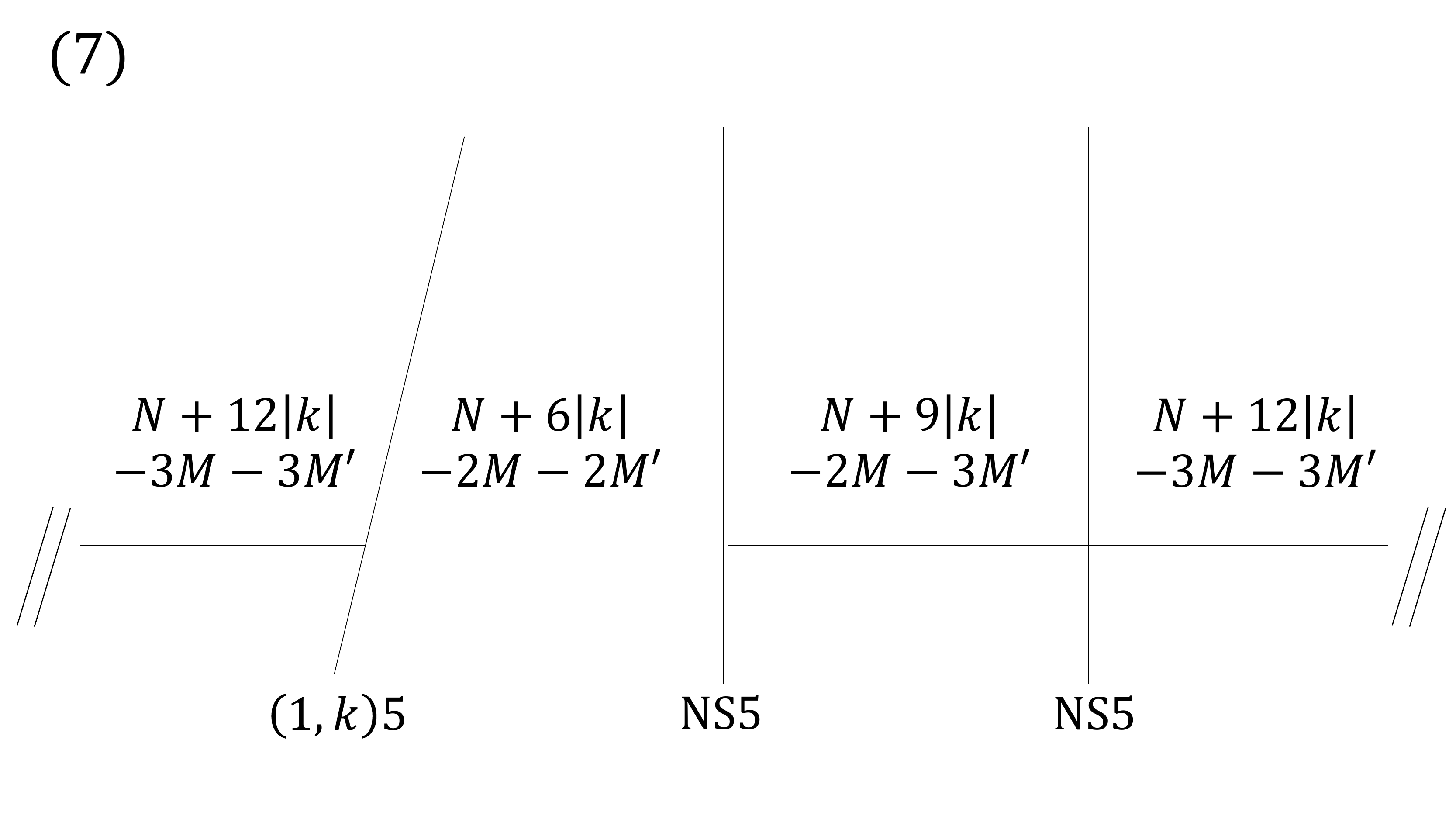}
\caption{
Illustration of the duality cascade 
for the $U(N)_k \times U(N+M+M')_{-k}\times U(N+M)_0$ gauge theory 
with $2|k|<M\leq 3|k|$ and $|k|<M' \leq 2|k|$ 
($N=5$, $M=3$, $M'=2$, $k=1$ in this figure).
Applying further Hanany-Witten moves to (7) gives (4), (5), (6) and (7) with the flipped $k$.
}
\label{fig:3quiver}
 \end{center}
\end{figure}

Here we emphasize that all the Seiberg-like dualities considered here are related by the sequence of single Hanany-Witten move.
To see this explicitly,
it is convenient to note that
the generic Hanany-Witten type brane configuration is 
a combination of the brane configurations
$\left\langle N_{1},\left(\ell,k\right),N_{2},\left(\ell',k'\right) ,N_{3}  \right\rangle$
given in fig.~\ref{fig:local2},
which we refer to as the local theories.
Then the dual theories are obtained by applying the Hanany-Witten move,
which amounts to apply the transformation 
\begin{\eq}
\left\langle N_{1},\left(\ell,k\right),N_{2},\left(\ell',k'\right) ,N_{3}  \right\rangle
\rightarrow
\left\langle N_{1},\left(\ell',k'\right),N_1+N_3-N_{2}+|\ell k'-k\ell'|,\left(\ell,k\right) ,N_{3}  \right\rangle,
\label{eq:HWgen}
\end{\eq}
to one of the local theories in a given brane configuration.
Therefore the dualities coming from Hanany-Witten type brane configurations
come from the duality of the local theory.
Fig.~\ref{fig:3quiver} shows an illustration for the 3-quiver case.

If anti-D3-branes remain after one Hanany-Witten move, the modified s-rule again implies that the SUSY is broken for the worldvolume theory as with the two nodes example.
Since all the dual theories are related by the sequence of the Hanany-Witten moves, this immediately means that SUSY is broken also for them.

Thus the problem of showing the duality cascade is boiled down
to show the result of the Hanany-Witten effect 
for the local theory shown in fig.~\ref{fig:local}.
This point will play a crucial role 
when we perform non-perturbative tests of the dualities in the next section.

\section{Non-perturbative tests of the  duality cascades}
\label{sec:test}
In this section,
we perform non-perturbative tests of the duality cascades
by analyzing the supersymmetric partition function on $S^3$.

\subsection{The prediction from the duality cascade on sphere partition function}
\label{subsec:LocalMM}
Let us consider a $\mathcal{N}=3$ supersymmetric theory $\mathcal{T}$
coming from a generic Hanany-Witten type brane configuration described in sec.~\ref{subsec:setup}.
Applying the Hanany-Witten moves to the brane configuration associated with the theory $\mathcal{T}$,
we can obtain various quantum field theories that are apparently different from each other.
Let us denote a set of such field theories by $\widehat{\rm HW} (\mathcal{T})$:
\begin{\eq}
\widehat{\rm HW} (\mathcal{T})
= \left\{
{\rm theories\ obtained\ by\ Hanany-Witten\ moves\ of\ } \mathcal{T}
\right\} .
\end{\eq}
Then according to the brane argument in the last section,
we expect that
all the theories in $\widehat{\rm HW} (\mathcal{T})$ are physically equivalent in IR.

Let us elaborate on the prediction from the duality cascade on $S^3$ partition functions.
It is known that 
$S^3$ partition functions of any $\mathcal{N}=2$ SUSY theories
are the same as their IR limits\footnote{
Technically this can be understood from the fact that 
the actions of the $\mathcal{N}=2$ super Yang-Mills and matters are $Q$-exact.
Therefore $\mathcal{N}=2$ sphere partition functions are
independent of Yang-Mills couplings and coupling constants appearing in superpotential.
}.
Therefore the duality cascade implies that
the $S^3$ partition functions of all the theories in $\widehat{\rm HW} (\mathcal{T})$ have the common absolute value\footnote{
It was argued in \cite{Kapustin:2009kz,Kapustin:2010mh} that the phase depending on the Chern-Simons levels or the FI-parameters is attributed to using a framing which is different from the standard one due to the localization technique. 
Therefore, we here do not take care of differences  
after the duality cascade.
It would be interesting if one can understand the phase from the view point of `t Hooft anomaly.
}:
\begin{\eq}
\left| Z_{\mathcal{T}_i } \right|  =\left| Z_{\mathcal{T}_j } \right|  \qquad 
{\rm for}\ \   ^\forall\mathcal{T}_i ,\mathcal{T}_j \in \widehat{\rm HW} (\mathcal{T})  .\label{eq:ZcascGen}
\end{\eq}
Furthermore,
we have seen that
supersymmetry may be broken due to the 
modified s-rule.
This implies that
the theory has fermionic zero modes and the partition function vanishes:
\begin{\eq}
 Z_{\mathcal{T}_i} = 0  \quad 
{\rm for}\ \   ^\forall\mathcal{T}_i  \in \widehat{\rm HW} (\mathcal{T})\quad
{\rm if}\  ^\exists  \mathcal{T}_j  \in \widehat{\rm HW} (\mathcal{T})\ 
{\rm whose\ brane\ configuration\ involves\ \overline{D3}}.\label{eq:ZcascSrule}
\end{\eq}
Below we prove that the above identities indeed hold.

\subsubsection{Reducing the problem to the local theory}
The brane argument in sec.~\ref{sec:generalization} suggests that
the problem of showing the duality cascade can be reduced
to show the relation 
\eqref{eq:HWgen}
(we further reduce it to \eqref{eq:HW} using $SL\left( 2,\mathbb{Z}\right)$ duality later)
coming from the Hanany-Witten effect
for the local theory.
This is clearly true for the cases with Lagrangians 
while it is less obvious for the non-Lagrangian cases.
Furthermore, even if this is true,
the problem is to show the relation \eqref{eq:HWgen} 
for arbitrary configurations of the background vector multiplets of the flavor symmetries in the local theory,
which are essentially gauge symmetries of the left and right branes in the whole system.
Therefore it sounds still difficult. 
However, 
it will turn out that
we do not have to consider generic configurations of the background multiplets
if we restrict ourselves to the supersymmetric partition functions on $S^3$.
Below we explicitly see that 
our problem is reduced to show the relation \eqref{eq:HWgen}
for the local theory with only real masses of the flavor symmetries.

\subsubsection*{Brane configurations giving Lagrangian theories and matrix models}
As we have seen in sec.~\ref{subsec:setup},
the Hanany-Witten type brane configurations
consist of $\left(\ell ,k\right)$5-branes and D3-branes. 
When the 5-branes are only $(1,k)$ or $(0,1)$ types,
the worldvolume theories of the brane configurations are Lagrangian theories
\cite{Hanany:1996ie, Kitao:1998mf}.
Although physical observables of these theories are originally expressed by infinite dimensional path integrals,
we can apply SUSY localization to the $S^3$ partition functions, and
then the path integrals are dominated by saddle points
and finally reduced to finite dimensional integrals \cite{Kapustin:2009kz,Hama:2010av,Jafferis:2010un}.
The saddle points for this case are so-called Coulomb branch configurations,
which are constant configurations of adjoint scalar fields 
in the vector multiplets (up to gauge transformation),
and the partition functions are expressed as integrations over the Coulomb branch
whose dimensions are the rank of the gauge group\footnote{
More details will be explained in sec.~\ref{sec:gluing_sub}.
}.

Let us write down the localization formula for the $S^3$ partition functions.
Although the localization formula exists for general Lagrangian $\mathcal{N}=2$ theories,
here we focus on the cicular quiver $\mathcal{N}=3$ YMCS theory
with the gauge group 
\[
U(N_1 )_{k_1 -k_n} \times U(N_2 )_{k_2 -k_1}\times \cdots \times U(N_n )_{k_n -k_{n-1}},
\]
which is generic Lagrangian theory 
coming from the Hanany-Witten brane configuration in fig.~\ref{fig:generic}
with $\ell_1 =\cdots =\ell_n =1$.
The localization formula of this theory is given by
\begin{\eq}
Z_{\rm{total}}
= \left( \prod_{j=1}^n \int \frac{d^{N_j} x^{(j)} }{N_j !} \right)  Z_{\rm cl} (x ) Z_{\rm 1loop} (x) ,
\end{\eq}
where $Z_{\rm cl}(x)$ is the classical contribution:
\begin{\eq}
Z_{\rm cl} (x)
=\prod_{j=1}^n  e^{i\pi (k_j -k_{j-1}) \sum_{n=1}^{N_j} ( x_n^{(j)} )^2 -2\pi i (\zeta_j-\zeta_{j-1}) \sum_{n=1}^{N_j} x_a^{(j)}    },
\end{\eq}
with $k_0 \equiv k_n$ and $\zeta_0 \equiv \zeta_n$.
$(\zeta_j-\zeta_{j-1})$ is the FI-parameter
associated with the $U(1)$ part of 
$U(N_j )_{k_j-k_{j-1}}$.
$Z_{\rm 1loop}(x)$ is the one-loop contribution given by
\begin{\eq}
 Z_{\rm 1loop}(x ) 
= \prod_{j=1}^n \frac{\prod_{m<n}^{N_j}   4\sinh^2{\left( \pi (x_m^{(j)} -x_n^{(j)}) \right) }  }
   {\prod_{m}^{N_j} \prod_{n}^{N_{j+1}} 2\cosh{\left( \pi (x_m^{(j)} -x_n^{(j+1)}) \right)}} , 
\end{\eq}
where $N_{N+1}\equiv N_1$, $x_n^{(N+1)}\equiv x_n^{(1)}$ and 
\begin{equation}
\prod_{n}^{N}\ldots=\prod_{n=1}^{N}\ldots\ ,\quad
\prod_{n<n'}^{N}\ldots=\prod_{n=1}^{N-1}\prod_{n'=n+1}^{N}\ldots \ .
\end{equation}
Note that the localization formula is independent of Yang-Mills coupling 
because of $Q$-exactness of the SYM terms.
This implies that
the localization formula captures the IR limit: $g_{\rm YM}\rightarrow \infty$ of theories under consideration
even if we start with the action including the SYM term.

We can recast the localization formula in a way that the connection to the brane interpretation is transparent.
An idea is that
there is a natural one-to-one correspondence between the branes and 
objects in the localization formula.
Noting that the theory is given by $(1,k_j )$5-branes 
between $N_j$ D3-branes on left and $N_{j+1}$ D3-branes on right,
we can rewrite the localization formula as
\begin{\eq}
Z_{\rm{total}}
= \left( \prod_{j=1}^n \int \frac{d^{N_j} x^{(j)} }{N_j !} \right) 
\prod_{j=1}^n Z_{\left( 1,k_j \right),\zeta_j}\left(N_{j},N_{j+1}; x^{(j)} ,x^{(j+1)}  \right),
\end{\eq}
where $Z_{\left( 1,k\right),\zeta}$ is the factor corresponding to the $(1,k)$5-brane\footnote{ 
We specify the parameters for the factor of single 5-brane because these parameters are important to show the Hanany-Witten effect.
}:
\begin{align}
Z_{\left( 1,k\right),\zeta}\left(N_{1},N_{2};\alpha,\beta\right)
= & 
e^{i\pi k \left(\sum_{m}^{N_{1}}\alpha_{m}^{2}-\sum_{n}^{N_{2}}\beta_{n}^{2}\right)}
e^{-2\pi i\zeta \left(\sum_{m}^{N_{1}}\alpha_{m}-\sum_{n}^{N_{2}}\beta_{n}\right)}\nonumber \\
 & \times
 \frac{\prod_{m<m'}^{N_{1}}2\sinh{\pi\left(\alpha_{m}-\alpha_{m'}\right)}
 \prod_{n<n'}^{N_{2}}2\sinh{\pi\left(\beta_{n}-\beta_{n'}\right)}  }
 {\prod_{m}^{N_{1}}\prod_{n}^{N_{2}}2\cosh{\pi\left(\alpha_{m}-\beta_{n}\right)}  }.
\label{eq:1k}
\end{align}

\subsubsection*{Generic brane configurations}
In general, the Hanany-Witten type brane configurations give non-Lagrangian theories\footnote{
Lagurangian is defined even in this case for abelian theories.
The Chern-Simons level on each interval between a $\left(\ell,k\right)$5-brane on the left side and a $\left(\ell',k'\right)$5-brane on the right side is $-\left( \frac{k}{\ell}-\frac{k'}{\ell'}\right)$ \cite{Kitao:1998mf,Bergman:1999na}.
} and
it seems harder to compute their partition functions.
Nevertheless, 
there is a reasonable proposal on generalization of the $(1,k)$5-brane factor \eqref{eq:1k} to $(\ell ,k)$5-brane \cite{Assel:2014awa} (see also \cite{Benvenuti:2011ga,Gulotta:2011si,Nishioka:2011dq}).
It has been obtained by considering $SL\left( 2,\mathbb{Z}\right)$ dualities 
and has passed various nontrivial consistency checks.

The proposal on the factor corresponding to
$\left\langle N_{1},\left(\ell,k\right),N_{2}\right\rangle $ with ${\rm gcd}(\ell ,k )=1$
is
\begin{align}
Z_{\left(\ell,k\right),\zeta}\left(N_{1},N_{2};\alpha,\beta\right)= & \frac{1}{\left|\ell\right|^{\frac{N_{1}+N_{2}}{2}}}e^{\frac{i\pi k}{\ell}\left(\sum_{m}^{N_{1}}\alpha_{m}^{2}-\sum_{n}^{N_{2}}\beta_{n}^{2}\right)}e^{-\frac{2\pi i\zeta}{\ell}\left(\sum_{m}^{N_{1}}\alpha_{m}-\sum_{n}^{N_{2}}\beta_{n}\right)}\nonumber \\
 & \times\frac{\prod_{m<m'}^{N_{1}}2\sinh\frac{\pi\left(\alpha_{m}-\alpha_{m'}\right)}{\ell}\prod_{n<n'}^{N_{2}}2\sinh\frac{\pi\left(\beta_{n}-\beta_{n'}\right)}{\ell}}{\prod_{m}^{N_{1}}\prod_{n}^{N_{2}}2\cosh\frac{\pi\left(\alpha_{m}-\beta_{n}\right)}{\ell}}.
\label{eq:Mfactor}
\end{align}
Thus, using the $(\ell ,k)$5-brane factor,
the $S^3$ partition function of the worldvoluome theory
of the brane configuration in fig.~\ref{fig:generic} is given by
\begin{\eq}
Z_{\rm total}
= \left( \prod_{j=1}^n \int \frac{d^{N_j} x^{(j)} }{N_j !} \right) 
\prod_{j=1}^n Z_{\left( \ell_j ,k_j \right),\zeta_j}\left(N_{j},N_{j+1}; x^{(j)} ,x^{(j+1)}  \right) .
\label{eq:total}
\end{\eq}
In this paper we assume that the proposal \eqref{eq:Mfactor} is true.

\subsubsection*{Extracting the local theory}
In sec.~\ref{sec:cascade}, we have seen that
it is sufficient to focus on the local brane structure upon applying the Hanany-Witten move once
and the problem of showing the duality is reduced
to show the relation \eqref{eq:HWgen} for the local theory.
Now we show that this is indeed true for the $S^3$ partition function.
Let us decompose the partition function \eqref{eq:total} into
the part involving $x^{(j')}$ and the others:
\begin{\eq}
Z_{\rm total}
= \left( \prod_{j \neq j'}^{n} \int \frac{d^{N_j} x^{(j)} }{N_j !} \right) 
\left( \prod_{j\neq j'-1 ,j'}^n Z_{\left( \ell_j ,k_j \right),\zeta_j}\left(N_{j},N_{j+1}; x^{(j)} ,x^{(j+1)}  \right) \right)
Z_{\rm local}(x^{(j'-1)},x^{(j'+1)} ) ,
\end{\eq}
where
\begin{\eqa}
Z_{\rm local} (x^{(j'-1)},x^{(j'+1)} )
&\equiv & \int \frac{d^{N_{j'}} x^{(j')} }{N_{j'} !} 
Z_{\left( \ell_{j'-1} ,k_{j'-1} \right),\zeta_{j'-1}}\left(N_{j'-1},N_{j'}; x^{(j'-1)} ,x^{(j')}  \right) \NN\\
&& \qquad\qquad ~~\cdot  Z_{\left( \ell_{j'} ,k_{j'} \right),\zeta_{j'}}\left(N_{j'},N_{j'+1}; x^{(j')} ,x^{(j' +1)}  \right) .
\label{eq:local_start}
\end{\eqa}
The local matrix model $Z_{\rm local}$ has a natural brane interpretation.
This can be regarded as the partition function of the worldvolume theory associated to the local theory
\[
\left\langle N_{j'-1},\left(\ell_{j'-1},k_{j'-1}\right),N_{j'},\left(\ell_{j'},k_{j'}\right) ,N_{j'+1}  \right\rangle  ,
\]
explained in sec.~\ref{sec:generalization}.
Therefore applying the Hanany-Witten move to this theory generates the duality of the original theory.

While the above argument has already simplified the original problem drastically,
we can further simplify the problem by applying the $SL\left( 2,\mathbb{Z}\right)$ dualities
to $Z_{\rm local}$ shown in \cite{Assel:2014awa}.
Using this duality,
one can fix either $\left(\ell_{j'-1},k_{j'-1}\right)$ or $\left(\ell_{j'},k_{j'}\right)$
to a specific value that we want without loss of generality.
Here we change $\left(\ell_{j'},k_{j'}\right)$ into $(1,0)$.
If $\ell_j =0$, we also use the remaining duality to set $\ell_j\neq 0$.
Thus the original problem to show \eqref{eq:ZcascGen} and \eqref{eq:ZcascSrule} has been reduced to show that
the local matrix model 
satisfies relations coming from \eqref{eq:HW} with the modified s-rule.

\subsubsection{The prediction for the local matrix model}
Here we explicitly write down the predictions on the local matrix models
coming from the Hanany-Witten effect \eqref{eq:HW} and the modified s-rule, 
which are sufficient to prove the relation \eqref{eq:ZcascGen} and \eqref{eq:ZcascSrule} as explained in the previous section.

The local matrix model \eqref{eq:local_start} corresponding to $\left\langle N_{\Left}\bullet N\circ N_{\Right}\right\rangle $ is
\begin{equation}
Z(x,y)
=\int\frac{d^{N}\mu}{N!}Z_{\left(\ell,k\right),\zeta}\left(N_{\Left},N;x,\mu\right)Z_{\left(1,0\right),\eta}\left(N,N_{\Right};\mu,y\right),
\label{eq:ZUVdef}
\end{equation}
and corresponding to $\left\langle N_{\Left}\circ\tilde{N}\bullet N_{\Right}\right\rangle $ where $\tilde{N}=N_{\Left}+N_{\Right}-N+|k|$ is
\begin{equation}
Z_{\HW}(x,y)=\int\frac{d^{\tilde{N}}\nu}{\tilde{N}!}Z_{\left(1,0\right),\eta}\left(N_{\Left},\tilde{N};x,\nu\right)Z_{\left(\ell,k\right),\zeta}\left(\tilde{N},N_{\Right};\nu,y\right).\label{eq:ZIRdef}
\end{equation}
We omitted the word of ``local'' since hereafter we focus only on these local matrix models.
The required relation is
\begin{equation}
Z(x,y)=\begin{cases}
e^{i\Theta}Z_{\HW}(x,y) & \left(\tilde{N}\geq0\right)\\
0 & \left(\tilde{N}<0\right)
\end{cases}.
\label{eq:ZcascLoc}
\end{equation}
The explicit form of the phase $e^{i\Theta}$ when $\ell>0$ and $k>0$ is in \eqref{eq:HWexact}.
In the next section, we prove this identity\footnote{
This has been partially done in \cite{Assel:2014awa} for $\tilde{N}>0$.
}.
Before that, we prepare for the proof and make a few comments.

First, we focus on the case when $\ell>0$ and $k>0$
because the results in the other cases can be easily derived from this case.
In fact, $\ell \rightarrow -\ell$ corresponds to take complex conjugate
and multiply a constant phase factor
while
$k\rightarrow -k $ corresponds to take complex conjugate with $\zeta \rightarrow -\zeta$.

Second, we introduce symbols $M$ and $L$ satisfying
\begin{\eq}
N-N_{\Right}=M=k+L, \quad
\tilde{N}=N_{\Left}-L .
\end{\eq}
Using these symbols, the Hanany-Witten move (\ref{eq:HW}) can be written as
\begin{align}
 & \left\langle N_{\Left}\bullet N\circ N_{\Right}\right\rangle =\left\langle N_{\Left}\bullet N_{\Right}+M\circ N_{\Right}\right\rangle =\left\langle N_{\Left}\bullet N_{\Right}+k+L\circ N_{\Right}\right\rangle \nonumber \\
 & \rightarrow\left\langle N_{\Left}\circ N_{\Left}-L\bullet N_{\Right}\right\rangle =\left\langle N_{\Left}\circ\tilde{N}\bullet N_{\Right}\right\rangle .
 \label{eq:D3label}
\end{align}
This notation helps us to prove smoothly.

Third, we are interested in the case when $N>N_{\Left}+k$ and $N>N_{\Right}+k$ because in this case $\tilde{N}<N_{\Right}$ and $\tilde{N}<N_{\Left}$, so that at least locally the lowest rank indeed decreases.
However, we loosen the former restriction because there is no reason to restrict from the technical point of view.
Therefore, we prove in the case when $M>k$, or equivalently, $L>0$.

\subsection{Proof of the prediction}
\label{subsec:Derivation}
In this section we prove (\ref{eq:ZcascLoc}) when $\ell>0$, $k>0$ and $L>0$.

We start with rewriting the partition function as
\begin{equation}
Z(x,y)=\frac{1}
       {N! \ell^{\frac{N_{\Left}+N}{2}}}\int d^{N} \mu\  
e^{\frac{i\pi k}{\ell}\left(\sum_{m}^{N_L}x_{m}^{2}-\sum_{n}^{N}\mu_{n}^{2}\right)}
Z^{\FIL}_{\ell,\zeta}\left(N_{\Left},N;x,\mu\right)
Z^{\FIL}_{1,\eta}\left(N,N_{\Right};\mu,y\right) ,
\label{eq:Z1}
\end{equation}
where $Z^{\FIL}_{\ell,\zeta}$ is the combination of the FI factors and the one-loop factors
\begin{equation}
Z^{\FIL}_{\ell,\zeta}\left(N_{1},N_{2};\alpha,\beta\right)
=e^{-\frac{2\pi i\zeta}{\ell}\left(\sum_{m}^{N_{1}}\alpha_{m}-\sum_{n}^{N_{2}}\beta_{n}\right)}\frac{\prod_{m<m'}^{N_{1}}2\sinh\frac{\pi\left(\alpha_{m}-\alpha_{m'}\right)}{\ell}\prod_{n<n'}^{N_{2}}2\sinh\frac{\pi\left(\beta_{n}-\beta_{n'}\right)}{\ell}}{\prod_{m}^{N_{1}}\prod_{n}^{N_{2}}2\cosh\frac{\pi\left(\alpha_{m}-\beta_{n}\right)}{\ell}}.
\label{eq:FILoopDef}
\end{equation}

\subsubsection{Quantum mechanics representation}
It is known that
for this type of integrals,
we can apply the technique developed in \cite{Marino:2011eh},
which regards the partition function as the one of an ideal Fermi gas system\footnote{
After the original work,the computation technique was gradually improved in \cite{Moriyama:2015asx,Moriyama:2016xin,Moriyama:2016kqi}.
This approach for local matrix model with general ranks was studied in \cite{Kubo:2019ejc,Kubo:2020qed}.
}.
In this technique,
we regard the factor \eqref{eq:FILoopDef}
as a combination of objects appearing in one particle quantum mechanics
with the canonical commutation relation
\begin{\eq}
\left[\hat{q},\hat{p}\right]=i\hbar ,\quad {\rm with}\ \  \hbar=\frac{\ell}{2\pi k} ,
\end{\eq}
where
$\hat{q}$ and $\hat{p}$ are Hermitian operators 
corresponding to position and momentum operators respectively.
We take normalizations of their eigenstates as
\begin{align}
 & \braket{q_{1}|q_{2}}=\delta\left(q_{1}-q_{2}\right),\quad\bbrakket{p_{1}|p_{2}}=\delta\left(p_{1}-p_{2}\right),\nonumber \\
 & \brakket{q|p}=\frac{1}{\sqrt{2\pi\hbar}}e^{\frac{i}{\hbar}pq},\quad\bbraket{p|q}=\frac{1}{\sqrt{2\pi\hbar}}e^{-\frac{i}{\hbar}pq} ,
 \label{eq:Normalization}
\end{align}
where $\ket{\cdot}$ denotes position eigenstate while $\kket{\cdot}$ denotes momentum one.

In terms of the quantum mechanics notation,
we prove in app.~\ref{app:det} that the factor \eqref{eq:FILoopDef} can be written as
\begin{\eq}
Z^{\FIL}_{\ell,\zeta}\left(N_{1},N_{2};\alpha,\beta\right)
=\tilde{Z}^{\FIL}_{\ell,\zeta}\left(N_{1},N_{2};\alpha,\beta\right),
\label{eq:CauchyDet}
\end{\eq}
where
\begin{align}
 & \tilde{Z}^{\FIL}_{\ell,\zeta}\left(N_{1},N_{2};\alpha,\beta\right)  \nonumber \\
 & \equiv\begin{cases}
\det\left(\begin{array}{c}
\left[A_{m,n}\right]_{m,n}^{N_{1}\times N_{2}}\\
\left[\sqrt{2\pi\hbar}\bbraket{\frac{2\pi\hbar}{\ell}\left(it_{N_2-N_1,j}-\zeta\right)|\beta_{n}}\right]_{j,n}^{(N_2-N_1)\times N_{2}}
\end{array}\right) & \left(N_{1}\leq N_{2}\right)\\
\det\left(\begin{array}{cc}
\left[A_{m,n}\right]_{m,n}^{N_{1}\times N_{2}} & \left[\sqrt{2\pi\hbar}\brakket{\alpha_{m}|-\frac{2\pi\hbar}{\ell}\left(it_{N_1-N_2,j}+\zeta\right)}\right]_{m,j}^{N_{1}\times (N_1-N_2)}\end{array}\right) & \left(N_{1}>N_{2}\right)
\end{cases},\nonumber\\
&A_{m,n}=\ell\braket{\alpha_{m}|\frac{1}{2\cosh\left(\frac{\ell}{2\hbar}\hat{p}+\frac{i\pi\left(N_{1}-N_{2}\right)}{2}+\pi\zeta\right)}|\beta_{n}},\nonumber\\
&t_{M,j}=\frac{M+1}{2}-j,
\end{align}
and $\left[f_{a,b}\right]_{a,b}^{A\times B}$ denotes an $A\times B$ matrix whose $\left(a,b\right)$ element is $f_{a,b}$.
We apply this  formula
to both $Z^{\FIL}$ in the integral \eqref{eq:Z1}.

Next, 
we rewrite the Gaussian 
factors in \eqref{eq:Z1} in terms of operators in the quantum mechanics:
\begin{align}
&Z(x,y)\nonumber\\
 &= \frac{1}{N!\ell^{\frac{N_{\Left}+N_{\Right}}{2}}k^{\frac{M}{2}}}\int d^{N_{\Left}}\alpha d^{N}\mu d^{N_{\Right}}\beta\prod_{n}^{N_{\Left}}\bra{x_{n}}e^{\frac{i}{2\hbar}\hat{q}^{2}}\ket{\alpha_{n}}\tilde{Z}^{\FIL}_{\ell,\zeta}\left(N_{\Left},N;\alpha,\mu\right)\nonumber \\
 &\ \ \  \times\det\left(\begin{array}{cc}
\left[\braket{\mu_{m}|e^{-\frac{i}{2\hbar}\hat{q}^{2}}\frac{1}{2\cosh\left(\frac{1}{2\hbar}\hat{p}+\frac{i\pi M}{2}+\pi\eta\right)}|\beta_{n}}\right]_{m,n}^{N\times N_{\Right}} & \left[\brakket{\mu_{m}|e^{-\frac{i}{2\hbar}\hat{q}^{2}}|-\frac{\ell}{k}\left(it_{M,j}+\eta\right)}\right]_{m,j}^{N\times M}\end{array}\right)\nonumber \\
 &\ \ \  \times\prod_{n}^{N_{\Right}}\braket{\beta_{n}|y_{n}} ,
\end{align}
where we have inserted
\begin{equation}
1=\int d^{N_{\Left}}\alpha\prod_{n}^{N_{\Left}}\braket{x_{n}|\alpha_{n}},\quad 1=\int d^{N_{\Right}}\beta\prod_{n}^{N_{\Right}}\braket{\beta_{n}|y_{n}} .
\label{eq:Z_QM}
\end{equation}

\subsubsection{Applying similarity transformations}
Now, all the integrals have the form of the identity
\begin{equation}
\int d\gamma\ket{\gamma}\bra{\gamma}= 1 ,
\end{equation}
and therefore 
the integral is invariant under the similarity transformations:
\begin{align}
\ket{\alpha_{n}}\bra{\alpha_{n}} & \rightarrow e^{\frac{i}{2\hbar}\hat{p}^{2}}\ket{\alpha_{n}}\bra{\alpha_{n}}e^{-\frac{i}{2\hbar}\hat{p}^{2}},\quad\ket{\mu_{n}}\bra{\mu_{n}}\rightarrow e^{\frac{i}{2\hbar}\hat{p}^{2}}\ket{\mu_{n}}\bra{\mu_{n}}e^{-\frac{i}{2\hbar}\hat{p}^{2}},\nonumber \\
\ket{\beta_{n}}\bra{\beta_{n}} & \rightarrow e^{\frac{i}{2\hbar}\hat{q}^{2}}e^{\frac{i}{2\hbar}\hat{p}^{2}}\ket{\beta_{n}}\bra{\beta_{n}}e^{-\frac{i}{2\hbar}\hat{p}^{2}}e^{-\frac{i}{2\hbar}\hat{q}^{2}}.
\label{eq:similarity}
\end{align}
The similarity transformations \eqref{eq:similarity}
slightly affect
the first FI and 1-loop factor $\tilde{Z}^{\FIL}_{\ell,\zeta}$ 
since it does not contain the position operator. 
It produces only the phase factor $e^{\frac{i\pi}{k\ell}\theta_{\zeta,N-N_{\Left}}}$, where
\begin{equation}
\theta_{\zeta,M}=-\frac{1}{12}\left(M^{3}-M\right)+M\zeta^{2}.
\end{equation}
On the other hand, they drastically change the components of the second determinant. 
Using 
\begin{\eq}
e^{-\frac{i}{2\hbar}\hat{p}^{2}}e^{-\frac{i}{2\hbar}\hat{q}^{2}}f\left(\hat{p}\right)e^{\frac{i}{2\hbar}\hat{q}^{2}}e^{\frac{i}{2\hbar}\hat{p}^{2}}=f\left(\hat{q}\right),
\quad
e^{-\frac{i}{2\hbar}\hat{p}^{2}}e^{-\frac{i}{2\hbar}\hat{q}^{2}}\kket p=\frac{1}{\sqrt{i}}e^{\frac{i}{2\hbar}p^{2}}\ket p,
\end{\eq}
all the momentum operators and the momentum eigenvectors become the position operators and the position eigenvectors, respectively. 
Note that this operation also provides the phase $i^{-\frac{1}{2}M}e^{\frac{i\pi \ell }{k}\theta_{\eta,M}}$.

Using the identity
\begin{align}
 & \frac{1}{N!}\int d^{N}\alpha\det\left(\begin{array}{cc}
\left[f_{m}\left(\alpha_{n}\right)\right]_{m,n}^{(N+M)\times N} & \left[f_{mj}'\right]_{m,j}^{(N+M)\times M}\end{array}\right)\det\left(\left[g_{n}\left(\alpha_{m}\right)\right]_{m,n}^{N\times N}\right)\nonumber \\
 & =\int d^{N}\alpha\det\left(\begin{array}{cc}
\left[f_{m}\left(\alpha_{n}\right)\right]_{m,n}^{(N+M)\times N} & \left[f_{mj}'\right]_{m,j}^{(N+M)\times M}\end{array}\right)\prod_{n}^{N}g_{n}\left(\alpha_{n}\right),\label{eq:Diagonalize}
\end{align}
we can diagonalize the second determinant,
and then we apply the determinant formula (\ref{eq:CauchyDet}) to $\tilde{Z}^{\FIL}_{\ell,\zeta}$ backwards. 
As a result, we find
\begin{align}
Z(x,y)= & \frac{i^{-\frac{1}{2}M}e^{\frac{i\pi }{k\ell}\theta_{\zeta,N-N_{\Left}}}e^{\frac{i\pi \ell }{k}\theta_{\eta,M}}}{\ell^{\frac{N_{\Left}+N_{\Right}}{2}}k^{\frac{M}{2}}}\nonumber \\
 & \times\int d^{N_{\Left}}\alpha d^{N_{\Right}}\beta\left(\prod_{n}^{N_{\Left}}\braket{x_{n}|e^{\frac{i}{2\hbar}\hat{q}^{2}}e^{\frac{i}{2\hbar}\hat{p}^{2}}|\alpha_{n}}\right)
 \mathcal{Z}(\alpha,\beta)
 \left(\prod_{n}^{N_{\Right}}\braket{\beta_{n}|e^{-\frac{i}{2\hbar}\hat{p}^{2}}e^{-\frac{i}{2\hbar}\hat{q}^{2}}|y_{n}}\right) ,
\label{eq:Z2}
\end{align}
where
\begin{align}
\mathcal{Z}(\alpha,\beta)= & \int d^{N}\mu\ Z^{\FIL}_{\ell,\zeta}\left(N_{\Left},N;\alpha,\mu\right)\nonumber \\
 & \times\left(\prod_{n}^{N_{\Right}}\braket{\mu_{n}|\frac{1}{2\cosh\left(\frac{1}{2\hbar}\hat{q}+\frac{i\pi M}{2}+\pi\eta\right)}|\beta_{n}}\right)
 \left( \prod_{j}^{M}\braket{\mu_{N_{\Right}+j}|-\frac{\ell}{k}\left(it_{M,j}+\eta\right)} \right).
\label{eq:Z0Def}
\end{align}

\subsubsection{Computation of $\mathcal{Z}(\alpha,\beta)$}
Here we rewrite $\mathcal{Z}(\alpha,\beta)$ 
in a way that a connection to the duality is transparent.
The final results will be (\ref{eq:SUSYbreak}) and (\ref{eq:Z0res}). 
In order to compute $\mathcal{Z}(\alpha,\beta)$, we should perform the integration over $\mu_{n}$. 
At first sight,
one might think that
we could perform the integration simply using the delta functions appeared in the second line. 
This is indeed true for $\mu_{n}$ $\left(1\leq n\leq N_{\Right}\right)$.
After the integration, all the $\mu_{n}$ are replaced with $\beta_{n}$.
On the other hand, it is no longer true for $\mu_{N_{\Right}+j}$ $\left(1\leq j\leq M\right)$ 
because the arguments of the delta functions are now complex numbers
and need to shift the integral contour appropriately.
We see this point carefully.

\subsubsection*{The subtlety on integrating over $\mu_{N_R +j}$}
The last factor in \eqref{eq:Z0Def} is
\[
\prod_{j}^{M}\braket{\mu_{N_{\Right}+j}|-\frac{\ell}{k}\left(it_{M,j}+\eta\right)} ,
\]
which gives $M$ delta functions with complex variables.
To make it the standard delta functions,
we should shift the integral contour from $\mathbb{R}$ to $\mathbb{R}-\frac{\ell}{k}t_{M,j}i$. 
If there are poles inside the region surrounded by the integral contour,
then we have to take account of their residues.

Let us elaborate on this point for a moment.
We are interested in the following type of integral\footnote{
A similar integral was studied in \cite{Assel:2014awa}.
}
\begin{equation}
\int_{\mathbb{R}}dxf\left(x\right)\braket{x|a+ib},
\end{equation}
where $a$ and $b$ are real numbers\footnote{
Strictly speaking, we should assume that $f\left(x\right)=o\left(e^{-\left|x\right|}\right)$,
otherwise we can not control the behavior at the infinity.
To avoid this, we implicitly give $k$ a small negative imaginary part
and take it to zero in the end.
}.
We consider the case when $f\left(x\right)$ has poles in $0<\mathrm{Im}\left(x\right)<b$ or $b<\mathrm{Im}\left(x\right)<0$ when $b>0$ or $b<0$ respectively. 
$z_{i}$ denotes the position of poles, and $R_{i}$ denotes its residue. Exceptionally, if $\mathrm{Im}\left(z_{j}\right)=b$, $R_{i}$ denotes the half of its residue. 
We can perform the integration by inserting the identity operator $\int_{\mathbb{R}}dp\kket p\bbra p$ and shifting the integration contour from $\mathbb{R}$ to $\mathbb{R}+ib$ so that we can use the property of the delta function:
\begin{align}
  \int_{\mathbb{R}}dxf\left(x\right)\braket{x|a+ib}\nonumber 
 & =\frac{1}{2\pi}\int_{\mathbb{R}}dpdxf\left(x\right)e^{ip\left(x-a-ib\right)}\nonumber \\
 & =\frac{1}{2\pi}\left(\int_{\mathbb{R}}dpdxf\left(x+ib\right)e^{ip\left(x-a\right)}+2\pi i\mathrm{sgn}\left(b\right)\sum_{j}R_{j}\int_{\mathbb{R}}dpe^{ip\left(z_{j}-a-ib\right)}\right)\nonumber \\
 & =f\left(a+ib\right)+2\pi i\mathrm{sgn}\left(b\right)\sum_{j}R_{j}\braket{z_{j}|a+ib}.
\label{eq:CompDelta}
\end{align}

Coming back to the integral \eqref{eq:Z0Def},
we should study the structure of the distribution of the poles.
The part of the function of $\mu_{N_{\Right}+j}$ having poles is 
\[
\prod_{n}^{N_{\Left}} \frac{1}{2\cosh\frac{\pi\left(\alpha_{n}-\mu_{N_{\Right}+j}\right)}{\ell} }, 
\]
which has poles at $\mu_{N_{\Right}+j}=\alpha_{n}+\left(m-\frac{1}{2}\right)i\ell $ with residues $\left(-1\right)^{m}\frac{i\ell }{2\pi}$ with an integer $m$.
Thus the residue terms actually appear since we are considering $M>k$.

\subsubsection*{Integrating over $\mu_{N_R +j}$ for even $L$}
First, we integrate over $\mu_{n}$ 
where
$t_{M,j}$ satisfies $\left|t_{M,j}\right|<\frac{k}{2}$. 
This is straightforward because the integral contour does not pass any poles. 
Therefore, all $\mu_{n}$ are replaced with $-\frac{\ell}{k}\left(it_{M,j}+\eta\right)$.
Second, we perform the integration over $\mu_{n}$ 
with $\frac{k}{2}\leq\left|t_{M,j}\right|<\frac{3k}{2}$. 
However, when there exist $t_{M,j}$ such that $\left|t_{M,j}\right|=\frac{k}{2}$, 
additional subtleties appear although it is not essential. 
To avoid this, we assume that $L$ is even for a while, and we study the odd $L$ case soon later. 
For $t$ satisfying $\frac{k}{2}<\left|t\right|<\frac{3k}{2}$,
using (\ref{eq:CompDelta}) and $\braket{x|y}=\braket{x+z|y+z}$,
we obtain
\begin{align}
 & \int d\mu\frac{f\left(\mu\right)}{\prod_{n}^{N_{\Left}}2\cosh\frac{\pi\left(\alpha_{n}-\mu\right)}{\ell}}\braket{\mu|-\frac{\ell}{k}\left( it +\eta\right)}\nonumber \\
 & =
 \frac{f\left(-\frac{\ell}{k}\left( it +\eta\right)\right)}
 {\prod_{n}^{N_{\Left}}2\cosh\left(\frac{\pi}{\ell}\alpha_{n}+\frac{\pi}{k}\left( it+\eta\right)\right)}
 +\ell\sum_{n}^{N_{\Left}}
 \frac{f\left(\alpha_{n}\mp\frac{i}{2}\ell \right)}
 {\prod_{n'\neq n}^{N_{\Left}}2\cosh\frac{\pi\left(\alpha_{n'}
    -\alpha_{n}\pm\frac{i}{2}\ell \right)}{\ell}}
 \braket{\alpha_{n}|-\frac{\ell}{k}\left(\left(t\mp\frac{k}{2}\right)i+\eta\right)},
 \label{eq:Recidue1}
\end{align}
where $\mathrm{sgn}\left(t\right)=\pm1$. 
This identity tells us that 
integrating over $\mu_n$ under consideration
is the same as summing over all the cases 
where $\mu_n$ is replaced with $-\frac{\ell}{k}\left( it_{M,j}+\eta\right)$ 
and $\alpha_{n'}\mp\frac{i}{2}\ell $
with multiplying the appropriate factors.
However, not all the terms in the summation survive.
We therefore focus on which terms survive for a while and postpone the concrete calculation until \eqref{eq:Z01}.

We first realize that 
even if one $\mu_n$ is replaced with $-\frac{\ell}{k}\left( it_{M,j} +\eta\right)$, 
the term vanishes thanks to the factor $\prod_{n<n'}^{N}2\sinh\frac{\pi\left(\mu_{n}-\mu_{n'}\right)}{\ell}$ at the numerator of $Z^{\FIL}_{\ell,\zeta}\left(N_{\Left},N;\alpha,\mu\right)$ in (\ref{eq:Z0Def}). 
This is because there must exist 
$\mu_{n'}$ that is replaced by $-\frac{\ell}{k}\left( it_{M,j\pm k}+\eta\right)$ 
with $\pm=\mathrm{sgn}\left(t_{M,j}\right)$ 
since $\left|t_{M,j\pm k}\right|<\frac{k}{2}$, and $\left|t_{M,j}-t_{M,j\pm k}\right|=k$. 
Therefore, it is sufficient to consider the cases
where all $\mu_n$'s are replaced by $\alpha_{n'}\mp\frac{i}{2}\ell $.
Among these cases, 
there are further cases that give vanishing contributions.
As a conclusion, nonzero contributions come from 
only the cases where
all the chosen $\alpha_{n'}$'s are different.
This is because when $\mu_{n}$ is replaced with $\alpha_{n'}\mp\frac{i}{2}\ell $, 
all the trigonometric functions including $\alpha_{n'}$ cancel with the denominator and the numerator of $Z^{\FIL}_{\ell,\zeta}\left(N_{\Left},N;\alpha,\mu\right)$ in (\ref{eq:Z0Def}),
thus there are no longer poles at $\alpha_{n'}\mp\frac{i}{2}\ell$.

Third, let us perform the integration over $\mu_{n}$ 
with $\frac{3k}{2}<\left|t_{M,j}\right|<\frac{5k}{2}$. 
The only difference from the previous case is that 
$\mu_{n}$ can be also replaced with $\alpha_{n'}\mp\frac{3i}{2}\ell $ 
because the integration contour passes new poles. 
However, these cases do not contribute because of the following reason. 
When $\mu_{n}$ is replaced with $\alpha_{n'}\mp\frac{3i}{2}\ell $, 
$\mu_{n\pm k}$ should have been replaced with $\alpha_{n''}\mp\frac{i}{2}\ell$ 
since corresponding $t_{M,j\pm k}$ satisfies 
$\frac{k}{2}<\left|t_{M,j\pm k}\right|<\frac{3k}{2}$. 
There is also the case
in which $\mu_{n}$ is replaced with $\alpha_{n''}\mp\frac{3i}{2}\ell$, 
$\mu_{n\pm k}$ is replaced with $\alpha_{n'}\mp\frac{i}{2}\ell $ and another part is the same. 
These two cases actually cancel out. In fact, the remaining $\alpha_{n'}$ dependence and the remaining $\alpha_{n''}$ dependence are
\begin{align}
 & \braket{\alpha_{n'}|-\frac{\ell}{k}\left(\left(t_{M,j}\mp\frac{3}{2}k\right)i+\eta\right)}\braket{\alpha_{n''}|-\frac{\ell}{k}\left(\left(t_{M,j\pm k}\mp\frac{1}{2}k\right)i+\eta\right)}\nonumber \\
 & =\braket{\alpha_{n'}|-\frac{\ell}{k}\left(\left(t_{M,j\pm k}\mp\frac{1}{2}k\right)i+\eta\right)}\braket{\alpha_{n''}|-\frac{\ell}{k}\left(\left(t_{M,j\pm k}\mp\frac{1}{2}k\right)i+\eta\right)},
\end{align}
so that this part is symmetric under the exchange of $\alpha_{n'}$ and $\alpha_{n''}$. 
On the other hand, the factor $\left(-1\right)$ appears 
from the trigonometric functions under the exchange.

Continuing the above arguments,
we conclude that $\mu_{n}$ 
with $\frac{k}{2}<\left|t_{M,j}\right|$ 
should be replaced with $\alpha_{n'}\mp\frac{i}{2}\ell $, 
and all the chosen $\alpha_{n'}$ should be different. 
In other words, cases not satisfying this condition vanish.
It is worth noting that the number of $\mu_n$ where corresponding $t_{M,j}$ satisfies $\frac{k}{2}<\left|t_{M,j}\right|$ is $L$.
Therefore, if $\tilde{N}<0$, the above condition is never satisfied, which means that $\mathcal{Z}(\alpha,\beta)=0$.
This is, in fact, also true for the odd $L$ case, as we will see just below.
We assume that $\tilde{N}\geq0$ for a while and return to this point soon later.

\subsubsection*{Integrating over $\mu_{N_R +j}$ for odd $L$}
When $L$ is odd, there exist $t_{M,j}=\pm\frac{k}{2}$ and 
we should be careful on these cases. 
We denote the corresponding $\mu_{N_{\Right}+j}$ by $\mu_{\pm}$. 
First, the cases where $\mu_{\pm}=-\frac{\ell}{k}\left(\pm\frac{i}{2}k +\eta\right)$ vanish 
by virtue of $\prod_{n<n'}^{N}2\sinh\frac{\pi\left(\mu_{n}-\mu_{n'}\right)}{\ell}$ at the numerator of $Z^{\FIL}_{\ell,\zeta}\left(N_{\Left},N;\alpha,\mu\right)$.
The terms where $\mu_{\pm}$ is replaced with $\alpha_{n_{\pm}}\mp\frac{i}{2}\ell $ 
also vanish even if all the chosen $\alpha_{n}$ are different. 
This is 
because there is cancellation between this type of term,
and the one where $\mu_{\pm}$ is replaced with $\alpha_{n_{\mp}}\mp\frac{i}{2}\ell $.
The only remaining terms are the terms where $\mu_{+}=\alpha_{n'}-\frac{i}{2}\ell $ and $\mu_{-}=+\frac{i}{2}\ell $ and the terms where $\mu_{+}=-\frac{i}{2}\ell $ and $\mu_{-}=\alpha_{n'}+\frac{i}{2}\ell $. 
We should also notice that the $\frac{1}{2}$ factor appears in this case as explained above \eqref{eq:CompDelta}.
However, the contributions from these two cases are equal, thus the weight of the contribution of the odd $L$ case is the same as the even $L$ case.

We show that the two types of contributions are indeed equal. 
The first case
\begin{equation}
\mu_{+}=\alpha_{n'}-\frac{i}{2}\ell ,\quad\mu_{-}=+\frac{i}{2}\ell ,
\end{equation}
accords with the second case
\begin{equation}
\mu_{+}=-\frac{i}{2}\ell ,\quad\mu_{-}=\alpha_{n'}+\frac{i}{2}\ell ,\label{eq:Lodd1}
\end{equation}
by two operations, 
namely $\mu_{\pm}\rightarrow\mu_{\mp}$ and $\pm\frac{i}{2}\ell \rightarrow\mp\frac{i}{2}\ell$. 
The first operation provides $\left(-1\right)$ since $\mathcal{Z}(\alpha,\beta)$ is anti-symmetric under the exchange of two $\mu_{n}$. 
The second operation also provides many $\left(-1\right)$ factors 
from the $\cosh$ and $\sinh$ factors in $Z^{\FIL}_{\ell,\zeta}$. 
Although the number of the $\cosh$ and $\sinh$ factors related to $\mu_{-}=\alpha_{n'}-\frac{i}{2}\ell \rightarrow\alpha_{n'}+\frac{i}{2}\ell $ and $\mu_{-}=+\frac{i}{2}\ell \rightarrow-\frac{i}{2}\ell $ is almost the same, 
the former is smaller than the latter by 1 
due to the absence of the $n=n'$ in the denominator in (\ref{eq:Recidue1}). 
Therefore, the two types of contributions are equal. 
Hereafter we focus on the second case (\ref{eq:Lodd1}) without the half factor.

\subsubsection*{The $\tilde{N}<0$ case}
Again, we should choose $L$ different $\alpha_{n'}$'s and replace $\mu_{n}$ with $\alpha_{n'}\pm\frac{i}{2}\ell $ 
as is the case with even $L$. 
Therefore, regardless of whether $L$ is even or odd, if $\tilde{N}<0$,
we find
\begin{equation}
\mathcal{Z}(\alpha,\beta)=0 \quad \left(\tilde{N}<0\right).
\label{eq:SUSYbreak}
\end{equation}
This immediately leads to
\begin{equation}
Z(x,y)=0\quad\left(\tilde{N}<0\right).
\end{equation}
This is nothing but the case when $\tilde{N}<0$ in (\ref{eq:ZcascLoc}), which embodies the modified s-rule.
Hereafter we again assume that $\tilde{N}\geq0$.

\subsubsection*{Substitution for $\mu_n$}
So far we have learned that 
there are contributions only from the terms where $\mu_{n}$ 
is replaced with $\beta_{n}$ for $1\leq n \leq N_\Right$, $-\frac{\ell}{k} \left( it_{M,j}+\eta\right)$ for $|t_{M,j}|<\frac{k}{2}$ 
(and one $|t_{M,j}|=\frac{k}{2}$ for odd $L$)
and $\alpha_{n}\mp\frac{i}{2}\ell $ for $|t_{M,j}|>\frac{k}{2}$ 
(and the other $|t_{M,j}|=\frac{k}{2}$ for odd $L$).
Therefore, we obtain
\begin{align}
\mathcal{Z}(\alpha,\beta)= & \frac{\ell^{L}}{\tilde{N}!}\sum_{\sigma\in S_{N_{\Left}}}e^{-\frac{2\pi i\zeta}{\ell}\left(\sum_{m}^{N_{\Left}}\alpha_{m}-\sum_{n}^{N}\mu_{n}\right)}\frac{\prod_{m<m'}^{N_{\Left}}2\sinh\frac{\pi\left(\alpha_{m}-\alpha_{m'}\right)}{\ell}\prod_{n<n'}^{N}2\sinh\frac{\pi\left(\mu_{n}-\mu_{n'}\right)}{\ell}}{\left(\prod_{m}^{N_{\Left}}\prod_{n}^{N}\right)'2\cosh\frac{\pi\left(\alpha_{m}-\mu_{n}\right)}{\ell}}\nonumber \\
 & \times\prod_{n}^{N_{\Right}}\frac{1}{2\cosh\left(\frac{1}{2\hbar}\beta_{n}+\frac{i\pi M}{2}+\pi\eta\right)}
 \prod_{j}^{L}\braket{\alpha_{\sigma\left(\tilde{N}+j\right)}| 
    -\frac{\ell}{k}\left( it_{L,j} +\eta\right)}
    ,
 \label{eq:Z01}
\end{align}
where
\begin{equation}
\mu_{n}=\begin{cases}
\beta_{n} & \left(1\leq n\leq N_{\Right}\right)\\
\alpha_{\sigma\left(n-N_{\Right}+\tilde{N}\right)}-\frac{i}{2}\ell  
      & \left(N_{\Right}+1\leq n\leq N_{\Right}+\frac{L}{2}-c_L\right)\\
-\frac{\ell}{k}\left(\left(t_{k,n-N_{\Right}-\frac{L}{2}+c_L}+c_L\right)i+\eta\right)
   & \left(N_{\Right}+\frac{L}{2}-c_L+1\leq n\leq N_{\Right}+\frac{L}{2}-c_L+k\right)\\
\alpha_{\sigma\left(n-N_{\Right}-k+\tilde{N}\right)}+\frac{i}{2}\ell  
  & \left(N_{\Right}+\frac{L}{2}-c_L+k+1\leq n\leq N\right)
\end{cases},\label{eq:Rep}
\end{equation}
and
\begin{align}
c_{L} & =\begin{cases}
0 & \left(\rm{even}\ L\right)\\
\frac{1}{2} & \left(\rm{odd}\ L\right)
\end{cases}.
\end{align}
The symbol $\left(\prod_{m}^{N_{\Left}}\prod_{n}^{N}\right)'$ means that 
$m$ and $n$ do not run where the factor
$2\cosh\frac{\pi\left(\alpha_{n}-\left(\alpha_{n}\mp\frac{i}{2}\ell \right)\right)}{\ell}$ appears. 
$S_{N_{\Left}}$ is the permutation group and 
we have divided by $\tilde{N}!$ because of the over-counting.

It is convenient to express all the subscripts of $\alpha$ in terms of $\sigma$:
\begin{align}
&\mathcal{Z}(\alpha,\beta) \nonumber\\
&=\frac{\ell^{L}}{\tilde{N}!}  \sum_{\sigma\in S_{N_{\Left}}}\mathrm{sgn}\left(\sigma\right)e^{-\frac{2\pi i\zeta}{\ell}\left(\sum_{m}^{N_{\Left}}\alpha_{\sigma\left(m\right)}-\sum_{n}^{N}\mu_{n}\right)}\frac{\prod_{m<m'}^{N_{\Left}}2\sinh\frac{\pi\left(\alpha_{\sigma\left(m\right)}-\alpha_{\sigma\left(m'\right)}\right)}{\ell}\prod_{n<n'}^{N}2\sinh\frac{\pi\left(\mu_{n}-\mu_{n'}\right)}{\ell}}{\left(\prod_{m}^{N_{\Left}}\prod_{n}^{N}\right)'2\cosh\frac{\pi\left(\alpha_{\sigma\left(m\right)}-\mu_{n}\right)}{\ell}}\nonumber \\
 &\ \ \ \ \ \  \times\prod_{n}^{N_{\Right}}\frac{1}{2\cosh\left(\frac{1}{2\hbar}\beta_{n}+\frac{i\pi M}{2}+\pi\eta\right)}
 \prod_{j}^{L}\braket{\alpha_{\sigma\left(\tilde{N}+j\right)}|-\frac{\ell}{k}\left( it_{L,j} +\eta\right)}.
 \label{eq:Z02}
\end{align}
Surprisingly, after substituting (\ref{eq:Rep}) into (\ref{eq:Z02}) and calculation, all the trigonometric functions including $\alpha_{N_{\Left}-L+j}$ at the denominator and the numerator completely cancel out, and all the FI factors including $\alpha_{N_{\Left}-L+j}$ also completely cancel out. 
Therefore, $\alpha_{n'}$ appears only at the argument of the delta functions coming from (\ref{eq:Recidue1}).

Although we have not treated the phases in detail, 
from now on we also study the phases carefully. 
First, the factor $i^{-kL}$ appears in the $\prod_{n<n'}^{N}2\sinh\frac{\pi\left(\mu_{n}-\mu_{n'}\right)}{\ell}$ 
with one $-\frac{\ell}{k}\left( it +\eta\right)$ and one $\alpha\pm\frac{i}{2}\ell $. 
Second, $\left(-1\right)^{-\frac{1}{4}L^{2}}$ appears for even $L$ 
in the same factor with one $\alpha-\frac{i}{2}\ell $ 
and one $\alpha+\frac{i}{2}\ell $.
$\left(-1\right)^{\frac{L-1}{2}\frac{L+1}{2}}$ also appears for odd $L$ though this is equal to $1$.
Third, $\left(-1\right)^{\frac{1}{2}L\left(L-1\right)}$ appears when all the trigonometric functions including only $\alpha_{N_{\Left}-L+j}$ at the denominator and the numerator cancel out. 
Fourth, for odd $L$, $i^{N_{\Left}-1}$ and $i^{-N_{\Right}}$ appear in the denominator and the numerator respectively since in this case the number of the trigonometric functions including only $\alpha-\frac{i}{2}\ell $ is one smaller than that of the trigonometric functions including only $\alpha+\frac{i}{2}\ell $. 
Using $\left(-1\right)^{-\frac{1}{4}L^{2}+\frac{1}{2}L\left(L-1\right)}=1$ for even $L$ and 
$\left(-1\right)^{\frac{1}{2}\left(L-1\right)+\frac{1}{2}L\left(L-1\right)}=1$ for odd $L$, 
all the phases are
\begin{equation}
\Omega=i^{-kL}\left(-1\right)^{c_L\left(\tilde{N}-N_{\Right}\right)}.
\end{equation}
Finally, we obtain the phase from the FI factor. Note that the summation of the imaginary part of $\mu_{n}$ completely cancels out.

We then obtain
\begin{align}
\mathcal{Z}(\alpha,\beta)= & 
\frac{\Omega e^{-2\pi i\zeta\eta}\ell^{L}}{\tilde{N}!}
\sum_{\sigma\in S_{N_{\Left}}}\mathrm{sgn}\left(\sigma\right)e^{-\frac{2\pi i\zeta}{\ell}\left(\sum_{m}^{\tilde{N}}\alpha_{\sigma\left(m\right)}-\sum_{n}^{N_{\Right}}\beta_{n}\right)}\nonumber \\
 & \times \frac{\prod_{m<m'}^{\tilde{N}}2\sinh\frac{\pi\left(\alpha_{\sigma\left(m\right)}-\alpha_{\sigma\left(m'\right)}\right)}{\ell}\prod_{n<n'}^{N_{\Right}}2\sinh\frac{\pi\left(\beta_{n}-\beta_{n'}\right)}{\ell}}{\prod_{m}^{\tilde{N}}\prod_{n}^{N_{\Right}}2\cosh\frac{\pi\left(\alpha_{\sigma\left(m\right)}-\beta_{n}\right)}{\ell}}\nonumber \\
 & \times\frac{\prod_{n}^{N_{\Right}}\prod_{j}^{k}2\sinh\left[\frac{\pi}{\ell}\beta_{n}-\frac{\pi}{k}\left(\left(t_{k,j}-c_{L}\right)i+\eta\right)\right]\prod_{j<j'}^{k}2\sinh\frac{\pi i}{k}\left(t_{k,j'}-t_{k,j}\right)}{\prod_{n}^{\tilde{N}}\prod_{j}^{k}2\cosh\left[\frac{\pi}{\ell}\alpha_{\sigma\left(n\right)}-\frac{\pi}{k}\left(\left(t_{k,j}-c_{L}\right)i+\eta\right)\right]}\nonumber \\
 & \times\prod_{n}^{N_{\Right}}\frac{1}{2\cosh\left(\frac{1}{2\hbar}\beta_{n}+\frac{i\pi M}{2}+\pi\eta\right)}\prod_{j}^{L}\braket{\alpha_{\sigma\left(\tilde{N}+j\right)}|-\frac{\ell}{k}\left( it_{L,j} +\eta\right)}.
\end{align}
We can rewrite this expression using the following identities shown in app.~\ref{sec:Formulas},
\begin{equation}
\prod_{j<j'}^{k}2\sinh\frac{2\pi i\left(t_{k,j'}-t_{k,j}\right)}{2k}=i^{-\frac{1}{2}k\left(k-1\right)}k^{\frac{k}{2}},\label{eq:ZCSdual}
\end{equation} 
and
\begin{align}
\prod_{j}^{k}2\sinh\left[\frac{\pi}{\ell}\beta-\frac{\pi}{k}\left(\left(t_{k,j}-c_{L}\right)i+\eta\right)\right] 
& =i^{-k}\left(-1\right)^{\frac{1}{2}L-c_{L}}2\cosh\left(\frac{1}{2\hbar}\beta+\frac{i\pi M}{2}+\pi\eta\right),\nonumber \\
\prod_{j}^{k}2\cosh\left[\frac{\pi}{\ell}\alpha-\frac{\pi}{k}\left(\left(t_{k,j}-c_{L}\right)i+\eta\right)\right] 
& =\left(-1\right)^{\frac{1}{2}L-c_{L}}2\cosh\left(\frac{1}{2\hbar}\alpha+\frac{i\pi L}{2}+\pi\eta\right),
\end{align}
coming from
\begin{\eq}
2\cosh\frac{x}{2}  =\prod_{j}^{k}2\cosh\frac{x-2\pi i t_{k,j}}{2k},
\quad
i^{-k}2\cosh\frac{x+i\pi k}{2}  =\prod_{j}^{k}2\sinh\frac{x-2\pi i t_{k,j}}{2k} .
\label{eq:TrigDual}
\end{\eq}
Then we find
\begin{align}
\mathcal{Z}(\alpha,\beta)= & 
\frac{\Omega'e^{-2\pi i\zeta\eta}\ell^{L}k^{\frac{1}{2}k} }{\tilde{N}!}
\NN\\
 & \times \sum_{\sigma\in S_{N_{\Left}}}\mathrm{sgn}\left(\sigma\right) \left(\prod_{n}^{\tilde{N}}\frac{1}{2\cosh\left(\frac{1}{2\hbar}\alpha_{\sigma\left(n\right)}+\frac{i\pi L}{2}+\pi\eta\right)}\right)
 \left(\prod_{j}^{L}\braket{\alpha_{\sigma\left(\tilde{N}+j\right)}|-\frac{\ell}{k}\left( it_{L,j} +\eta\right)}\right)\nonumber \\
 & \times Z^{\FIL}_{\ell,\zeta}\left(\tilde{N},N_{\Right};\alpha_{\sigma},\beta\right),
\end{align}
where
\begin{equation}
\Omega'
=i^{-kL-kN_{\Right}-\frac{1}{2}k\left(k-1\right)+L\left(\tilde{N}-N_{\Right}\right)}
=i^{\frac{1}{2}\left( N_\Left^2-N^2-\tilde{N}^2+N_\Right^2+k \right)}.
\end{equation}
We insert the integration
\begin{equation}
\prod_{n}^{\tilde{N}}\frac{1}{2\cosh\left(\frac{\alpha_{\sigma\left(n\right)}}{2\hbar}+\frac{i\pi L}{2}+\pi\eta\right)}=\int d^{\tilde{N}}\nu\prod_{n}^{\tilde{N}}
\braket{\alpha_{\sigma\left(n\right)}|\frac{1}{2\cosh\left(\frac{\hat{q}}{2\hbar}+\frac{i\pi L}{2}+\pi\eta\right)}|\nu_{n}}.
\end{equation}
The summation over the permutation can be expressed by the determinant. Using (\ref{eq:CauchyDet}) to the second factor, we obtain 
\begin{align}
\mathcal{Z}(\alpha,\beta)= &
\frac{\Omega'e^{-2\pi i\zeta\eta}k^{\frac{1}{2}M}\ell^{\frac{1}{2}L}}{\tilde{N}!}
\int d^{\tilde{N}}\nu\nonumber \\
 & \times\det\left(\begin{array}{cc}
\left[
\braket{\alpha_{m}|\frac{1}{2\cosh\left(\frac{\hat{q}}{2\hbar}+\frac{i\pi L}{2}+\pi\eta\right)}|\nu_{n}}
\right]_{m,n}^{N_{\Left}\times\tilde{N}} & 
\left[
\sqrt{2\pi \hbar}
\braket{\alpha_{m}|-\frac{\ell}{k}\left( it_{L,j}+\eta\right)}\right]_{m,j}^{N_{\Left}\times L}\end{array}\right)\nonumber \\
 & \times\tilde{Z}^{\FIL}_{\ell,\zeta}\left(\tilde{N},N_{\Right};\nu,\beta\right) 
 \quad\quad\quad\quad\quad\quad\quad\quad\quad\quad\quad\quad\quad\quad\quad\quad\quad\quad\quad\quad\quad\left(\tilde{N}\geq0\right).
 \label{eq:Z0res}
\end{align}

\subsubsection{Return to the whole part}
Now we return to $Z(x,y)$ in (\ref{eq:Z2}). 
By substituting the above result, we obtain
\begin{align}
Z(x,y)= & \frac{\Omega'i^{-\frac{1}{2}M}e^{\frac{i\pi }{k\ell}\theta_{\zeta,N-N_{\Left}}}e^{\frac{i\pi \ell}{k}\theta_{\eta,M}}e^{-2\pi i\zeta\eta}}{\tilde{N}!\ell^{\frac{\tilde{N}+N_{\Right}}{2}}}
\int d^{N_{\Left}}\alpha d^{\tilde{N}}\nu d^{N_{\Right}}\beta
\prod_{n}^{N_{\Left}}\braket{x_{n}|e^{\frac{i}{2\hbar}\hat{q}^{2}}e^{\frac{i}{2\hbar}\hat{p}^{2}}|\alpha_{n}}\nonumber \\
 & \times\det\left(\begin{array}{cc}
\left[
\braket{\alpha_{\sigma\left(n\right)}|\frac{1}{2\cosh\left(\frac{\hat{q}}{2\hbar}+\frac{i\pi L}{2}+\pi\eta\right)}|\nu_{n}}
\right]_{m,n}^{N_{\Left}\times\tilde{N}} & 
\left[
\sqrt{2\pi \hbar}
\braket{\alpha_{m}|-\frac{\ell}{k}\left( it_{L,j} +\eta\right)}\right]_{m,j}^{N_{\Left}\times L}\end{array}\right)
\nonumber \\
 & \times\tilde{Z}^{\FIL}_{\ell,\zeta}\left(\tilde{N},N_{\Right};\nu,\beta\right)
 \prod_{n}^{N_{\Right}}\braket{\beta_{n}|e^{-\frac{i}{2\hbar}\hat{p}^{2}}e^{-\frac{i}{2\hbar}\hat{q}^{2}}|y_{n}}.
\end{align}
After the similarity transformations
\begin{align}
\ket{\alpha_{n}}\bra{\alpha_{n}} & \rightarrow e^{-\frac{i}{2\hbar}\hat{p}^{2}}e^{-\frac{i}{2\hbar}\hat{q}^{2}}\ket{\alpha_{n}}\bra{\alpha_{n}} e^{\frac{i}{2\hbar}\hat{q}^{2}}e^{\frac{i}{2\hbar}\hat{p}^{2}},\quad
\ket{\nu_{n}}\bra{\nu_{n}}\rightarrow e^{-\frac{i}{2\hbar}\hat{p}^{2}}e^{-\frac{i}{2\hbar}\hat{q}^{2}}\ket{\nu_{n}}\bra{\nu_{n}}e^{\frac{i}{2\hbar}\hat{q}^{2}}e^{\frac{i}{2\hbar}\hat{p}^{2}},\nonumber \\
\ket{\beta_{n}}\bra{\beta_{n}} & \rightarrow e^{-\frac{i}{2\hbar}\hat{p}^{2}}e^{-\frac{i}{2\hbar}\hat{q}^{2}}\ket{\beta_{n}}\bra{\beta_{n}}e^{\frac{i}{2\hbar}\hat{q}^{2}}e^{\frac{i}{2\hbar}\hat{p}^{2}},
\end{align}
and using 
\begin{\eq}
e^{\frac{i}{2\hbar}\hat{q}^{2}}e^{\frac{i}{2\hbar}\hat{p}^{2}}f\left(\hat{q}\right)e^{-\frac{i}{2\hbar}\hat{p}^{2}}e^{-\frac{i}{2\hbar}\hat{q}^{2}}=f\left(\hat{p}\right),
\quad
e^{\frac{i}{2\hbar}\hat{q}^{2}}e^{\frac{i}{2\hbar}\hat{p}^{2}}\ket{q} =\sqrt{i}e^{-\frac{i}{2\hbar}q^{2}}\kket q,
\end{\eq}
we finally arrive at
\begin{align}
Z(x,y)= 
& i^{\frac{1}{2}\left( N_\Left^2-N^2-\tilde{N}^2+N_\Right^2 \right)}
e^{\frac{i\pi }{k\ell}\left(
\theta_{\zeta,N-N_{\Left}}-\theta_{\zeta,N_{\Right}-\tilde{N}}
\right)}
e^{\frac{i\pi \ell }{k}\left(
\theta_{\eta,M}-\theta_{\eta,L}
\right)}
e^{-2\pi i\zeta\eta }\nonumber \\
 & \times\int\frac{d^{\tilde{N}}\nu}{\tilde{N}!}Z_{\left(1,0\right),\eta}\left(N_{\Left},\tilde{N};x,\nu\right)Z_{\left(\ell,k\right),\zeta}\left(\tilde{N},N_{\Right};\nu,y\right)\quad\quad\quad\left(\tilde{N}\geq0\right).
 \label{eq:HWexact}
\end{align}
This identity and (\ref{eq:SUSYbreak}) are what we expected (\ref{eq:ZcascLoc}).

\section{Gluing SUSY theories and new dualities}
\label{sec:gluing}
In this section,
we propose new dualities
motivated by the structures of supersymmetric partition functions and brane construction.

\subsection{Gluing $\mathcal{N}=2$ SUSY theories}
\label{sec:gluing_sub}
Let us consider a 3d $\mathcal{N}=2$ supersymmetric theory $\mathcal{T}$ with a flavor symmetry $F$
which may or may not have a Lagrangian.
We also turn on the background vector multiplet $V_F$ of the symmetry $F$.
If the theory $\mathcal{T}$ has Lagrangian,
then the partition function of $\mathcal{T}$ is schematically written as
\begin{\eq}
Z_\mathcal{T}[V_F] = \int D\Phi\  e^{-S_\mathcal{T} [\Phi ,V_F ]} ,
\end{\eq}
where $\Phi$ is dynamical fields in $\mathcal{T}$
and $S_\mathcal{T} [\Phi ,V_F ]$ is the action of $\mathcal{T}$ 
in the presence of the background vector multiplet $V_F$.
When the theory $\mathcal{T}$ is a non-Lagrangian theory,
we do not have a path integral representation of the partition function
but we can still denote the partition function as a functional of $V_F$.

\begin{figure}[t]
\begin{center}
\includegraphics[width=80mm]{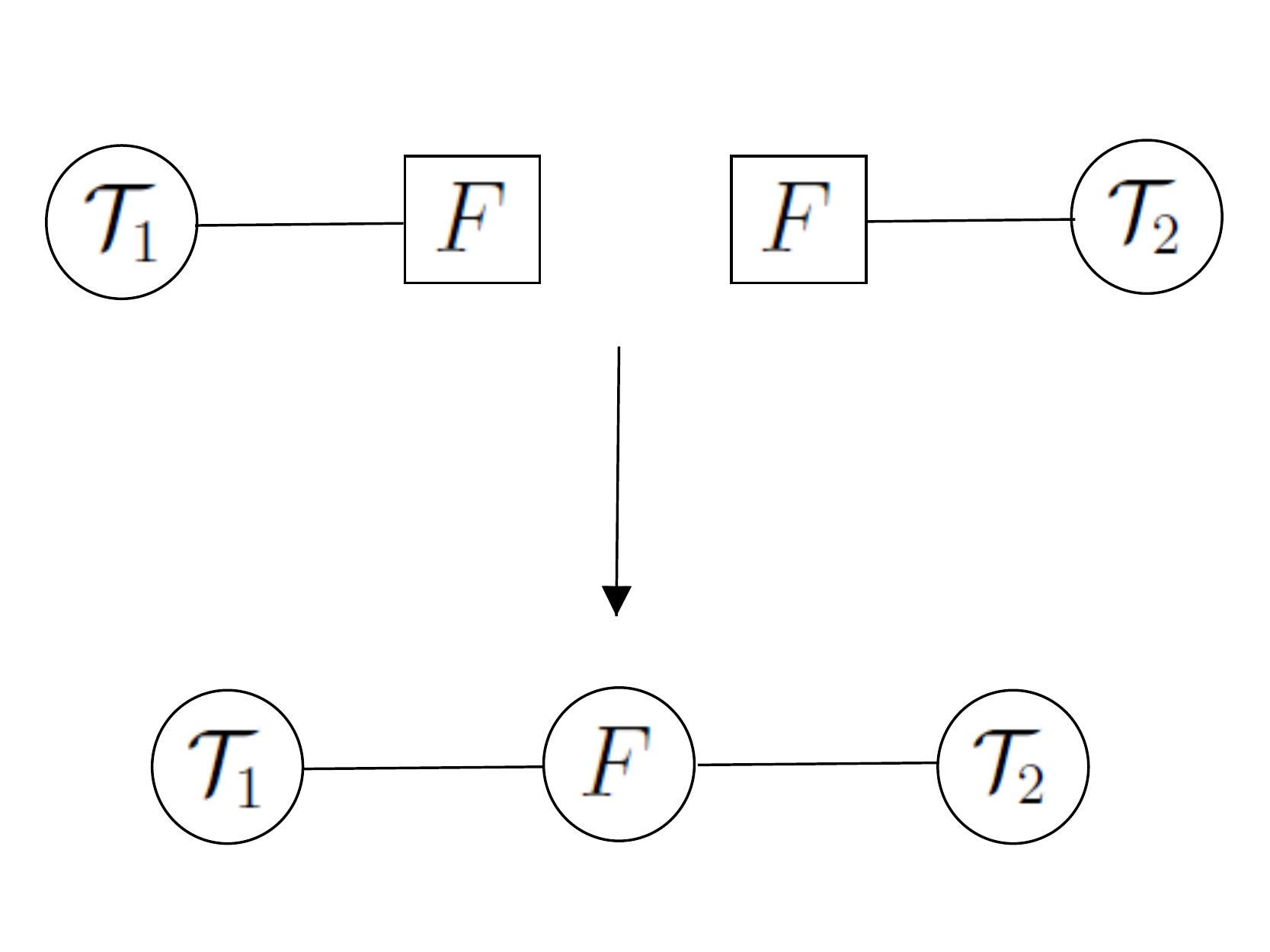}
\caption{
Gluing two theories $\mathcal{T}_1$ and $\mathcal{T}_2$ 
by identifying and gauging the common flavor symmetry $F$.
}
 \label{fig:gluing}
  \end{center}
\end{figure}
Suppose that we have two theories $\mathcal{T}_1$ and $\mathcal{T}_2$
with a common flavor symmetry $F$.
Then we can glue the two theories by identifying and gauging $F$ (see also fig.~\ref{fig:gluing}):
\begin{\eq}
Z_{\mathcal{T}_1 \circ \mathcal{T}_2 }  
=\int DV_F\ e^{-S_{\rm YMCS} [V_F] }
   Z_{\mathcal{T}_1} [V_F ] \cdot  Z_{\mathcal{T}_2} [V_F ] ,
\end{\eq}
where $S_{\rm YMCS} [V_F]$ is the action of 3d $\mathcal{N}=2$ SUSY YM CS theory 
with the gauge group $F$ promoted from global symmetry\footnote{
If we would like to preserve larger supersymmety,
then we replace the YMCS theory with the one with that supersymmetry.
For example, if we would like to preserve $\mathcal{N}=3$ SUSY,
then we should also add the adjoint chiral multiplet of $F$ with $U(1)_R$ charge 1.
}.
The path integral is over all the possible configuration of the vector multiplet $V_F$
and therefore gluing the two theories in general seems to require 
information on the two theories for any configuration of $V_F$.
This sounds technically difficult to do
since $V_F$ may take non-supersymmetric configuration.
However, it is known that for a class of problems including $S^3$ partition functions,
we can restrict $V_F$ to a class of supersymmetric configurations. 

\subsubsection*{Localization formula for $S^3$ partition functions of 3d $\mathcal{N}=2$ theories}
Let us focus on $S^3$ partition functions of 3d $\mathcal{N}=2$ SUSY theories with Lagrangians.
Using SUSY localization,
the path integral is dominated by a saddle point
where matter fields are trivial and vector multiplet takes the configuration 
\begin{\eq}
\sigma = {\rm const.} ,\quad D= -\sigma ,\quad ({\rm others})=0,
\quad ({\rm up\ to\ gauge\ transformation}) ,
\label{eq:saddle}
\end{\eq}
where $\sigma$ is the adjoint scalar and 
$D$ is the auxiliary field in the vector multiplet\footnote{
More precisely, here we are applying so-called Coulomb branch localization.
There is also another type of localization called Higgs branch localization \cite{Fujitsuka:2013fga,Benini:2013yva}.
}.
Therefore the exact result can be written as integration over
constant configurations of $\sigma$.

As a result,
the localization formula of the theory with the gauge group $G$ is given by \cite{Kapustin:2009kz,Hama:2010av,Jafferis:2010un}
\begin{\eq}
Z_{S^3}
=\frac{1}{|\mathcal{W}|} \int d^{|G|}x\ Z_{\rm cl} (x) Z_{\rm 1loop} (x) ,
\end{\eq}
where $|G|$ is the rank of $G$
and $|\mathcal{W}|$ is rank\footnote{
In particular, $|\mathcal{W}|=N!$ for $G=U(N)$.
} of the Weyl group of $G$.
If the gauge group has the structure $G=G_1 \times G_2 \times \cdots$,
then the classical contribution $Z_{\rm cl}$ is
\begin{\eq}
Z_{\rm cl} (x)
= \prod_j e^{i\pi k_j  {\rm tr}_{G_j} (x^2 )  -2\pi i\zeta_j {\rm tr}_{G_j}(x)   } ,
\end{\eq}
where $k_j$ and $\zeta_j$ are the Chern-Simons level and FI-parameter\footnote{
If $G_j$ does not contain a $U(1)$ part, then $\zeta_j$ is zero.
} of the gauge subgroup $G_j$,
and this part is independent of matters.
When the theory is coupled to chiral multiplets in representations $\{ \mathbf{R_1}, \mathbf{R_2},\cdots \}$
with $U(1)_R$ charges $\{ \Delta_1 ,\Delta_2 ,\cdots \} $, 
the one-loop determinant $Z_{\rm 1loop}$ is given by
\begin{\eq}
 Z_{\rm 1loop}(x ) 
= \frac{\prod_{\alpha \in {\rm root}_+ }
 4\sinh^2{(\pi \alpha \cdot \sigma )}  }
   {\prod_{p} \prod_{\rho_p \in \mathbf{R_p}}  
   s_1 \left( \rho_p \cdot \sigma -i(1-\Delta_p ) \right)} , 
\end{\eq}
where $\alpha$ is root vector of $G$,
$\rho_m$ is weight vector of representation $\mathbf{R}_m$,
and
\begin{\eq}
s_b (z)
\equiv \prod_{p=0}^\infty \prod_{q=0}^\infty
\frac{pb +qb^{-1}+(b+b^{-1})/2 -iz}{pb +qb^{-1}+(b+b^{-1})/2 +iz} .
\end{\eq}
Note that the localization formula is independent of Yang-Mills couplings 
because of $Q$-exactness of the SYM terms.
This implies that
the localization formula captures the IR limit corresponding to the large YM coupling limit
of the theories 
even if we start with actions including the SYM term.

When we have a flavor symmetry $F$,
one can also turn on the background vector multiplet $V_F$ of the symmetry $F$ 
while preserving SUSY by taking the fields in $V_F$
to be the configuration \eqref{eq:saddle}.
For this case, 
the one-loop determinant becomes
\begin{\eq}
 Z_{\rm 1loop}(x ,m) 
= \frac{\prod_{\alpha \in {\rm root}_+ }
 4\sinh^2{(\pi \alpha \cdot \sigma )}  }
   {\prod_{p} \prod_{\rho_p \in \mathbf{R_p}}  
   s_1 \left( \rho_p \cdot \sigma +\rho_p^F \cdot m -i(1-\Delta_p ) \right)} , 
\end{\eq}
where $m$ is the background value of $\sigma$ in $V_F$ (often called real mass)
and $\rho_p^F$ is the weight vector of a representation in $F$ 
to which the $p$-th chiral multiplet belongs.
Then the partition function becomes dependent on $m$.

The above formula is for Lagrangian theories.
For non-Lagrangian theories, it is harder to compute their partition functions
but we can still denote them as functions of $m$.
For instance, this is true for the partition function \eqref{eq:local_start} of the local theory
even for the cases without Lagrangians.

\subsubsection*{Gluing two theories for $S^3$ partition functions}
The above localization formula indicates that
for the $S^3$ partition functions of 3d $\mathcal{N}=2$ SUSY theories,
we can glue two theories by integrating over only
the SUSY configurations \eqref{eq:saddle}:
\begin{\eq}
Z_{\mathcal{T}_1 \circ \mathcal{T}_2 } 
=\int d^{|F|} m\  Z_{V_F}(m)   Z_{\mathcal{T}_1} (m) \cdot  Z_{\mathcal{T}_2} (m) ,
\label{eq:glue_S3}
\end{\eq}
where $Z_{V_F}(m) $ is the classical and one-loop contribution of the 3d $\mathcal{N}=2$ YMCS theory.
Thus the gluing procedure is much simplified for the supersymmetric $S^3$ partition functions.

\subsection{New dualities from gluing the local theory to another theory}
\begin{figure}[t]
\begin{center}
\includegraphics[width=120mm]{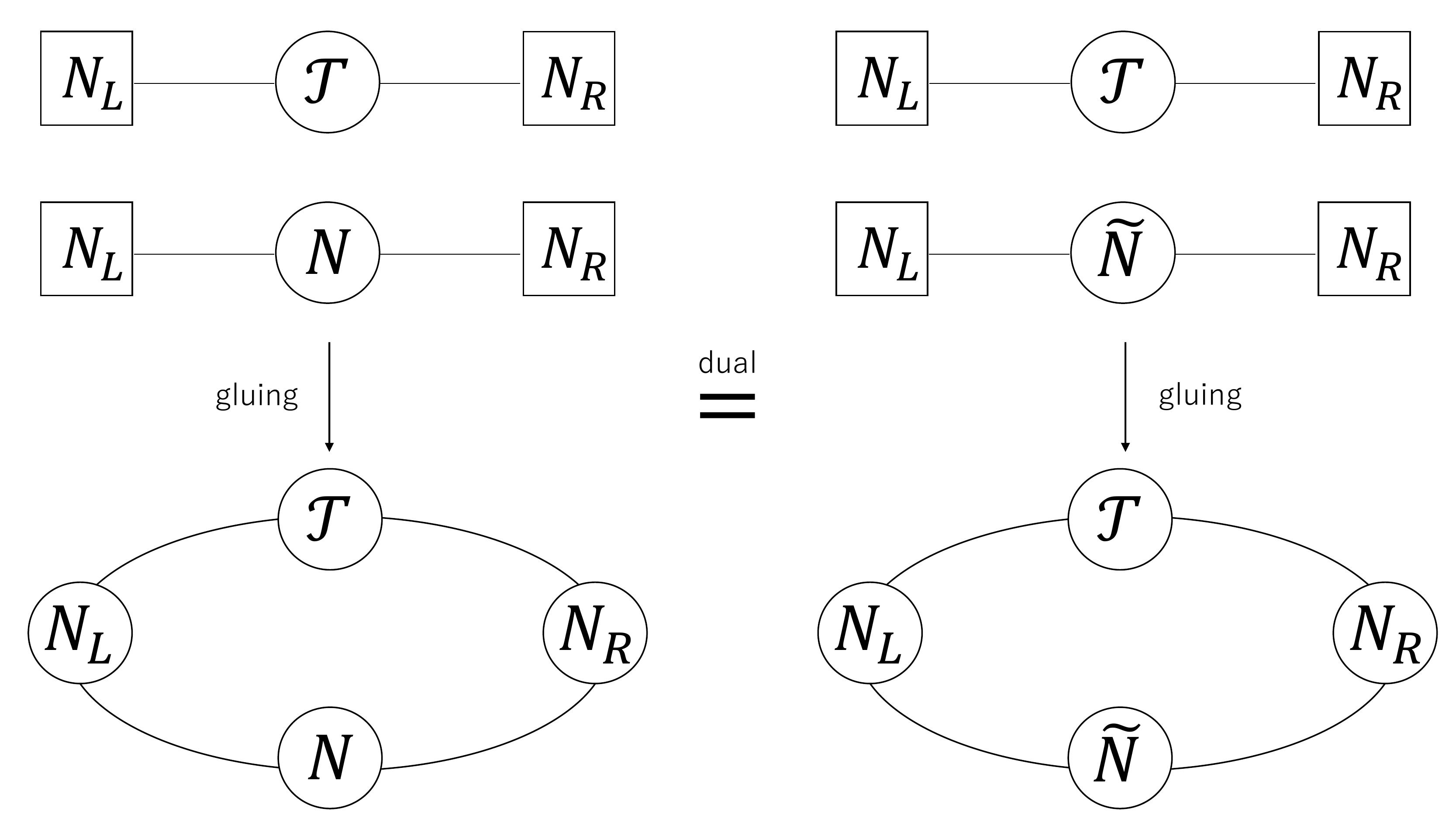}
\caption{
Illustration of generating new dualities when the local theory has a Lagrangian.
The duality of the local theory implies
new dualities by gluing theories with the common flavor symmetries.
Gluing the two theories with two common flavor symmetries.
}
\label{fig:new_duality1}
 \end{center}
\end{figure}

The above structure motivates us to find new dualities as follows.
Let us prepare the two theories:
a theory $\mathcal{T}$ with $U(N_L ) \times U(N_R )$ global symmetry
and the local theory described in fig.~\ref{fig:local}.
Then we glue them and obtain the $S^3$ partition function of the glued theory by the procedure \eqref{eq:glue_S3}.
This procedure is illustrated on the left of fig.~\ref{fig:new_duality1}.
Let us replace the local theory with its dual theory as illustrated on the right of fig.~\ref{fig:new_duality1}.
Then the duality for the local theory proven in the last section and the gluing structure implies that
the two glued theories are dual to each other in IR.

\begin{figure}[t]
\begin{center}
\includegraphics[width=120mm]{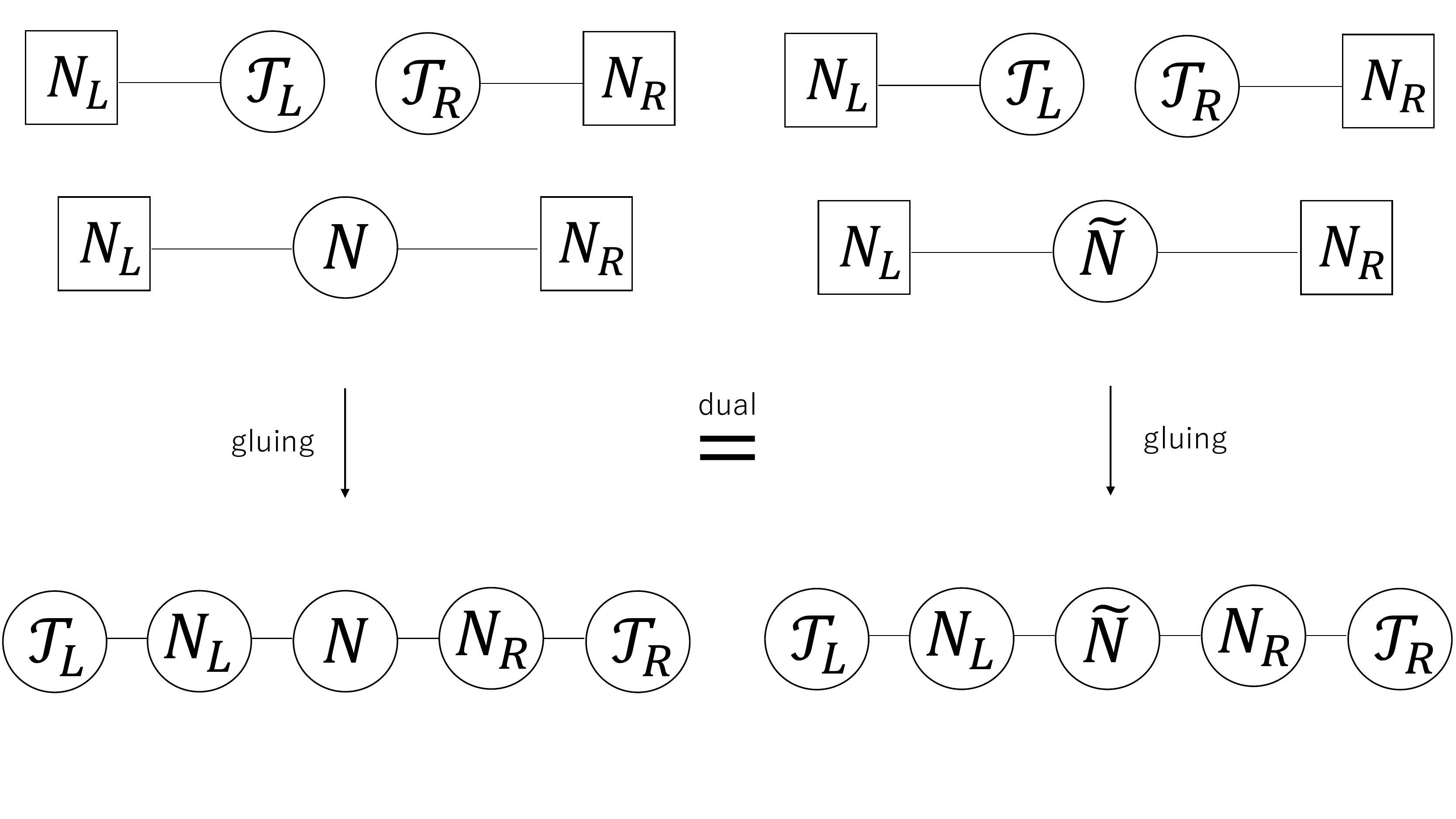}
\caption{
Gluing the three theories.
The duality of the local theory implies the duality of the glued theory as well.
}
\label{fig:new_duality2}
 \end{center}
\end{figure}

We can generate another type of new dualities in a similar way.
For this time, we glue the three theories:
a theory $\mathcal{T}_L$ with $U(N_L )$ global symmetry, a theory $\mathcal{T}_R$ with $U(N_R )$ global symmetry and the local theory.
We can obtain the $S^3$ partition function of the glued theory as well
Then the duality of the local theory implies the new duality for the glued theory.
This is illustrated on the left of fig.~\ref{fig:new_duality2}.
Although we have shown the above two types of dualities only for the $S^3$ partition functions,
we conjecture that these are true for more general observables.

\subsubsection*{Brane viewpoint}
In the brane configuration shown in fig.~\ref{fig:brane},
we have taken the 5-branes to be oblique in the $x^3-x^7$, $x^4-x^8$ and $x^5-x^9$ planes with the same angle.
The condition on the angles is required to preserve $\mathcal{N}=3$ supersymmetry.
If we relax this condition, then the preserved supersymmetry is reduced \cite{Kitao:1998mf,Bergman:1999na}:
We have $\mathcal{N}=2$ SUSY if two of the three angles are the same, and
$\mathcal{N}=1$ SUSY if all of them are independent.

Let us consider the less supersymmetric type of brane configurations which includes the one for the local theory
drawn in fig.~\ref{fig:local2}.
Although the total system generically has only 3d $\mathcal{N}=1$ SUSY,
applying the Hanany-Witten move to the local theory part implies another duality for the total system.
For the $\mathcal{N}\geq 2$ cases,
the duality for the $S^3$ partition functions of the total system is reduced to
the one of the local theory because of the structure of the $S^3$ partition functions which we have explained.
For the $\mathcal{N}=1$ cases,
it is unclear whether or not the $S^3$ partition functions have similar structures 
as in the $\mathcal{N}\geq 2$ cases,
and therefore showing the duality for the local theory may not be sufficient to show the one for the total system.
So we leave it as only a conjecture.
Similarly, if there remain anti-D3-branes after the Hanany-Witten move of the local theory,
we expect that supersymmetry of the total system is broken.

\subsection{New duality cascades?}
In sec.~\ref{sec:cascade}, we have seen that
generic Hanany-Witten type brane configuration is a combination of the brane configuration for the local theory and
applying the Hanany-Witten move for 5-branes along the circle leads us to the duality cascades for a range of the parameters.
In sec.~\ref{subsec:LocalMM}, we have seen the similar structures for the $S^3$ partition functions 
purely from the viewpoint of the field theories.

In the current case, generic theory cannot be regarded as a combination of the local theory
and therefore only the duality for the local theory is not sufficient to discuss duality cascades.
In other words, we may be able to find examples of new duality cascades
if we have some duality relations beyond the local theory.
A natural example of such dualities is the one coming from the Hanany-Witten move generalized 
to the less supersymmetric brane configurations explained above.
We expect that theories coming from the generalized brane configurations 
can be regarded as a combination of something like the local theory extended to the less supersymmetric case.
A first step to see this would be to focus on the $\mathcal{N}=2$ case and
extract a kind of $\mathcal{N}=2$ version of the local theory.

\section{Conclusion and Discussions}
\label{sec:conclusion}
In this paper 
we have studied the conjectural duality cascades 
in the 3d $\mathcal{N}=3$ supersymmetric gauge theories coming from the Hanany-Witten brane configurations.
We have made the non-perturbative tests of the duality cascades using supersymmetry localization.
We have focused on the $S^3$ partition functions and show that 
the duality cascades are generated by the duality relation coming from the Hanany-Witten effect for the local theory.
Then we have proven the fundamental duality relation for the local theory and
this amounts to show the predictions from the duality cascades for the $S^3$ partition functions.
We have also discussed that
our result for the local theory can be used to generate the new dualities for more general theories including less supersymmetric theories.
It would be interesting to extend our results to other observables.

Our non-perturbative tests for the partition functions have been done for the brane configurations consisting of 5-branes and D3-branes.
However, there are more general configurations, such as including orientifold 3-planes  \cite{Hosomichi:2008jb,Aharony:2008gk}.
It would be interesting 
if one can generalize our analysis to these cases 
(see e.g.~\cite{Argurio:2017upa} for the 4d case).
In addition, 
the authors in \cite{Karasik:2019bxn} have recently proposed that
a duality cascade occurs in a 4d non-supersymmetric gauge theory.
It would be illuminating to seek a 3d counterpart of this story.

In this paper, we have focused primarily on the IR structure of the dual theories.
On the other hand, the duality cascade for the Klebanov-Strassler theory was originally studied by directly seeing the behavior of the mass and the coupling constants along the RG flow \cite{Klebanov:2000hb}.
They also found its gravity dual, which indeed describes the smooth RG flow.
The similar analysis was performed for 3d ABJ-type theories from the string and the supergravity viewpoint \cite{Aharony:2009fc,Evslin:2009pk}.
Therefore, it is important to perform the same analysis for generic cases we studied in this paper to find the structure of RG flow in detail.

It is known that the matrix models associated to a class of brane configurations possessing $\mathcal{N}=4$ SUSY are characterized by operators called quantum curves \cite{Kashaev:2015wia, Kubo:2020qed} by using the Fermi gas formalism \cite{Marino:2011eh}.
The quantum curves with genus one are expected to have affine Weyl symmetry of exceptional type Lie algebra because this type of quantum curves is related to $q$-Painlev\'{e} equations \cite{Bonelli:2017gdk}.
In fact, the Weyl symmetry was explicitly constructed for these quantum curves \cite{Kubo:2018cqw, Moriyama:2020lyk}, and they are related to dualities between different rank theories related by the Hanany-Witten effect \cite{Kubo:2019ejc,Furukawa:2020cjp}.
On the other hand, the translation symmetry, which plays a crucial role in lifting the Weyl symmetry to affine version, has not been found yet.
Because the duality cascade relates an infinite number of ranks, there is a possibility that the duality cascade is related to the translation symmetry.

\subsection*{Acknowledgement}
M.H. is partially supported by MEXT Q-LEAP.
N.K. is supported by Grant-in-Aid for JSPS Fellows No.20J12263.

\appendix
\section{Proof of the determinant formula \eqref{eq:CauchyDet}}
\label{app:det}
In this appendix we prove the determinant identity \eqref{eq:CauchyDet}.
We write down the both side for convenience:
\begin{align}
 &e^{-\frac{2\pi i\zeta}{\ell}\left(\sum_{m}^{N_{1}}\alpha_{m}-\sum_{n}^{N_{2}}\beta_{n}\right)}\frac{\prod_{m<m'}^{N_{1}}2\sinh\frac{\pi\left(\alpha_{m}-\alpha_{m'}\right)}{\ell}\prod_{n<n'}^{N_{2}}2\sinh\frac{\pi\left(\beta_{n}-\beta_{n'}\right)}{\ell}}{\prod_{m}^{N_{1}}\prod_{n}^{N_{2}}2\cosh\frac{\pi\left(\alpha_{m}-\beta_{n}\right)}{\ell}}  \nonumber \\
 & =\begin{cases}
\det\left(\begin{array}{c}
\left[A_{m,n}\right]_{m,n}^{N_{1}\times N_{2}}\\
\left[\sqrt{2\pi\hbar}\bbraket{\frac{2\pi\hbar}{\ell}\left(it_{N_2-N_1,j}-\zeta\right)|\beta_{n}}\right]_{j,n}^{\left(N_2-N_1\right)\times N_{2}}
\end{array}\right) & \left(N_{1}\leq N_{2}\right)\\
\det\left(\begin{array}{cc}
\left[A_{m,n}\right]_{m,n}^{N_{1}\times N_{2}} & \left[\sqrt{2\pi\hbar}\brakket{\alpha_{m}|-\frac{2\pi\hbar}{\ell}\left(it_{N_1-N_2,j}+\zeta\right)}\right]_{m,j}^{N_{1}\times \left(N_1-N_2\right)}\end{array}\right) & \left(N_{1}>N_{2}\right)
\end{cases},
\label{eq:DetForm2}
\end{align}
where
\begin{equation*}
t_{M,j}=\frac{M+1}{2}-j,
\quad
A_{m,n}=\ell\braket{\alpha_{m}|\frac{1}{2\cosh\left(\frac{\ell}{2\hbar}\hat{p}+\frac{\pi\left(N_{1}-N_{2}\right)}{2}i+\pi\zeta\right)}|\beta_{n}}.
\end{equation*}
Note that 
although we relate $\ell$ and $\hbar$ in (\ref{eq:FILoopDef}), 
they are independent of each other in this formula.
We only derive the first case $N_{1}\leq N_{2}$ because the second case $N_{1}>N_{2}$ can be easily derived from the first identity. 

We start with the formula which can be regarded as the combination of the Cauchy determinant formula and the Vandermonde determinant formula \cite{Matsumoto:2013nya,Assel:2014awa}:
\begin{align}
\frac{\prod_{m<m'}^{N_{1}}2\sinh\frac{\pi\left(\alpha_{m}-\alpha_{m'}\right)}{\ell}\prod_{n<n'}^{N_{2}}2\sinh\frac{\pi\left(\beta_{n}-\beta_{n'}\right)}{\ell}}{\prod_{m}^{N_{1}}\prod_{n}^{N_{2}}2\cosh\frac{\pi\left(\alpha_{m}-\beta_{n}\right)}{\ell}} & =\left(-1\right)^{N_{1}\left(N_{2}-N_{1}\right)}\det\left(\begin{array}{c}
\left[\frac{e^{\frac{\pi M}{\ell}\left(\alpha_{m}-\beta_{n}\right)}}{2\cosh\frac{\pi\left(\alpha_{m}-\beta_{n}\right)}{\ell}}\right]_{m,n}^{N_{1}\times N_{2}}\\
\left[e^{\frac{2\pi}{\ell}t_{M,j}\beta_{n}}\right]_{j,n}^{\left(N_{2}-N_{1}\right)\times N_{2}}
\end{array}\right).\label{eq:Det0}
\end{align}
Using the Fourier transform formula
\begin{equation}
\frac{1}{2\cosh\pi x}=\frac{1}{2\pi}\int_{\mathbb{R}}dp\frac{e^{ipx}}{2\cosh\frac{p}{2}},
\end{equation}
the upper part of the matrix in the right-hand side can be written as
\begin{align}
\frac{e^{\frac{\pi M}{\ell}\left(\alpha-\beta\right)}}{2\cosh\frac{\pi\left(\alpha-\beta\right)}{\ell}} & =\frac{1}{2\pi}\int_{\mathbb{R}}dp\frac{e^{\frac{i}{\ell}\left(p-i\pi M\right)\left(\alpha-\beta\right)}}{2\cosh\frac{p}{2}}\nonumber \\
 & =\frac{1}{2\pi}\int_{\mathbb{R}}dp\frac{e^{\frac{i}{\ell}p\left(\alpha-\beta\right)}}{2\cosh\frac{p+i\pi M}{2}}+\sum_{j}^{\left\lfloor \frac{M+1}{2}\right\rfloor }\left(-1\right)^{j+1}e^{\frac{2\pi}{\ell}t_{M,j}\left(\alpha-\beta\right)}.
\end{align}
At the second line, we shifted the integration contour from $\mathbb{R}$ to $\mathbb{R}+i\pi M$.
In this process, the residues of $\left(2\cosh\frac{p}{2}\right)^{-1}$,
which has poles at $p=\left(2j-1\right)\pi i$ with residues $\left(-1\right)^{j}i$,
appeared.
However, these terms do not contribute to the determinant in (\ref{eq:Det0}) because these terms are just the linear combinations of the lower part of the matrix. 
Now, all the elements can be expressed in terms of vectors and operators in the quantum mechanics using (\ref{eq:Normalization}):
\begin{align}
\frac{1}{2\pi}\int_{\mathbb{R}}dp\frac{e^{\frac{i}{\ell}p\left(\alpha-\beta\right)}}{2\cosh\frac{p+i\pi M}{2}} & =\ell\braket{\alpha_{n}|\frac{1}{2\cosh\left(\frac{\ell}{2\hbar}\hat{p}+\frac{i\pi M}{2}\right)}|\beta_{m}},\nonumber \\
e^{\frac{2\pi}{\ell}t_{M,j}\beta_{m}} & =\sqrt{2\pi\hbar}\bbraket{\frac{2\pi\hbar}{\ell}t_{M,j}i|\beta_{m}}.
\end{align}

Next, we take the FI terms $e^{-\frac{2\pi i\zeta}{\ell}\left(\sum_{m}^{N_{1}}\alpha_{m}-\sum_{n}^{N_{2}}\beta_{n}\right)}$ into account.
We put all the FI terms between the bras and the kets, which is equal to change all the position eigenvectors as follows:
\begin{equation}
\bra{\alpha_{m}}\rightarrow\bra{\alpha_{m}}e^{-\frac{2\pi i\zeta}{\ell}\hat{q}},\quad\ket{\beta_{n}}\rightarrow e^{\frac{2\pi i\zeta}{\ell}\hat{q}}\ket{\beta_{n}}.
\end{equation}
Since these operators can be used to shift the momentum operators and the momentum eigenvectors using
\begin{equation}
e^{-\frac{ia}{\hbar}\hat{q}}f\left(\hat{p}\right)e^{\frac{ia}{\hbar}\hat{q}}=f\left(\hat{p}+a\right),
\quad
e^{\frac{ia}{\hbar}\hat{q}}\kket p=\kket{p+a},
\end{equation}
we finally arrive at (\ref{eq:DetForm2}), or equivalently, (\ref{eq:CauchyDet}).

\section{Useful formula on trigonometric functions}
\label{sec:Formulas}
In this section, we prove the two types of the identities of the trigonometric functions. The first identities are \eqref{eq:TrigDual}:
\begin{align}
2\cosh\frac{x}{2} & =\prod_{j}^{k}2\cosh\frac{x-2\pi it_{k,j}}{2k}, \nonumber \\
i^{-k}2\cosh\frac{x+i\pi k}{2} & =\prod_{j}^{k}2\sinh\frac{x-2\pi i t_{k,j}}{2k},
\label{eq:TrigDual0}
\end{align}
and the second identity is \eqref{eq:ZCSdual}:
\begin{equation}
\prod_{j<j'}^{k}2\sinh\frac{2\pi i\left(t_{k,j'}-t_{k,j}\right)}{2k}=i^{-\frac{1}{2}k\left(k-1\right)}k^{\frac{k}{2}},
\label{eq:ZCSdual0}
\end{equation}
where $k$ is a positive integer.

We start with the well-known identity
\begin{equation}
z^{k}-1=\prod_{j}^{k}\left(z-e^{\frac{2\pi j}{k}i}\right).
\end{equation}
By substituting $z=e^{\frac{1}{k}\left(x+\pi i\right)}$, we obtain 
\begin{equation}
-e^{\frac{x}{2}}2\cosh\frac{x}{2}=\prod_{j}^{k}\left(i^{-1}e^{\frac{1}{2k}\left(x+\pi i+ 2\pi ij\right)}2\cosh\frac{x+2\pi it_{k,j}}{2k}\right).
\end{equation}
This identity immediately leads to the first line of (\ref{eq:TrigDual0}). The second line of (\ref{eq:TrigDual0}) can be obtained by replacing $x$ to $x+i\pi k$.

Next, we focus on (\ref{eq:ZCSdual0}). The second line of (\ref{eq:TrigDual0}) leads to
\begin{equation}
1=i^k\frac{\prod_{j}^{k}2\sinh\frac{x-2\pi it_{k,j}}{2k}}{2\cosh\frac{x+i\pi k}{2}}.
\end{equation}
Substituting $x=2\pi it_{k,j'}+\epsilon$ and multiplying $j'=1,2,\ldots,k$, we obtain
\begin{equation}
1=i^{k^2}\left(\prod_{j'}^{k}\frac{2\sinh\frac{\epsilon}{2k}}{2\cosh\frac{2\pi t_{k,j'}i+i\pi k +\epsilon}{2}}\right) \times (-1)^{\frac{1}{2}k(k+1)}\left(\prod_{j<j'}^{k}2\sinh\frac{2\pi i \left(t_{k,j'}-t_{k,j}\right)+\epsilon}{2k}\right)^2.
\end{equation}
Taking $\epsilon\rightarrow 0$, we arrive at (\ref{eq:ZCSdual0}).

\bibliographystyle{utphys}
\bibliography{References}

\end{document}